\documentstyle[psfig]{mn}

\newif\ifAMStwofonts
\AMStwofontstrue

\newcommand{\mv}{\mbox{$M_{V}$}}
\newcommand{\mi}{\mbox{$M_{I}$}}
\newcommand{\mimax}{\mbox{$M_{I}^{\rm RC}$}}
\newcommand{\vi}{\mbox{$V\!-\!I$}}
\newcommand{\feh}{\mbox{[Fe/H]}}
\newcommand{\Msun}{\mbox{$M_{\odot}$}}
\newcommand{\sub}[1]{\mbox{$_{\rm #1}$}}
\newcommand{\Mi}{\mbox{$m\sub{i}$}}
\newcommand{\Mto}{\mbox{$m\sub{TO}$}}
\newcommand{\Teff}{\mbox{$T\sub{eff}$}}
\newcommand{\diff}{\mbox{d}}

\title[Distances from the Red Giant Clump]
        {Population effects on the Red Giant Clump absolute magnitude,
and distance determinations to nearby galaxies}
\author[L\'eo Girardi \& Maurizio Salaris]
       {L\'eo Girardi$^{1,3}$ and Maurizio Salaris$^{2,3}$ \\
$^1$Dipartimento di Astronomia, Universit\'a di Padova, 
	Vicolo dell'Osservatorio 5, I-35122 Padova, Italy \\
$^2$Astrophysics Research Institute, Liverpool John Moores
        University, Twelve Quays House, Egerton Wharf, Birkenhead CH41 1LD, 
	UK \\
$^3$Max-Planck-Institut f\"ur Astrophysik, 
	Karl-Schwarzschild-Str.\ 1, D-85741 Garching bei M\"unchen,
	Germany } 
\date{Accepted 2000 ???.
      Received 2000, July 19;
      in original form 2000 ???}

\pagerange{\pageref{firstpage}--\pageref{lastpage}}
\pubyear{2000}

\begin{document}

\maketitle

\label{firstpage}

\begin{abstract}

The red giant clump has been recently argued to be a reliable
distance indicator for the galaxies in the Local Group.
The accuracy of distance determinations based on this method, 
however, depends on the possible presence of systematic magnitude
differences ($\Delta M_I^{\rm RC}$) between the local clump revealed by the 
{\em Hipparcos} colour-magnitude diagram, and the clump stars 
observed in distant galaxies.

In this paper, we re-address the problem of these systematic 
`population' effects.
First, we present tables with the theoretically-predicted 
$I$-band clump magnitude as a function of age and metallicity. 
Simple equations, taken from basic population synthesis theory, are 
provided for the easy computation of the mean clump magnitude for 
any given galaxy model. We use our models to explain in some detail 
what determines the distribution of masses, ages, and 
metallicities of clump stars in a galaxy. Such an approach has so 
far been neglected in the analysis of clump data related with
distance determinations. We point out that, in galaxies 
with recent/ongoing star formation (e.g.\ the disks of spirals), 
the age distribution of clump stars 
is strongly biased toward younger ($\sim1-3$~Gyr) ages, and hence toward
higher metallicities. Obviously, this does not happen in galaxies
with predominantly old stellar populations (e.g.\ ellipticals and 
bulges). 

We construct detailed models for the clump population in the local 
({\em Hipparcos}) sample, the Bulge, Magellanic Clouds, and
Carina dSph galaxy. In all cases, star formation rates 
and chemical enrichment histories are taken from the literature. 
The {\em Hipparcos} model is shown to produce 
distributions of metallicities, colours, and magnitudes, that are
similar to those derived from spectroscopic and {\em Hipparcos}
data. The Bulge, Magellanic Clouds, and
Carina dSph models are used to analyse the values of 
$\Delta M_I^{\rm RC}$ for these different stellar systems.
We show how the clump--RR Lyrae data from 
Udalski (1998a) is well reproduced by the models. 
However, despite the similarity between the models and data, 
the models indicate that the linear $\Delta M_I^{\rm RC}$ 
versus \feh\ relations that have been derived from the same data  
(Udalski 1998a, 2000; Popowski 2000) are not general. 
In fact, the distribution of clump stars
has several factors hidden in it -- e.g.\ the 
age-metallicity relation, the rate of past star formation -- 
that cannot be described by such relations.

The model behaviour is also supported by empirical data for open 
clusters by Sarajedini (1999) and Twarog et al.\ (1999). 
We argue that Udalski's (1998b) data for LMC and SMC star 
clusters do not allow a good assessment of 
the age dependence of the clump magnitude. 
Moreover, we remark that similar analyses of cluster data
should better include clump stars with ages $1-2$~Gyr, 
which turn out to be very important in determining the 
mean clump in galaxies with recent star formation.

Finally, we provide 
revised clump distances to the Bulge, Magellanic Clouds, 
and Carina dSph, and further comment on their reliability. 
The largest $\Delta M_I^{\rm RC}$ values are found for the 
Magellanic Clouds and Carina dSph, which turn out to be located 
at distance moduli $\sim0.2-0.3$~mag longer than indicated by works which 
ignore population effects. 
The Galactic Bulge, instead, may be slightly closer (up to
$0.1$~mag in distance modulus) than indicated by
previous works based on the red clump, the exact result depending
on the use of either scaled-solar or $\alpha$-enhanced stellar 
models.

\end{abstract}


\section{Introduction}
\label{sec_intro}

The red giant clump has been recently claimed to provide a very 
accurate standard candle.
Once the mean $I$-band\footnote{Throughout this paper we will 
refer to the Cousins $I$-band}
magnitude of the red clump, $I^{\rm RC}$, 
is measured in a nearby galaxy, its absolute distance modulus 
$\mu_0=(m-M)_0$ is easily derived by means of
	\begin{equation}
\mu_0 = I^{\rm RC} - M_I^{\rm RC} - A_I + \Delta M_I^{\rm RC}.
	\label{eq_mu}
	\end{equation}
In this equation, $M_I^{\rm RC}$ is the reference value (zero point)
provided by nearby clump stars whose distances are known from 
{\em Hipparcos} trigonometric parallaxes, $A_I$ is the 
interstellar extinction to the 
red clump population of an external galaxy, and 
$\Delta M_I^{\rm RC} = M_I^{\rm RC}({\rm Hipp}) - 
M_I^{\rm RC}({\rm galaxy})$ is the population effect, i.e.\ the 
difference of the mean red clump absolute magnitude between 
the local and external
samples of stars.

Determining $M_I^{\rm RC}$ and $I^{\rm RC}$ is, usually, less of a
problem. Both in the {\em Hipparcos} database of nearby stars,
and in CMDs covering even a small fraction of a nearby galaxy, 
one typically finds several hundreds of clump stars, clearly 
identifiable by their position in the CMD. Then,
by performing a non-linear least-square fit of the function 
	\begin{equation}
N(I) = a + b I + c I^2 + 
	d \exp\left[-\frac{(I^{\rm RC}-I)^2}{2\sigma_I^2}\right]
	\label{eq_fit}
	\end{equation}
to the histogram of stars in the clump region per magnitude bin 
[i.e.\ the luminosity function $N(I)$], the value of 
$I^{\rm RC}$ and its associated standard error 
are easily determined (Stanek \& Garnavich 1998). 
The parameter $\sigma_I$ gives a good indication of how sharp the 
luminosity distribution of clump stars is, whereas $a$, $b$, $c$ 
(also derived from the fitting procedure) are constants of less interest. 

By applying this procedure to the {\em Hipparcos} database of 
nearby stars (closer than, say, 70~pc),
$M_I^{\rm RC}$ has been determined with accuracy of hundredths of
magnitude (Paczy\'nski \& Stanek 1998; 
Stanek, Zaritsky \& Harris 1998). 
Similar accuracies are obtained for
$I^{\rm RC}$ in nearby systems like the Galactic Bulge
(Paczy\'nski \& Stanek 1998), the Magellanic Clouds 
(Udalski et al.\ 1998), and M~31 (Stanek et al.\ 1998). 
In all cases, the accuracy is limited mainly by the 
calibration of the photometry, rather than by number
statistics of clump stars.

Therefore, the main concerns in the use of clump stars as standard
candles are in the determination of the extinction $A_I$, and the
population effects $\Delta M_I^{\rm RC}$. The Magellanic Clouds 
provide emblematic examples of the problems associated with
both determinations. In fact, 
the first applications of the red clump method to the LMC
have provided extremely short distances (e.g.\ 
$\mu_0^{\rm LMC}=18.1-18.2$~mag; see Udalski et al.\ 
1998 and Stanek et al.\ 1998) if compared with the values commonly 
accepted a few years ago ($\mu_0^{\rm LMC}\simeq18.5\pm0.1$~mag;
see Westerlund 1997), 
and in strong disagreement with the latest values derived from 
classical methods as the Cepheids P-L relation (e.g.\ 
$\mu_0^{\rm LMC}=18.7\pm0.1$~mag 
cf.\ Feast \& Catchpole 1997). These `short' LMC distances have been 
suspected to arise from errors in estimating either $A_I$, or 
$\Delta M_I^{\rm RC}$, or both.

With respect to $A_I$, Romaniello et al.\ (1999) and Zaritsky (1999) 
recently claimed that the $A_I$ values used for LMC clump stars
in Stanek et al.\ (1998) are overestimated by about 0.2~mag, 
thus resulting in an underestimate of $\mu_0^{\rm LMC}$ by 
the same amount. Both argue to 
obtain more reliable estimates of $A_I$, either using high-quality
multi-band HST photometry as Romaniello et al.\ (1999), or 
determining the reddening directly from red stars 
(from 5500 to 6500~K) in the observed fields, instead of using 
mean reddening values (Zaritsky 1999). 

In this paper, we address the subtle problem of
determining the population effect $\Delta M_I^{\rm RC}$. 
This factor has been initially neglected (Paczy\'nski \& Stanek 
1998; Udalski et al.\ 1998; Stanek et al.\ 1998), after noticing 
that $I^{\rm RC}$ is remarkably constant, as a 
function of the \vi\ colour, in several stellar systems like the
Bulge (Paczy\'nski \& Stanek 1998), M~31 (Stanek et al.\ 1998),
LMC and SMC (Udalski et al.\ 1998; Stanek et al.\ 1998), 
especially in the colour range $0.8<(\vi)_0<1.5$ which defines the
local clump from {\em Hipparcos} (Paczy\'nski \& Stanek 1998). 
However, Cole (1998) and Girardi 
et al.\ (1998, hereafter GGWS98) called attention to the 
non-negligible $\Delta M_I^{\rm RC}$ values 
(up to 0.6~mag according to Cole 1998) expected 
from theoretical models of clump stars, and claimed a red clump 
distance to the LMC larger by about 
$\Delta M_I^{\rm RC}({\rm LMC})=0.2 - 0.3$~mag. GGWS98, and 
later Girardi (1999),
provided simulations of the red clump in model galaxies which
naturally showed the effect of an almost constant 
$M_I^{\rm RC}$ with \vi, and presented significant 
$\Delta M_I^{\rm RC}$ values at the same time.

In two successive papers, Udalski (1998ab) aimed at empirically 
determining the dependence of $M_I^{\rm RC}$ on stellar parameters. 
The result was a very modest dependence on both 
metallicity and age. Namely, 
$M_I^{\rm RC}=(0.09\pm0.03)\,\feh + \mbox{const.}$, was
obtained by comparing red clump and RR Lyrae stars in the Bulge,
LMC, SMC, and Carina dSph, and assuming 
$M_V^{\rm RRLy}=(0.18\pm0.03)\,\feh + \mbox{const.}$ (Udalski 1998a). 
Moreover, by comparing the clump in several LMC and SMC clusters, 
Udalski (1998b) claimed negligible changes in $M_I^{\rm RC}$ 
for cluster ages ranging from 2 to 10 Gyr. More recently
Udalski (2000) fits a 
$M_I^{\rm RC}=(0.13\pm0.07)\,\feh + \mbox{const.}$ relationship 
to the nearby clump stars with spectroscopic \feh\ determinations.
All together, these results would imply quite modest 
$\Delta M_I^{\rm RC}$ values (less than 0.1~mag for the LMC), 
thus still supporting a `short distance scale', as opposed to the 
`long distance scale' provided by, e.g.\ Cepheids
and RR Lyrae calibrated by subdwarf fitting to globular clusters
(see Feast 1999 and Carretta et al.\ 2000 for recent reviews).

Udalski's (1998ab, 2000) results have been extensively used in 
subsequent works regarding the clump method for distance determinations. 
Zaritsky (1999) and Popowski (2000), for instance, attempted to 
reduce the errors in the red clump method associated with 
reddening estimates, but assumed that Udalski's (1998ab, 2000) 
conclusions were valid. However,
there are some potential drawbacks in Udalski's analyses
that we think should be considered in more detail. These drawbacks,
in fact, are among the main points we wish to address in this paper.
 
A subtle one is that $M_I^{\rm RC}$ is assumed either not to depend 
on ages (as in Udalski 1998a, 2000), or not to depend on 
metallicities (as in Udalski 1998b). Thus, the presence of an 
age-metallicity relation (AMR) in the observational data
could be masking a possible dependence of 
$M_I^{\rm RC}$ on both parameters, provided
that this dependence were in the sense of producing brighter clumps
at both lower metallicities and younger 
ages. And these are exactly the trends indicated by theoretical
models. 

We should also mention the particular selection of Magellanic Cloud 
clusters in Udalski (1998b). They are limited to the $2-10$~Gyr 
age interval, that is not the complete range of possible ages of 
clump stars. Moreover, quite large `distance corrections' were
adopted for some SMC clusters in Udalski (1998b; see comments in 
Girardi 1999), which could also be
the source of systematic errors in his analysis of
the $M_I^{\rm RC}$ dependence on age. It is worth noticing 
that, although Udalski (1998b) finds evidence that the clump
fades by $0.3-0.4$~mag at later ages, this particular result has 
been ignored in later works, where the clump is simply
assumed not to depend on age. At the same time Twarog et al.\ 
(1998) and Sarajedini (1999), using data from open clusters with
main-sequence fitting distances, reached conclusions apparently 
in contradiction with Udalski's ones, i.e.\ favouring larger 
dependences of $M_I^{\rm RC}$ on either age or metallicity.

Another point of perplexity to us, is that Udalski (1998ab, 2000)
and later Popowski (2000) decide to express the population 
dependence of the clump magnitude by means of simple linear 
relationships between $M_I^{\rm RC}$ and \feh, 
and between $M_I^{\rm RC}$ and age. In our opinion, these choices 
are not justified. To this respect, we notice that similar 
relations are largely used in stellar astrophysics; 
but they are, in most cases, well supported by 
theory. A classical example 
of this kind comes from RR Lyrae stars, for which a linear $M_V$ versus 
\feh\ relation is well documented and widely accepted. 
However, we should keep in mind that RR Lyrae are very particular 
objects, which occupy a quite narrow range in both physical (\Teff)
and population (age, metallicity) properties. Clump stars, on the
contrary, do not have such tight constraints: they are almost 
ubiquitous, with a very large range of ages, metallicities, 
and \Teff s. Clearly, nature has much more freedom to play with 
the luminosity of clump stars, than it has with RR Lyrae.
Why should we expect both types of stars to behave in a similar 
way\,? Then, why should we expect linear relations to hold
for the properties of clump stars\,?

To clarify these points, in the following we explore the 
several systematic effects
on $M_I^{\rm RC}$ indicated by stellar evolutionary models.
In Sect.~\ref{sec_agemet},
the metallicity and age effects are described with some detail.
We present simple formulas for computing the mean clump magnitude
in a galaxy, then allowing the readers to determine 
the appropriate value for the population correction 
$\Delta M_I^{\rm RC}$ for any given history of star 
formation and chemical enrichment.
We then present a careful comparison between models and the local
clump from {\em Hipparcos} (Sect.~\ref{sec_hipparcos}), which is 
particularly helpful to illustrate the different effects involved 
in the problem, and to understand Udalski's (2000) results. 
The clump behaviour in Galactic, LMC and SMC clusters 
is briefly commented on in Sect.~\ref{sec_clusters}.
We proceed by modelling the clump in several nearby 
galaxies, and comparing the results with the data discussed by 
Udalski (1998a) and Popowski (2000) (Sect.~\ref{sec_galaxies}). 
Overall, we find that {\em systematic effects 
(at the level of $\la0.3$~mag) are not only present in
any determination of red clump distances to nearby galaxies, 
but are also hidden in many of the previous analyses of 
clump data discussed in this regard}.
Our final conclusions regarding the red clump distance scale
to nearby galaxies are presented in Sect.~\ref{sec_conclu}. 

\section{The clump as a function of age and metallicity}
\label{sec_agemet}

As in previous works, we will base the discussion of model 
behaviour mostly on the Girardi et al.\ (2000) set of
evolutionary tracks and isochrones. These models have 
been extensively discussed in GGWS98 and Girardi (1999). 
It is worth remarking that different models in the literature
-- although presenting systematic luminosity differences for 
the core helium burning (CHeB) stars that have passed through
the helium flash -- present similar behaviours for this 
luminosity as a function of either age or metallicity 
(Castellani et al.\ 2000). The model behaviour is also
supported by empirical data for clusters 
(see Sect.~\ref{sec_clusters}).
Moreover, the formulas provided in this section allow 
the reader to check our results using any alternative 
set of stellar models. 

\subsection{Simple models for the mean clump magnitude}
\label{sec_howto}

In our previous works (GGWS98 and Girardi 1999) we have 
discussed the behaviour of the clump as a function of
mass, age and metallicity, considering mainly the 
sequences of zero-age horizontal branch models (ZAHB), 
i.e.\ of the stellar configurations at the beginning of 
quiescent CHeB. 
This time, however, we prefer to discuss the behaviour 
of the {\em mean} clump as a function of age and metallicity.

The properties
of the mean clump are defined as follows. For a given isochrone
of age and metallicity $(t,Z)$, we perform the following integrals
over the isochrone section corresponding only to CHeB stars:
	\begin{equation}
\langle M_\lambda(t,Z) \rangle = -2.5\log \left[ 
	\frac{1}{N_{\rm cl}(t, Z)}
	\int^{\rm CHeB} \!\!\!\! \phi(\Mi)\, 10^{-0.4 M_\lambda}
		\diff\Mi 
	\right] 
	\label{eq_inte}
	\end{equation}
where $M_\lambda$ is the absolute magnitude in the pass-band 
$\lambda$, \Mi\ is the initial mass of the star at each isochrone 
point, and $\phi(\Mi)$ is the Salpeter IMF (number of stars by 
initial mass interval $[\Mi,\Mi+\diff\Mi]$). $N_{\rm cl}$ is the 
{\em number of clump stars (at age $t$) per unit mass of stars 
initially born}. It is simply given by the integral
of the IMF by number, along the CHeB isochrone section, i.e.\
	\begin{equation}
N_{\rm cl}(t, Z) = \int^{\rm CHeB} \phi(\Mi)\, \diff\Mi \;.
	\label{eq_ncl}
	\end{equation}
In our case, the IMF is normalised such as to produce a 
single-burst stellar population of total initial
mass of 1~\Msun\ (i.e.\ $\int \Mi\,\phi(\Mi)\,\diff\Mi=1$~\Msun), 
and a mass-to-light ratio of $M/L_V=0.2$ at an age 
of $10^8$~yr. The details of this normalisation 
can be found in Girardi \& Bica (1993) and 
Salasnich et al.\ (2000). It is worth remarking
that none of the results presented in this paper depends on 
the particular choice of IMF normalisation. 
However, having an IMF normalised to unit mass turns out to be 
a convenient choice.

\begin{figure}
	\psfig{file=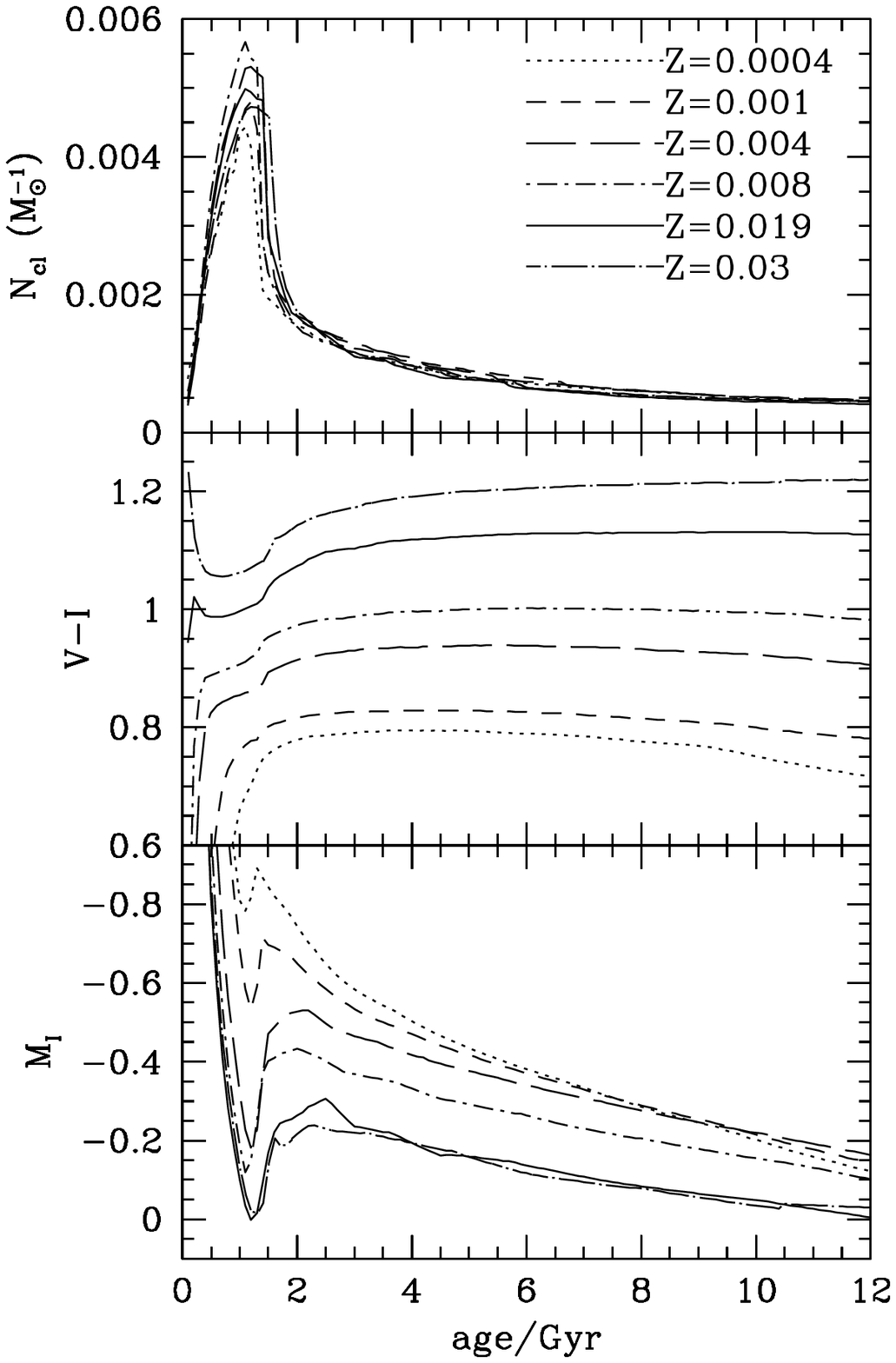,width=8.3cm}
        \caption{The behaviour of the clump as a function of age, 
for several metallicities. The top panel shows the current number 
of clump stars per unit mass of a stellar population, 
$N_{\rm cl}$ (see eq.~\protect\ref{eq_ncl}). 
The mean $\langle\vi\rangle$ and $\langle\mi\rangle$ 
(from eq.~\protect\ref{eq_inte} and Table~\protect\ref{tab_mi}) 
are shown in the middle and bottom panels, respectively.}
	\label{fig_mcore}
	\end{figure}

From eq.~\ref{eq_inte}, accurate values for the 
mean clump colours and magnitudes can be obtained.
Table~\ref{tab_mi} presents the values of $\langle\mi\rangle$, 
$\langle\vi\rangle=\langle\mv\rangle-\langle\mi\rangle$, 
and $N_{\rm cl}$ obtained for ages ranging 
from 0.5 to 12~Gyr, and metallicities from $Z=0.0004$ to 0.03. 
More extensive tables, suitable to perform 
accurate interpolation in the quantities $(t,Z)$, are available in
computer-readable form in \verb$http://pleiadi.pd.astro.it$. 
Part of this information is also illustrated in 
Fig.~\ref{fig_mcore}.
\begin{table*}
\caption{Mean clump properties, 
as a function of age and metallicity, from 
Girardi et al.\ (2000) isochrones.}
\label{tab_mi}
\begin{tabular}{l llll llll llll}
\noalign{\smallskip}\hline\noalign{\smallskip}
  & \multicolumn{4}{l}{$Z=0.0004$} & \multicolumn{4}{l}{$Z=0.001$} & \multicolumn{4}{l}{$Z=0.004$} \\
\noalign{\smallskip}\cline{2-5} \cline{6-9}  \cline{10-13}\noalign{\smallskip}
$t$ & 
$N_{\rm cl}$ & $\langle\Mi\rangle$ & $\langle\mv\rangle$ & $\langle\mi\rangle$ &
$N_{\rm cl}$ & $\langle\Mi\rangle$ & $\langle\mv\rangle$ & $\langle\mi\rangle$ &
$N_{\rm cl}$ & $\langle\Mi\rangle$ & $\langle\mv\rangle$ & $\langle\mi\rangle$ 
\\
(Gyr) & 
$(10^{-4})$ & (\Msun) &  &  &
$(10^{-4})$ & (\Msun) &  &  & 
$(10^{-4})$ & (\Msun) &  &  
\\
\noalign{\smallskip}\hline\noalign{\smallskip}
0.5 & 26.30 & $2.358$ & $-1.186$ & $-1.521$    & 25.70 & $2.415$ & $-1.061$ & $-1.591$	      & 31.60 & $2.560$ & $-0.435$ & $-1.259$ \\
0.6 & 29.40 & $2.191$ & $-0.929$ & $-1.307$    & 30.60 & $2.245$ & $-0.737$ & $-1.375$	      & 37.00 & $2.386$ & $-0.118$ & $-0.955$ \\
0.7 & 32.30 & $2.065$ & $-0.685$ & $-1.149$    & 33.70 & $2.115$ & $-0.467$ & $-1.166$	      & 41.40 & $2.250$ & $0.114$ & $-0.729$ \\
0.8 & 36.80 & $1.962$ & $-0.466$ & $-1.020$    & 37.60 & $2.013$ & $-0.259$ & $-0.988$	      & 44.30 & $2.140$ & $0.285$ & $-0.562$ \\
0.9 & 38.00 & $1.880$ & $-0.318$ & $-0.933$    & 40.90 & $1.927$ & $-0.073$ & $-0.822$	      & 47.20 & $2.047$ & $0.417$ & $-0.435$ \\
1.0 & 43.50 & $1.807$ & $-0.148$ & $-0.809$    & 44.40 & $1.855$ & $0.074$ & $-0.686$	      & 50.10 & $1.967$ & $0.534$ & $-0.321$ \\
1.1 & 44.30 & $1.750$ & $-0.099$ & $-0.783$    & 47.00 & $1.793$ & $0.185$ & $-0.585$	      & 52.80 & $1.897$ & $0.632$ & $-0.227$ \\
1.2 & 41.50 & $1.699$ & $-0.114$ & $-0.816$    & 48.00 & $1.739$ & $0.245$ & $-0.532$	      & 53.10 & $1.839$ & $0.679$ & $-0.182$ \\
1.3 & 31.00 & $1.644$ & $-0.164$ & $-0.892$    & 44.00 & $1.691$ & $0.187$ & $-0.592$	      & 52.20 & $1.789$ & $0.639$ & $-0.227$ \\
1.4 & 20.70 & $1.595$ & $-0.121$ & $-0.867$    & 28.00 & $1.630$ & $0.084$ & $-0.715$	      & 51.60 & $1.744$ & $0.514$ & $-0.361$ \\
1.5 & 19.40 & $1.559$ & $-0.089$ & $-0.844$    & 23.00 & $1.588$ & $0.105$ & $-0.696$	      & 29.60 & $1.676$ & $0.422$ & $-0.471$ \\
1.6 & 18.70 & $1.526$ & $-0.059$ & $-0.821$    & 21.40 & $1.553$ & $0.115$ & $-0.691$	      & 22.50 & $1.632$ & $0.411$ & $-0.488$ \\
1.7 & 17.80 & $1.495$ & $-0.037$ & $-0.805$    & 20.10 & $1.520$ & $0.127$ & $-0.682$	      & 19.70 & $1.597$ & $0.400$ & $-0.503$ \\
1.8 & 16.90 & $1.466$ & $-0.016$ & $-0.788$    & 19.00 & $1.491$ & $0.139$ & $-0.673$	      & 18.20 & $1.566$ & $0.388$ & $-0.519$ \\
1.9 & 16.00 & $1.439$ & $0.008$ & $-0.768$     & 18.00 & $1.463$ & $0.150$ & $-0.665$	      & 16.60 & $1.538$ & $0.387$ & $-0.524$ \\
2.0 & 15.50 & $1.414$ & $0.035$ & $-0.744$     & 16.90 & $1.437$ & $0.167$ & $-0.650$	      & 16.20 & $1.512$ & $0.387$ & $-0.527$ \\
2.2 & 14.80 & $1.371$ & $0.081$ & $-0.703$     & 16.20 & $1.392$ & $0.198$ & $-0.622$	      & 15.50 & $1.466$ & $0.390$ & $-0.530$ \\
2.4 & 13.50 & $1.332$ & $0.121$ & $-0.665$     & 14.90 & $1.352$ & $0.225$ & $-0.597$	      & 14.80 & $1.425$ & $0.415$ & $-0.508$ \\
2.6 & 13.00 & $1.298$ & $0.155$ & $-0.633$     & 14.20 & $1.317$ & $0.251$ & $-0.573$	      & 14.10 & $1.389$ & $0.439$ & $-0.487$ \\
2.8 & 12.50 & $1.269$ & $0.184$ & $-0.605$     & 13.60 & $1.287$ & $0.273$ & $-0.552$	      & 13.30 & $1.357$ & $0.454$ & $-0.474$ \\
3.0 & 11.50 & $1.242$ & $0.207$ & $-0.584$     & 12.20 & $1.260$ & $0.293$ & $-0.533$	      & 12.90 & $1.328$ & $0.465$ & $-0.465$ \\
3.2 & 11.10 & $1.218$ & $0.226$ & $-0.567$     & 11.90 & $1.235$ & $0.307$ & $-0.519$	      & 12.50 & $1.301$ & $0.476$ & $-0.457$ \\
3.4 & 10.70 & $1.196$ & $0.243$ & $-0.550$     & 11.50 & $1.213$ & $0.321$ & $-0.506$	      & 11.80 & $1.278$ & $0.484$ & $-0.450$ \\
3.6 & 10.30 & $1.176$ & $0.259$ & $-0.536$     & 11.10 & $1.192$ & $0.333$ & $-0.494$	      & 11.50 & $1.256$ & $0.496$ & $-0.438$ \\
3.8 & 9.65 & $1.157$ & $0.275$ & $-0.520$      & 10.50 & $1.173$ & $0.346$ & $-0.482$	      & 11.10 & $1.235$ & $0.508$ & $-0.428$ \\
4.0 & 9.38 & $1.139$ & $0.292$ & $-0.503$      & 10.30 & $1.156$ & $0.358$ & $-0.470$	      & 10.80 & $1.216$ & $0.518$ & $-0.417$ \\
4.3 & 8.98 & $1.115$ & $0.314$ & $-0.480$      & 9.93 & $1.132$ & $0.376$ & $-0.452$	      & 9.55 & $1.190$ & $0.534$ & $-0.403$ \\
4.6 & 8.58 & $1.094$ & $0.335$ & $-0.459$      & 9.57 & $1.109$ & $0.392$ & $-0.436$	      & 9.21 & $1.167$ & $0.546$ & $-0.391$ \\
4.9 & 8.04 & $1.074$ & $0.354$ & $-0.441$      & 9.17 & $1.089$ & $0.407$ & $-0.422$	      & 8.87 & $1.146$ & $0.558$ & $-0.380$ \\
5.2 & 7.80 & $1.056$ & $0.370$ & $-0.423$      & 8.74 & $1.071$ & $0.421$ & $-0.407$	      & 8.54 & $1.126$ & $0.569$ & $-0.369$ \\
5.5 & 7.55 & $1.039$ & $0.385$ & $-0.407$      & 8.44 & $1.054$ & $0.434$ & $-0.393$	      & 7.72 & $1.108$ & $0.580$ & $-0.359$ \\
6.0 & 7.15 & $1.014$ & $0.409$ & $-0.381$      & 7.94 & $1.029$ & $0.455$ & $-0.371$	      & 7.36 & $1.082$ & $0.598$ & $-0.340$ \\
6.5 & 6.75 & $0.991$ & $0.430$ & $-0.358$      & 7.44 & $1.006$ & $0.474$ & $-0.350$	      & 7.01 & $1.058$ & $0.613$ & $-0.324$ \\
7.0 & 6.46 & $0.971$ & $0.451$ & $-0.334$      & 6.63 & $0.985$ & $0.492$ & $-0.329$	      & 6.65 & $1.037$ & $0.628$ & $-0.308$ \\
7.5 & 6.22 & $0.952$ & $0.471$ & $-0.309$      & 6.36 & $0.966$ & $0.509$ & $-0.309$	      & 6.23 & $1.018$ & $0.643$ & $-0.292$ \\
8.0 & 5.97 & $0.935$ & $0.490$ & $-0.286$      & 6.08 & $0.949$ & $0.526$ & $-0.289$	      & 5.99 & $1.000$ & $0.657$ & $-0.276$ \\
9.0 & 5.49 & $0.905$ & $0.524$ & $-0.245$      & 5.55 & $0.919$ & $0.555$ & $-0.254$	      & 5.52 & $0.969$ & $0.682$ & $-0.246$ \\
10.0 & 4.85 & $0.880$ & $0.549$ & $-0.202$     & 5.06 & $0.893$ & $0.585$ & $-0.215$	      & 5.14 & $0.942$ & $0.704$ & $-0.219$ \\
11.0 & 4.71 & $0.857$ & $0.574$ & $-0.159$     & 4.80 & $0.870$ & $0.612$ & $-0.177$	      & 4.91 & $0.919$ & $0.723$ & $-0.191$ \\
12.0 & 4.56 & $0.837$ & $0.595$ & $-0.122$     & 4.55 & $0.850$ & $0.638$ & $-0.143$	      & 4.69 & $0.897$ & $0.742$ & $-0.164$ \\
\noalign{\smallskip}\hline\noalign{\smallskip}
\end{tabular}

\end{table*}
\setcounter{table}{0}
\begin{table*}
\caption{(continued)}
\label{tab_mi}
\begin{tabular}{l llll llll llll}
\noalign{\smallskip}\hline\noalign{\smallskip}
  & \multicolumn{4}{l}{$Z=0.008$} & \multicolumn{4}{l}{$Z=0.019$} & \multicolumn{4}{l}{$Z=0.03$} \\
\noalign{\smallskip}\cline{2-5} \cline{6-9}  \cline{10-13}\noalign{\smallskip}
$t$ & 
$N_{\rm cl}$ & $\langle\Mi\rangle$ & $\langle\mv\rangle$ & $\langle\mi\rangle$ &
$N_{\rm cl}$ & $\langle\Mi\rangle$ & $\langle\mv\rangle$ & $\langle\mi\rangle$ &
$N_{\rm cl}$ & $\langle\Mi\rangle$ & $\langle\mv\rangle$ & $\langle\mi\rangle$ 
\\
(Gyr) & 
$(10^{-4})$ & (\Msun) &  &  &
$(10^{-4})$ & (\Msun) &  &  & 
$(10^{-4})$ & (\Msun) &  &  
\\
\noalign{\smallskip}\hline\noalign{\smallskip}
0.5 & 34.30 & $2.670$ & $-0.113$ & $-1.001$    & 30.60 & $2.819$ & $0.161$ & $-0.827$	      & 29.40 & $2.821$ & $0.231$ & $-0.827$ \\
0.6 & 40.10 & $2.492$ & $0.174$ & $-0.718$     & 36.10 & $2.634$ & $0.399$ & $-0.589$	      & 33.90 & $2.643$ & $0.444$ & $-0.612$ \\
0.7 & 44.90 & $2.352$ & $0.358$ & $-0.538$     & 40.20 & $2.491$ & $0.578$ & $-0.410$	      & 37.30 & $2.504$ & $0.609$ & $-0.446$ \\
0.8 & 48.50 & $2.236$ & $0.508$ & $-0.392$     & 43.90 & $2.374$ & $0.707$ & $-0.283$	      & 40.20 & $2.391$ & $0.735$ & $-0.321$ \\
0.9 & 51.80 & $2.139$ & $0.627$ & $-0.276$     & 46.50 & $2.275$ & $0.812$ & $-0.181$	      & 42.90 & $2.292$ & $0.846$ & $-0.212$ \\
1.0 & 54.80 & $2.056$ & $0.719$ & $-0.188$     & 48.30 & $2.191$ & $0.897$ & $-0.100$	      & 45.40 & $2.208$ & $0.936$ & $-0.125$ \\
1.1 & 56.70 & $1.985$ & $0.791$ & $-0.120$     & 49.90 & $2.118$ & $0.966$ & $-0.035$	      & 46.60 & $2.137$ & $1.003$ & $-0.061$ \\
1.2 & 54.30 & $1.930$ & $0.770$ & $-0.148$     & 49.40 & $2.056$ & $1.006$ & $0.001$	      & 47.20 & $2.076$ & $1.051$ & $-0.018$ \\
1.3 & 53.50 & $1.874$ & $0.674$ & $-0.251$     & 48.40 & $2.001$ & $0.994$ & $-0.016$	      & 47.10 & $2.021$ & $1.061$ & $-0.014$ \\
1.4 & 27.80 & $1.788$ & $0.575$ & $-0.371$     & 48.20 & $1.950$ & $0.927$ & $-0.091$	      & 46.40 & $1.971$ & $1.040$ & $-0.041$ \\
1.5 & 23.20 & $1.739$ & $0.552$ & $-0.401$     & 28.20 & $1.872$ & $0.869$ & $-0.168$	      & 45.80 & $1.925$ & $0.956$ & $-0.142$ \\
1.6 & 20.60 & $1.698$ & $0.548$ & $-0.410$     & 25.10 & $1.827$ & $0.828$ & $-0.221$	      & 31.60 & $1.862$ & $0.913$ & $-0.205$ \\
1.7 & 18.80 & $1.662$ & $0.542$ & $-0.421$     & 22.20 & $1.787$ & $0.814$ & $-0.243$	      & 24.30 & $1.815$ & $0.937$ & $-0.185$ \\
1.8 & 17.10 & $1.629$ & $0.542$ & $-0.424$     & 19.50 & $1.750$ & $0.812$ & $-0.250$	      & 20.90 & $1.777$ & $0.940$ & $-0.187$ \\
1.9 & 16.30 & $1.600$ & $0.541$ & $-0.429$     & 17.40 & $1.717$ & $0.814$ & $-0.254$	      & 18.90 & $1.743$ & $0.931$ & $-0.203$ \\
2.0 & 15.30 & $1.574$ & $0.540$ & $-0.433$     & 16.80 & $1.688$ & $0.810$ & $-0.263$	      & 17.40 & $1.713$ & $0.926$ & $-0.215$ \\
2.2 & 14.10 & $1.526$ & $0.556$ & $-0.422$     & 15.50 & $1.635$ & $0.800$ & $-0.285$	      & 16.00 & $1.660$ & $0.915$ & $-0.236$ \\
2.4 & 13.30 & $1.484$ & $0.573$ & $-0.409$     & 14.70 & $1.591$ & $0.794$ & $-0.299$	      & 14.40 & $1.614$ & $0.925$ & $-0.233$ \\
2.6 & 12.80 & $1.447$ & $0.591$ & $-0.393$     & 13.50 & $1.551$ & $0.806$ & $-0.293$	      & 12.90 & $1.573$ & $0.938$ & $-0.225$ \\
2.8 & 12.30 & $1.413$ & $0.610$ & $-0.375$     & 12.30 & $1.515$ & $0.839$ & $-0.263$	      & 12.00 & $1.537$ & $0.947$ & $-0.221$ \\
3.0 & 11.00 & $1.383$ & $0.619$ & $-0.369$     & 11.00 & $1.483$ & $0.868$ & $-0.235$	      & 11.50 & $1.505$ & $0.954$ & $-0.218$ \\
3.2 & 10.70 & $1.357$ & $0.628$ & $-0.363$     & 10.80 & $1.454$ & $0.877$ & $-0.230$	      & 11.00 & $1.475$ & $0.959$ & $-0.219$ \\
3.4 & 10.50 & $1.332$ & $0.635$ & $-0.357$     & 10.70 & $1.427$ & $0.887$ & $-0.225$	      & 10.80 & $1.448$ & $0.970$ & $-0.211$ \\
3.6 & 10.30 & $1.310$ & $0.643$ & $-0.352$     & 10.20 & $1.402$ & $0.893$ & $-0.222$	      & 10.60 & $1.423$ & $0.980$ & $-0.204$ \\
3.8 & 10.00 & $1.289$ & $0.653$ & $-0.343$     & 9.54 & $1.380$ & $0.908$ & $-0.208$	      & 9.99 & $1.400$ & $0.987$ & $-0.201$ \\
4.0 & 9.67 & $1.269$ & $0.665$ & $-0.332$      & 9.09 & $1.359$ & $0.924$ & $-0.194$	      & 9.63 & $1.378$ & $0.998$ & $-0.192$ \\
4.3 & 9.14 & $1.241$ & $0.681$ & $-0.316$      & 8.42 & $1.331$ & $0.945$ & $-0.174$	      & 9.23 & $1.349$ & $1.011$ & $-0.182$ \\
4.6 & 8.29 & $1.216$ & $0.696$ & $-0.303$      & 7.88 & $1.305$ & $0.958$ & $-0.164$	      & 8.83 & $1.322$ & $1.023$ & $-0.173$ \\
4.9 & 7.93 & $1.194$ & $0.705$ & $-0.295$      & 7.72 & $1.281$ & $0.963$ & $-0.161$	      & 8.07 & $1.298$ & $1.037$ & $-0.162$ \\
5.2 & 7.68 & $1.174$ & $0.714$ & $-0.286$      & 7.56 & $1.259$ & $0.969$ & $-0.156$	      & 7.69 & $1.276$ & $1.052$ & $-0.148$ \\
5.5 & 7.43 & $1.155$ & $0.724$ & $-0.277$      & 7.39 & $1.238$ & $0.977$ & $-0.150$	      & 7.31 & $1.256$ & $1.067$ & $-0.135$ \\
6.0 & 6.47 & $1.127$ & $0.738$ & $-0.264$      & 6.38 & $1.207$ & $0.991$ & $-0.137$	      & 6.46 & $1.226$ & $1.088$ & $-0.116$ \\
6.5 & 6.16 & $1.101$ & $0.756$ & $-0.245$      & 6.04 & $1.181$ & $1.007$ & $-0.122$	      & 6.06 & $1.199$ & $1.103$ & $-0.103$ \\
7.0 & 5.87 & $1.080$ & $0.771$ & $-0.230$      & 5.71 & $1.159$ & $1.022$ & $-0.108$	      & 5.82 & $1.175$ & $1.115$ & $-0.093$ \\
7.5 & 5.59 & $1.061$ & $0.784$ & $-0.216$      & 5.32 & $1.138$ & $1.035$ & $-0.094$	      & 5.59 & $1.154$ & $1.126$ & $-0.083$ \\
8.0 & 5.32 & $1.044$ & $0.795$ & $-0.205$      & 5.12 & $1.119$ & $1.046$ & $-0.084$	      & 5.31 & $1.134$ & $1.135$ & $-0.077$ \\
9.0 & 5.02 & $1.013$ & $0.819$ & $-0.179$      & 4.77 & $1.086$ & $1.065$ & $-0.065$	      & 4.98 & $1.099$ & $1.159$ & $-0.054$ \\
10.0 & 4.71 & $0.985$ & $0.840$ & $-0.155$     & 4.48 & $1.056$ & $1.083$ & $-0.048$	      & 4.66 & $1.069$ & $1.181$ & $-0.033$ \\
11.0 & 4.65 & $0.961$ & $0.859$ & $-0.131$     & 4.31 & $1.030$ & $1.103$ & $-0.026$	      & 4.53 & $1.043$ & $1.184$ & $-0.034$ \\
12.0 & 4.42 & $0.939$ & $0.880$ & $-0.102$     & 4.09 & $1.007$ & $1.122$ & $-0.005$	      & 4.42 & $1.019$ & $1.191$ & $-0.028$ \\
\noalign{\smallskip}\hline\noalign{\smallskip}
\end{tabular}

\end{table*}
%

Another useful quantity included in Table~\ref{tab_mi} is 
the mean initial mass of clump stars, 
	\begin{equation}
\langle \Mi(t,Z) \rangle = 
	\frac{1}{N_{\rm cl}(t, Z)} \int^{\rm CHeB} \Mi\, 
	\phi(\Mi)\, \diff\Mi \;.
	\label{eq_mini}
	\end{equation}

Figure~\ref{fig_lf} presents
the $I$-band luminosity function (LF) for several 
single-burst stellar populations, 
compared to the mean magnitudes calculated according to 
eq.~\ref{eq_inte}. It is worth noticing that:
	\begin{figure}
	\psfig{file=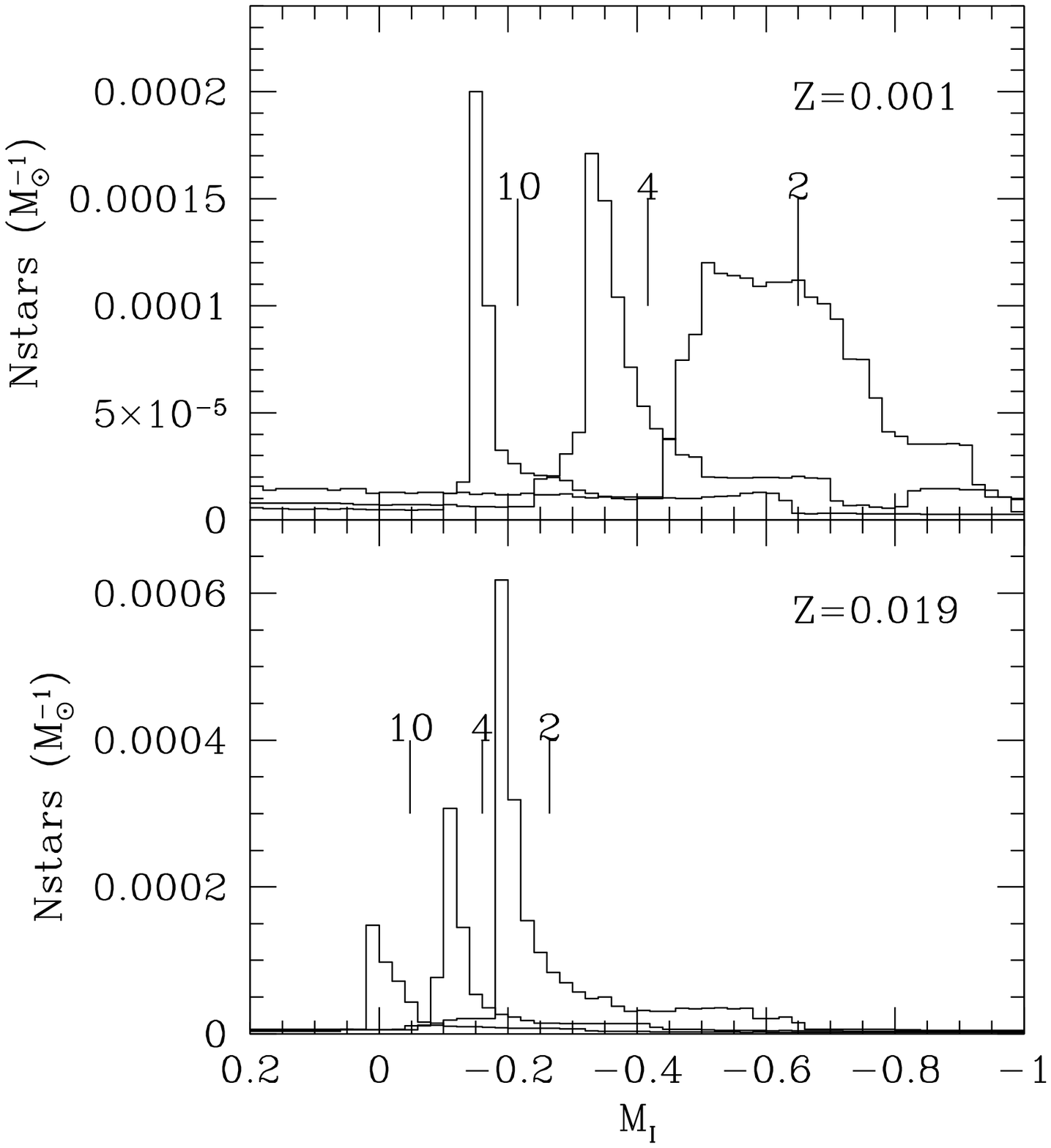,width=8.3cm}
        \caption{$I$-band luminosity functions 
around the clump region, computed with bins 0.02~mag wide, 
for several single-burst stellar populations. 
They correspond to isochrones of 2, 4, and 10 Gyr, and
metallicities $Z=0.001$ (top panel) and $Z=0.019$ (bottom panel).
For each isochrone, the mean $\langle M_\lambda\rangle$, as 
determined from eq.~\protect\ref{eq_inte}, is indicated by a 
vertical line labelled with the age in Gyr.}
	\label{fig_lf}
	\end{figure} 
 
(i) Since the range of clump magnitudes at a given $(t,Z)$ is small, 
it does not really matter whether the $\langle M_\lambda\rangle$
integral is performed over luminosities (as in eq.~\ref{eq_inte}) 
or magnitudes. The resulting $\langle M_\lambda\rangle$ values 
are always accurate to within $\sim0.01$~mag. In fact, what limits the 
accuracy of eq.~\ref{eq_inte} is, mainly, the coarseness of the 
library of stellar evolutionary tracks used to construct the 
isochrones.

(ii) The intrinsic dispersion of clump magnitudes is 
slightly larger for younger ages and lower metallicities. 
For high metallicities and old ages, the LF 
of clump stars is characterized by a sharp spike that 
coincides with the ZAHB bin,
and a decreasing tail for higher luminosities. Thus, the mean
values $\langle M_\lambda\rangle$ are always brighter than the 
maximum of the LF. This offset is of order 0.1~mag, for the 
several ages and metallicities considered in the figure.

Therefore, reducing the clump of a single generation of stars
to a single point of magnitude $\langle M_\lambda(t,Z)\rangle$,
should be seen as a useful approximation, rather than the
detailed behaviour indicated by models.

So far, the equations and data 
refer to the clump as expected in single-burst
stellar populations, i.e.\ in star clusters of given age and
metallicity $(t,Z)$. In order to compute the mean clump 
magnitude for a given galaxy model, of total age $T$, we need to perform 
the following integral:
	\begin{equation}
\langle M_\lambda({\rm gal}) \rangle = 
	\frac{1}{N_{\rm cl}({\rm gal})}
	\int_{t=0}^T \psi(t)\, \langle M_\lambda(t,Z)\rangle
	\,\diff t \;,
	\label{eq_integal}
	\end{equation}
where
	\begin{equation}
N_{\rm cl}({\rm gal}) = 
	\int_{t=0}^T N_{\rm cl}(t, Z)\, \psi(t)\, \diff t \;.
	\label{eq_nclgal}
	\end{equation}
The function $\psi(t)$ is the star formation rate (SFR,
in \Msun\ by unit time), at a moment $t$ in the past,
for the galaxy model considered.
Last but not least, also the age-metallicity relation (AMR) $Z(t)$ should
be specified. Notice that we average the magnitudes in 
eq.~\ref{eq_integal}, instead of luminosities as in the previous
eq.~\ref{eq_inte}. The reason for this is that, in this way,
we get a quantity similar to the $I^{\rm RC}$ derived in 
empirical works by means of eq.~\ref{eq_fit}.

\subsection{Comparison between different methods}

From eq.~\ref{eq_integal},
and the numbers tabulated in Table~\ref{tab_mi}, the reader can easily
derive the theoretical mean clump magnitudes for any galaxy model, once
the SFR $\psi(t)$ and AMR $Z(t)$
are provided. This is a simple and almost direct way of deriving 
$M_I^{\rm RC}$ from theoretical models. We will refer to it, hereafter, 
as {\bf method 1}. There are some additional details about this method, 
which are worth mentioning: 

(i) Some interpolation among the age and 
metallicity values presented in the tables (e.g.\ Tab.~\ref{tab_mi})
may be required.
In this case, our experience is that the most accurate interpolations 
in metallicity are obtained using $\log Z$ [or $\feh\simeq\log(Z/0.019)$] 
as the independent variable, instead of $Z$. 

(ii) For galaxy models with ongoing star formation, the integral of
eq.~\ref{eq_integal} should be given a lower-age cut-off of at least
0.5~Gyr. This because CHeB stars of younger ages, although few, have
much higher luminosities, so that they cannot be considered clump 
stars (see Fig.~\ref{fig_mcore}). 
The final results for $M_I^{\rm RC}$ depend slightly on
the choice of this cut-off. In the models discussed in this section, 
we will assume it equal to 0.5~Gyr.

A second, more refined approach, comes from
a complete population synthesis algorithm: 
synthetic CMDs (or simply the stellar LF) are 
simulated for a given galaxy model, and then $M_I^{\rm RC}$ 
is derived by fitting eq.~\ref{eq_fit} to the synthetic data. 
This is the approach followed by GGWS98, and will hereafter
be referred as {\bf method 2}. Its advantage is 
that the synthetic CMDs contain
all the information about the distribution of stellar luminosities 
and colours, which can then be easily compared to actual observations
of clump stars. (Additional effects such as sample size,
photometric errors, or differential reddening, could also be 
simulated if required.)
However, method 2 is much more demanding,
because it requires tools -- a population synthesis code, and a 
non-linear least squares fitting routine -- which are not needed in 
method 1. 

For our purposes, the interesting point is to test whether
the complexity of method 2 pays for the additional effort with 
a higher accuracy than method 1. 
Thus, we verify whether both methods lead 
to similar results for $M_I^{\rm RC}$. Using our population synthesis 
code (Girardi, unpublished), we generate two different galaxy models.
The first, {\bf model A}, assumes a constant SFR from $T=10$~Gyr ago 
until now, and a metallicity only slightly increasing with the galaxy 
age $T-t$, i.e.\ $Z(t)=0.008+0.011\,[1-0.1\,t({\rm Gyr})]$. 
This represents a relatively 
young and metal-rich galaxy population, as found in the discs of 
spirals. The second, {\bf model B}, assumes a predominantly old population, 
i.e.\ with constant SFR from 8 to 12 Gyr ago, and the complete
range of metallicities given by 
$Z(t)=0.0004+0.0296\,[1-0.25\,(t({\rm Gyr})-8)]$.

\begin{figure}
	\psfig{file=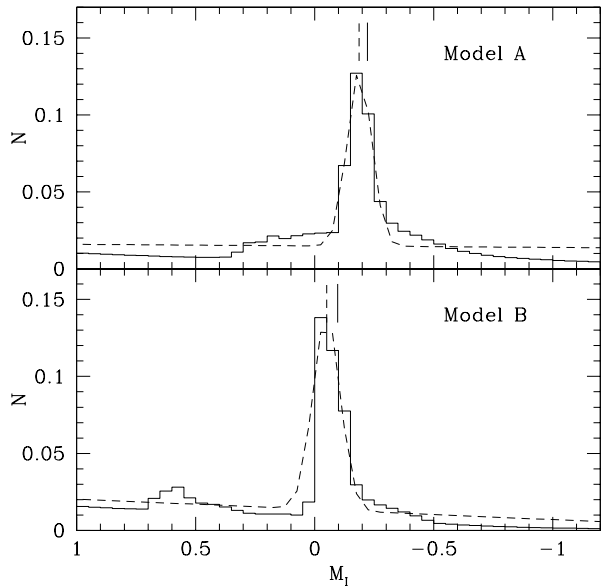,width=8.3cm}
        \caption{$I$-band luminosity functions 
for galaxy Models A and B (see text). The continuous histogram 
represents the synthetic LF as obtained from a population
synthesis code, whereas the dashed line is the result of a 
least-squares fit of eq.~\protect\ref{eq_mu}. For both models
vertical lines indicate the mean $M_I^{\rm RC}$ as obtained either
with method 1 (continuous) or with method 2 (dashed). }
	\label{fig_lf_models}
	\end{figure}

For both models A and B, the mean clump magnitude is computed either with 
method 1 (table~\ref{tab_mi} plus eq.~\ref{eq_integal}) or with 
method 2 (synthetic CMD plus eq.~\ref{eq_fit}). The results are 
presented in Table~\ref{tab_teste}, and in Fig.~\ref{fig_lf_models}. 
\begin{table}
\caption{$M_I^{\rm RC}$ obtained with different methods.}
\label{tab_teste}
\begin{tabular}{lll}
\noalign{\smallskip}\hline\noalign{\smallskip}
Model   & method   & $M_I^{\rm RC}$ \\
\noalign{\smallskip}\hline\noalign{\smallskip}
Model A & method 1 & $-0.222$ \\ 
	& method 2 & $-0.187$ \\
Model B & method 1 & $-0.096$ \\ 
	& method 2 & $-0.050$ \\
\noalign{\smallskip}\hline\noalign{\smallskip}
\end{tabular}
\end{table}
%

As seen, method 1 provides $M_I^{\rm RC}$ values about 0.05~mag
brighter than method 2. This offset results, mainly, from the 
luminosity of the clump LF maximum always being slightly fainter than 
the mean clump brightness (see Fig.~\ref{fig_lf}): whereas the maximum
is more easily accessed by method 2, method 1 always
accesses the mean.

Anyway, the important result is that 
both methods provide almost the same magnitude difference
between the two galaxy models, i.e.\ $\Delta M_I^{\rm RC}({\rm A-B})$ 
is equal to $-0.153$ cf.\ Method 1, and $-0.137$ cf.\ Method 2
(already on the base of of this simple exercise, we
should reasonably expect that the red clump is about 
$0.14-0.15$~mag brighter in a late spiral's disk than in an old 
elliptical galaxy). 
Thus, the results from the two methods turn out to be very similar.
In the following we will always apply method 2, based
on synthetic CMDs, since it permits to compare in more detail
the LF of the synthetic red clump for a given stellar system with the
corresponding observational data.
Moreover, any value of $\Delta M_I^{\rm RC}$ 
will refer to the difference between the `local' clump simulation 
and a synthetic model for the stellar system under scrutiny. 
Our results can be checked by any reader, by using the more simple 
method 1.

\section{An analysis of the Hipparcos clump}
\label{sec_hipparcos}

The clump stars in the Solar Neighbourhood 
(or `the {\em Hipparcos} clump' 
as hereafter referred) are fundamental for distance determinations 
based on clump stars, because they provide the only empirical zero point 
for eq.~\ref{eq_mu} that can be measured with good accuracy. 
The ESA (1997) catalog contains $\sim1500$ clump stars with 
parallax error lower than 10 per cent, 
and hence standard errors in absolute magnitude 
lower than 0.21~mag. The sample defined by this accuracy limit
is complete up to a distance of about $125$~pc. 
Accurate $BV$ photometry is available for these stars
(and also $I$ for $\sim1/3$ of them), and the interstellar 
absorption is small enough to be neglected 
(Paczy\'nski \& Stanek 1998). 
The Lutz-Kelker bias should not affect their mean absolute 
magnitude determination by more than 0.03~mag (GGWS98). 
Therefore, the intrinsic photometric 
properties of the {\em Hipparcos} 
clump can be said to be known with high accuracy,
when aspects such as data quality and number 
statistics are considered.

However, the same can not be said about the population parameters
of these stars, i.e.\ their distributions of masses, ages, and 
metallicities. Since it was only after {\em Hipparcos} that 
a significant number of clump stars was identified in
the Solar Neighbourhood (Perryman et al.\ 1997), 
few works attempted to describe these stars in terms of their parent
populations. This aspect, of course, is of fundamental importance 
in the present work. 

A comprehensive but short discussion of the nearby clump stars'
masses, ages, and metallicities, is presented by Girardi 
(2000). From this latter work, we select and develop the following
points which are more relevant to the present study. 
More specifically, we discuss general aspects in the mass, age, 
and CMD distribution of clump stars that may be applied to any 
galaxy, then focusing on the specific case of the {\em Hipparcos} 
clump.

\subsection{The mass, age and metallicity distributions:
general aspects}
\label{sec_mass}

	\begin{figure*}
	\begin{minipage}{12cm}
	\psfig{file=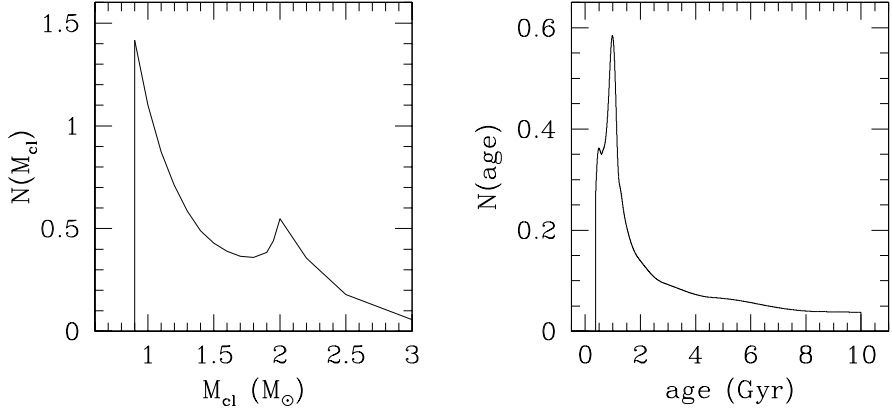,width=\textwidth}
	\end{minipage}
	\hfill
	\begin{minipage}{4.6cm}
      	\caption{
	Mass (left panel) and age (right panel) distribution of 
clump stars for the case of a constant star formation rate (see text). 
       }	 
	\label{fig_mass} 
	\label{fig_age} 
	\end{minipage}
	\end{figure*} 
The mass distribution of core-helium burning (CHeB) stars
in a galaxy of total age $T$, is roughly proportional to the IMF 
$\phi(\Mi)$, to the core-helium burning lifetime $t\sub{He}(\Mi)$ 
of each star, and to the star
formation rate (SFR) at the epoch of its birth, $\psi[t(\Mi)]$
(where $t$ is the stellar age, and not the galaxy age). 
There is also a low-mass cut-off, given by the lowest mass to leave the
main sequence at ages lower than $T$. Since
$t\sub{He}(\Mi)$ presents a peak at about 2~\Msun\ (the transition from
low to intermediate masses), and the IMF shows a peak at the lowest
masses, a {\em double-peaked mass distribution} turns out for 
CHeB stars (Girardi 1999). This is shown in the left panel of 
Fig.~\ref{fig_mass}, for the case of a constant SFR from 0.1 to 
10~Gyr ago, and a Salpeter IMF. This distribution contrasts with
the more natural idea of a clump mass distribution roughly 
following the IMF -- which would present only a single  
peak at the lowest possible masses, i.e.\ at $0.8-1.2$~\Msun. 

Moreover, from these simple considerations, it turns out that the 
intermediate-mass stars from say 2 to 2.5~\Msun, are not severely 
under-represented in the mass distribution, with respect to the
low-mass ones. Stars with $2-2.5$~\Msun\ are close enough to the 
clump region in the CMD, to be considered as genuine clump stars
(GGWS98). For the case of a constant SFR from now up to ages of 10~Gyr, 
they would make about 20 per cent of the clump. Again, this contrasts
with the common idea that the clump is formed {\em only} by 
low-mass stars.

Let us now consider the age distribution expected for
clump stars. The 
number of evolved stars of a certain type and age is proportional 
to: (1) their birth rate at the present time, and (2) 
their lifetime at the evolved stage under consideration 
(see e.g.\ Tinsley 1980). 
The birth rate of clump stars,
$\diff N_{\rm cl}/\diff\tau_{\rm H}$, can be estimated as
\begin{equation}
\frac{\diff N_{\rm cl}}{\diff\tau_{\rm H}} = 
	\psi(t=\tau_{\rm H}) \,
	\phi (\Mto) \,
	\left| \frac{\diff \Mto}{\diff\tau_{\rm H}} \right|
\end{equation}
where \Mto\ is the turn-off mass corresponding to a given
main sequence lifetime $\tau_{\rm H}$, and $\psi(t=\tau_{\rm H})$ is
the star formation rate at the epoch of stellar birth. 

For clump stars, the age distribution is obtained by multiplying 
this birth rate by the CHeB lifetime $\tau_{\rm He}$.
In the right panel of Fig.~\ref{fig_age}, we show the result
for the case of a galaxy with a constant SFR over all its
lifetime. This age distribution turns out to be far from constant: 
it peaks at an age of $1$~Gyr, and
decreases monotonically afterwards. In the particular case here
illustrated, half of the clump stars have ages lower than 2~Gyr. 
This result is in sharp contrast with the common idea that clump 
stars trace equally well the intermediate-age and old components 
of a galaxy. 

These aspects of the age distribution are so important that it
is worth commenting on them in more detail: 

(i) The continuous decline in the clump age distribution at 
ages larger than about $2$~Gyr, comes, essentially, from the continuous 
decrease, with the stellar age, of the birth rate of post-main 
sequence stars.  It can be easily understood as follows:
between 0.8 and 2~\Msun, the main sequence lifetime 
scales approximately as $\tau_{\rm H} \propto \Mto^{-3.5}$ (cf.\
Girardi et al.\ 2000 tables), whereas the lifetime of CHeB stars
$\tau_{\rm He}$ is roughly constant at about $10^8$~yr. Together with a 
Salpeter IMF ($\propto \Mto^{-2.35}$), this implies
that $\diff N_{\rm cl}/\diff\tau_{\rm H}$ (and also the age 
distribution function) scales as $\propto t^{-0.6}$ for ages 
$t\ga2$~Gyr. 

(ii) At ages of about 1~Gyr, $\tau_{\rm He}$ has a local maximum (about 
$2.5\times10^8$~yr, corresponding to the star with $\Mto=2~\Msun$), 
which sensibly increases the amplitude of the age distribution at that age. 

(iii) At younger and decreasing ages,  
$\tau_{\rm He}$ decreases slightly faster than the birth rate increase
of CHeB stars, then causing a decrease in the age
distribution function. Even if the number distribution of CHeB stars
does not become negligible as we go to younger ages, the ones with
ages lower than 0.5~Gyr can hardly be classified as
clump stars, since they become much brighter than the clump. This
justifies defining a cut-off in the age distribution function for 
ages $\tau_{\rm H}<0.5$~Gyr.

Finally, it should be noticed that, for the 
case of a constant SFR, the number of clump stars per unit mass 
of born stars $N_{\rm cl}$, 
as obtained from complete population synthesis models 
(eq.~\ref{eq_ncl}), is simply
proportional to the age distribution function of clump stars we have here 
obtained from simple considerations about stellar lifetimes and 
birth rates:
\[
N_{\rm cl} \propto \phi(\Mi) \, \tau_{\rm He} \, N_{\rm cl}/\diff t \,,
\]
This result is just expected, and can
be well appreciated by comparing the right panel of
Fig.~\ref{fig_age} with the upper panel of Fig.~\ref{fig_mcore}.
The similarity between these two figures, gives us even more
confidence in the numbers derived from eq.~\ref{eq_ncl}.

The hypothesis of constant SFR, above illustrated, is just a 
particular case of a most common one, namely that of 
{\em continued SFR for most of a galaxy's history} which in 
general applies to the disc of spiral and irregular galaxies.
When the SFR has not been constant, the age distribution function of 
clump stars can be
simply evaluated by multiplying the SFR, at any given age, by 
the curves in Fig.~\ref{fig_mcore}. 

\subsection{Simulating the {\em Hipparcos} clump} 

After these introductory considerations of general validity, let us
consider the specific case of the local {\em Hipparcos} clump. 
The local SFR is almost certainly not constant, but there are
good indications that it has been {\em continuous} from today back 
to at least 9~Gyr ago (e.g.\ Carraro 2000). This continuity renders
the age and mass distribution of clump stars qualititavely similar to
the cases illustrated in Fig.~\ref{fig_mass}.

\begin{figure}
	\psfig{file=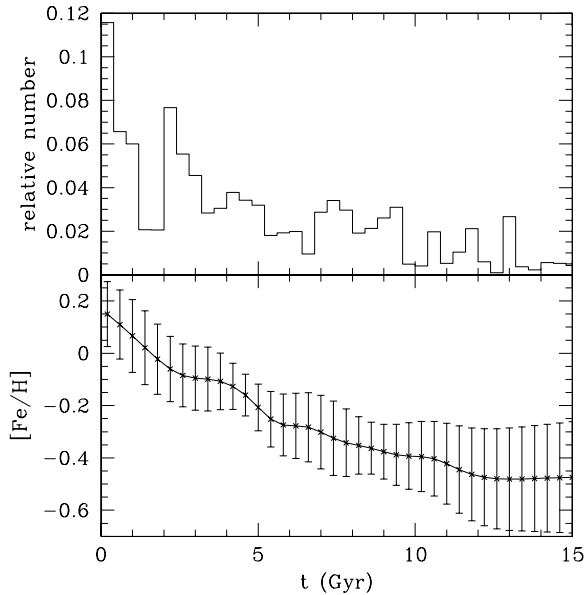,width=8.3cm}
      	\caption{
Adopted SFR and AMR for the Solar Neighbourhood. Upper panel: 
the volume-limited age distribution of local dwarfs as derived
from Rocha-Pinto et al.\ (2000b), sampled at age bins 0.4~Gyr
wide. Lower panel: the corresponding values of \feh\ and its
standard dispersion, as derived from interpolation of table~3
in Rocha-Pinto et al.\ (2000a).  
       }	 
	\label{fig_sfr_helio} 
	\end{figure}

Rocha-Pinto et al.\ (2000b), from a sample of nearby dwarfs 
with ages determined from their chromospheric activity, find a
SFR history marked by several (and statistically significant) 
`bursts' up to the oldest ages. This work also indicates
a volume-corrected age-distribution function which is higher
for the youngest stars, as can be seen in 
Fig.~\ref{fig_sfr_helio}. These properties are also 
supported by other recent works, based on the analysis of
{\em Hipparcos} CMD. Namely, Bertelli \& Nasi (2000) favour 
increasing rates of star formation in the last few 
Gyr, whereas Hernandez, Valls-Gabaud \& Gilmore (2000b) also
find evidences for different bursts in a sample of 
{\em Hipparcos} stars limited to ages lower than 3~Gyr. 

In the following, we adopt the results from Rocha-Pinto et al.'s 
(2000ab) data in order to simulate the local CMD. This work 
covers the complete range of ages we are interested in. 
Before proceeding, we should remind that that stars born at 
older ages in the local disc, are now distributed over 
higher scale heights in relation to the Galactic plane. 
Therefore, in our simulations of the {\em Hipparcos} sample, we should
use the volume-limited age distribution of unevolved dwarfs 
{\em not corrected} by any scale-height factor, 
rather than the cumulative SFR in the so-called `solar cillinder'. 
Rocha-Pinto et al.\ (2000b) kindly provided us with the 
uncorrected age distribution we need. 
Moreover, we have also the possibility of using the local
AMR, derived by Rocha-Pinto et al.\ (2000a) from the same data. 
Both functions are illustrated in Fig.~\ref{fig_sfr_helio}.

\begin{figure}
	\psfig{file=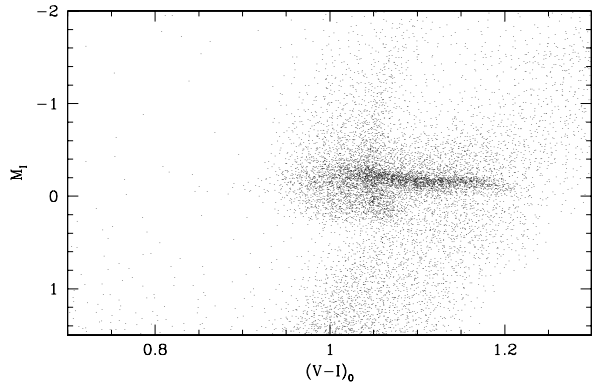,width=8.3cm}
      	\caption{
The synthetic CMD, around the clump region, derived from the data 
of Fig.~\protect\ref{fig_sfr_helio}. We assume a total number of 
10000 clump stars. 
       }	 
	\label{fig_cmd_helio} 
	\end{figure}

\begin{figure}
	\psfig{file=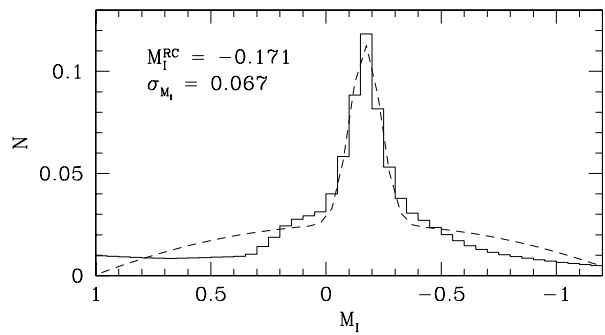,width=8.3cm}
      	\caption{
The synthetic LF, around the clump region, derived from the data 
of Fig.~\protect\ref{fig_sfr_helio}. A Gaussian curve (dashed line)
has been fitted to the LF with the aid of eq.~\ref{eq_mu}. 
       }	 
	\label{fig_lf_helio} 
	\end{figure}

Figure~\ref{fig_cmd_helio} shows the final CMD simulated from 
these data. For the sake of simplicity, we have not simulated --
the same applies also to the simulations presented in 
Sect.~\ref{sec_galaxies} --
observational errors (as discussed in GGWS98 the inclusion of observational
errors increases the width $\sigma_I$ of the clump stars' LF,
leaving almost unchanged the value of $M_I^{\rm RC}$).
Fig.~\ref{fig_lf_helio} shows
the resulting LF of clump stars from this simulation, 
together with the result of fitting eq.~\ref{eq_mu}. 

Even if in this paper we are discussing population effects 
(i.e., the value of $\Delta M_I^{\rm RC}$ for different stellar systems) 
on the red clump brightness,
and we will use theoretical models only in a differential way,
it is nevertheless interesting to note that the absolute value of 
$M_I^{\rm RC}=-0.171$ for the local clump
as derived from our simulations is satisfactorily close to the observed one,
$M_I^{\rm RC}=-0.23\pm0.03$ (Stanek et al.~1998).

%

\subsection{The observed metallicity distribution}
\label{sec_colour}

Let us now compare the results from our {\em Hipparcos} 
simulation, with additional data for nearby clump stars.
Their metallicity distribution has been the subject of some unexpected 
results in the recent past.

There have been few attempts to derive typical metallicities 
for {\em Hipparcos} clump stars. The first one has been a quite indirect
method by Jimenez et al.\ (1998). They applied the concept that red 
giants become redder at higher metallicities to derive, solely from the
colour range of {\em Hipparcos} clump stars, an estimate of their
metallicity range, obtaining $-0.7<\feh<0.0$. 
This approach was based on the fact that in Jimenez et al.\ (1998) 
models the clump at a given metallicity is very concentrated 
on the CMD, and has a mean colour very well correlated with 
metallicity. In this regard, however, we should consider that 
the galaxy models considered by Jimenez et al.\ were characterized 
by a SFR strongly decreasing with the galaxy age (i.e.\ increasing 
with the stellar age). This implies that 
they were considering, essentially, the behaviour of the {\em old} 
clump stars, with masses of about $0.8-1.4$~\Msun. 
Intermediate-mass clump stars with mass $\ga1.7$~\Msun\ 
were even absent in their simulations. As discussed in the previous 
Sect.~\ref{sec_mass}, this turns out to be an incomplete description 
of the clump stars, at least in galaxy systems with recent star 
formation as the Solar Neighbourhood.

In contrast, GGWS98 considered all the interesting mass
range of clump stars, and models with constant SFR up to 10~Gyr ago, 
obtaining clumps with a somewhat more extended distribution in 
colour than Jimenez et al.\ (1998). They
demonstrate that a galaxy model with mean solar metallicity and a very 
small metallicity dispersion ($\sigma_{\rm [Fe/H]}=0.1$~dex), 
shows a clump as wide in colour as the observed {\em Hipparcos} one -- 
i.e.\ with $\Delta(\vi)\simeq0.2$~mag.
This means that a significant fraction of the colour spread of the 
local clump could be due to an age spread, rather than to a metallicity
spread. Considering the SFR we discussed in the previous subsection,
GGWS98 results seem more realistic than 
Jimenez et al.\ (1998). It follows that {\em the colour spread of nearby 
clump stars cannot be interpreted just as being the result of
a metallicity spread} as suggested by Jimenez et al.; the age (mass) 
spread (from $\sim1$ to 10 Gyr) might well be playing an important role. 

The above aspect becomes clear when we consider clump metallicities 
not inferred from broad-band colours. The metallicities of 581 nearby 
K giants (mainly clump stars) have been derived by H\o g \& Flynn (1998), 
based on DDO photometry.  From their data it turns out that the \vi\ 
colour does not correlate with \feh\ (cf.\ Paczy\'nski 1998), contrarily 
to what would be expected from the Jimenez et al.\ (1998) analysis. 

One may suggest that the problem is in the \feh\ values inferred 
from photometry, rather than in the interpretation of clump
\vi\ colours. However, the same problem appears when we look at
the most reliable metallicity data, i.e.\ those derived from
spectroscopy. Udalski (2000) selected clump stars with spectroscopic 
abundance determinations from McWilliam (1990). They 
mention a non negligible range of clump metallicities ($-0.7<\feh<0.0$),
and again confirm the lack of \feh\ versus colour correlation. 

We have constructed a similar sample of red giants with spectroscopic 
abundance determinations. From the {\em Hipparcos} catalog, we select
all red giants (from their position in the colour--absolute magnitude 
diagram) marked as single stars, 
with lower than 10 per cent relative error in the parallax, 
with direct measurements of the \vi\ colour, and
\feh\ determinations according to the Cayrel de Strobel et al.\ (1997) 
catalog. All the \feh\ values are quoted as being relative to the Sun's. 
Nearly half of the selected stars ($\sim 160$ out of 334) are clump stars, 
the bulk of the remaining ones being in their first-ascent of the
red giant branch.  

	\begin{figure*}
	\begin{minipage}{12.2cm}
\psfig{file=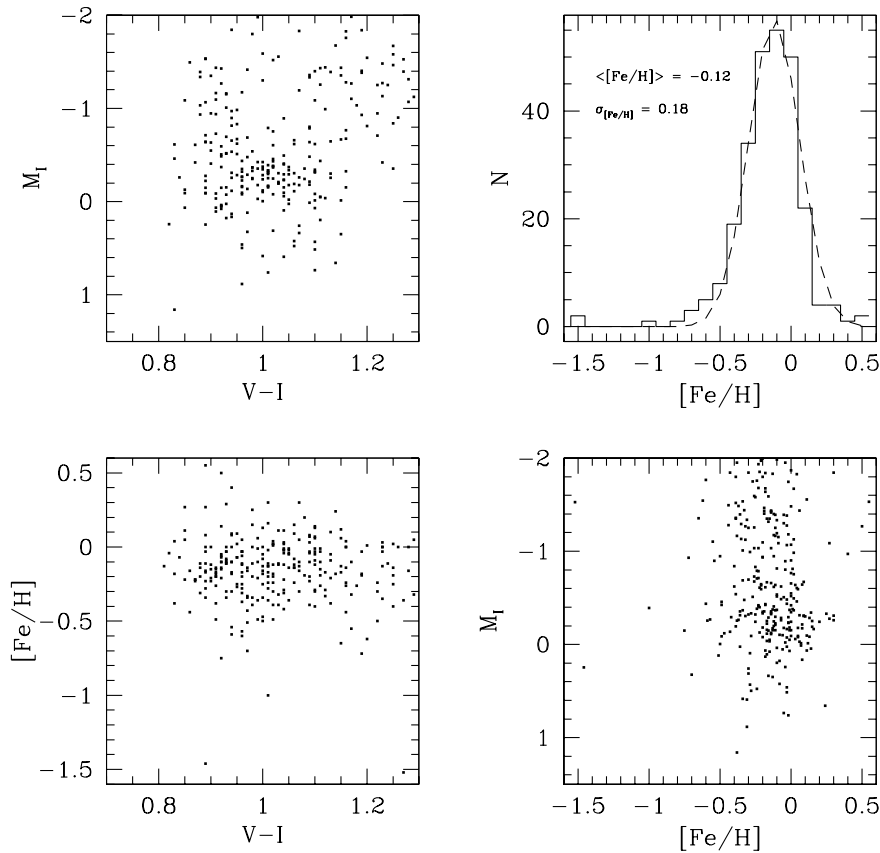,width=\textwidth}
	\end{minipage}
	\hfill
	\begin{minipage}{4.6cm}
       	\caption{
Metallicities of clump stars in the {\em Hipparcos} catalogue
(see text). For the stars shown in the CMD of the upper left panel,
the \feh\ distribution (upper right panel) turns out to be well 
described by a Gaussian of mean $-0.12$~dex and dispersion
0.18~dex. The two lower panels show the
same stars in the \feh\ versus \vi, and \mi\ versus \feh\ diagrams.
}	 
	\label{fig_feh_hipp} 
	\end{minipage}
	\end{figure*} 
%
The results can be appreciated in Fig.~\ref{fig_feh_hipp}. It presents,
in three panels, the quantities \mi, \vi, and \feh, plotted one against 
each other. 
We notice again that, also in this data,  
\feh\ and \vi\ colour seem not to be correlated.  
A fourth panel of Fig.~\ref{fig_feh_hipp} 
presents the histogram of \feh\ values.
Interestingly, we find that the \feh\ distribution of clump stars 
is fairly well represented by a Gaussian curve of mean
$\langle\feh\rangle=-0.12$~dex and standard deviation
of $\sigma_{\rm [Fe/H]}=0.18$~dex, as derived by means of
a least-squares fit. The surprising feature is {\em the very 
small dispersion of metallicities in the data}. It seems
to disagree with the wide metallicity ranges mentioned by 
Jimenez et al.\ (1998; i.e.\ $-0.7\le\feh\le0.0$) and Udalski 
(2000; i.e.\ $-0.6\le\feh\le0.2$). Actually, in the 
latter case, there is no disagreement at all: the total range 
of \feh\ values in our histogram of Fig.~\ref{fig_feh_hipp}, 
is similar to that mentioned by Udalski (2000). 

We notice that the small dispersion of clump 
metallicities, $\sigma_{\rm [Fe/H]}=0.18$~dex, 
has the same order of magnitude as 
the one found among coeval Galactic open clusters, after correction 
for the disc metallicity gradient
(figure~4 in Carraro \& Chiosi 1994), 
or among field stars of same age (Rocha-Pinto
et al.\ 2000a; with $\sigma_{\rm [Fe/H]}=0.13$~dex).
But how can we understand such a small metallicity dispersion,
coming out for clump stars in a so complex stellar environment 
as the Solar Neighbourhood\,? The key to the answer is in our 
discussion of the age distribution of clump stars in
Sect.~\ref{sec_mass}: Actually, {\em 
nearby clump stars are (in the mean) 
relatively young objects, reflecting mainly the near-solar 
metallicities developed in the local disc during the last few Gyr of 
its history}, rather than its complete chemical evolution history. 

It is also interesting 
to notice that the age distribution of nearby clump stars (mostly
K giants), turns out to be very different from that of low-main 
sequence stars (e.g.\ the G dwarfs). This because the long-lived G 
dwarfs have an age distribution simply proportional to the SFR, whereas 
K giants have it `biased' towards intermediate-ages ($1-3$~Gyr). 
This difference reflects into their metallicities: since younger stars 
are normally more metal rich, giants should necessarily be, in the
mean, more metal-rich than G dwarfs. 
G dwarfs in the Solar Neighbourhood
are already known to present a relatively narrow 
distribution of \feh\ -- the relative lack of low-metallicity stars, 
with respect to the predictions from simple closed-box models 
of chemical evolution, being known as the G-dwarf problem. 
Then, even more narrow should be the distribution of \feh\ 
among K giants. And this is, in fact, exactly what is suggested by the
observations of Fig.~\ref{fig_feh_hipp}.

\subsection{The magnitude and colour distribution}
\label{sec_mag}

In the above subsection, we presented arguments and data which go
against the interpretation of clump \vi\ colours 
as being mainly due to metallicity differences. This point is also
crucial for distance determinations. Were the clump colour determined
by metallicity only, the observed constancy of the $I$-band magnitude
with colour inside the clump, in different galaxies, could be
indicating that $M_I$ is virtually independent of metallicity, and
hence an excellent standard candle (cf.\ Paczy\'nski \& Stanek 1998;
Stanek et al.\ 1998; Udalski 1998a). This latter conclusion 
comes from a simplified analysis, where the possibility that the 
clump magnitude and colour depend significantly also on the age, has 
been neglected.

Let us briefly discuss the specific results we get from our models.
In them, \mimax\ presents a non negligible dependence on both age and 
metallicity. Nonetheless, our simulations of the {\em Hipparcos}
clump turn out to present an almost-horizontal clump feature, without
any detectable systematic effect on \mi\ as a function of colour or
\feh.  In fact, dividing our simulation of Fig.~\ref{fig_cmd_helio} 
into `blue' ($\vi<1.1$) and `red' ($\vi>1.1$) samples, we obtain values
of $M_I^{\rm RC}=-0.159$ and $M_I^{\rm RC}=-0.170$, respectively.

How can the galaxy models present almost-horizontal clumps, whereas
$M_I^{\rm RC}$ changes systematically as a function of both age and
metallicity\,? The answer is not straightforward, since a number
of effects enter into the game. First, \mi\ does not change monotonically
with age: it is fainter than the mean (by up to 0.4~mag) at
{\em both} the $\sim1$~Gyr and $\ga10$~Gyr age intervals, which represent 
the bluest and reddest clump stars for a given metallicity 
(provided that $Z$ and $t$ are not, respectively, too low and too old,
to cause the appearance of a blue HB). Due to this effect alone, the 
clump for a given metallicity would describe an arc in the CMD, 
without a detectable mean slope in the \mi\ versus \vi\ CMD (see GGWS98). 
Second, an age-metallicity relation tends to flatten the \mi\ vs.\ age 
relation in the age interval $t\ga2$~Gyr, since older clump stars 
tend to become dimmer due to their larger age, but brighter due to the 
lower metallicity. Third, the presence of a $\sim0.2$~dex
metallicity dispersion among stars of the same age, may cause
substantial blurring in the diagrams of Fig.~\ref{fig_feh_hipp}. 
Which effect prevails depends on the exact shape of
the SFR and AMRs. The important point is that,
due to these effects, the possible correlations between
colour, magnitude, and metallicity of clump stars, may become 
small enough to escape detection.

Therefore, clump models can assume a quite large variety of 
shapes, depending on the SFR and AMR. The position of an individual 
star in the CMD, in general cannot be unequivocally interpreted 
as the result of a given age or metallicity. There are, however,
important exceptions to this rule:
A quite striking feature of the models is the presence 
of a `secondary clump' feature located about 0.4~mag below the 
blue extremity of the clump, accompanied by a bright plume of clump
stars directly above it. These features are described in detail in
GGWS98 and Girardi (1999). They are caused by the
intermediate-mass clump stars, i.e.\ those just massive enough for
starting to burn helium in non-degenerate conditions, and are
the signature of $\la1$-Gyr old populations with metallicities 
$Z\ga0.004$ (Girardi 1999). The important point here is that these
structures are present in the {\em Hipparcos} CMD, and about as 
populated as they are in the models presented in 
Fig.~\ref{fig_cmd_helio}. 
It evidences that some key aspects of the formalism and stellar
models we use are correct, and that the adopted SFR and AMR 
constitute a reasonably good approximation.

\subsection{The magnitude as a function of \feh}

\begin{figure}
	\psfig{file=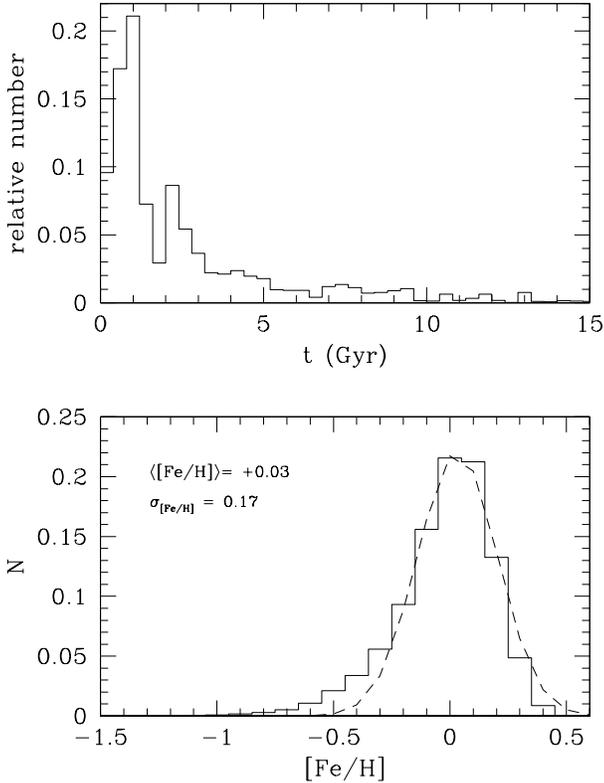,width=8.3cm}
      	\caption{
The distribution of ages (upper panel) and metallicities 
(lower panel) for clump stars in our {\em Hipparcos} model
(continuous lines).
Their mean age turns out to be $2.5$~Gyr.
The \feh\ distribution is well described by a Gaussian 
(dashed line) of mean $+0.03$~dex and dispersion 0.17~dex. 
       }	 
	\label{fig_helio_fehdist} 
	\end{figure}

From our simulation, we can also derive the metallicity 
and age distribution of local clump stars. They are presented 
in Fig.~\ref{fig_helio_fehdist}. 

Although the total range of
metallicities allowed by the model is very large
($-0.7\la\feh\la0.3$, see Fig.~\ref{fig_sfr_helio}), 
the distribution for clump stars turns out to be very narrow:
a Gaussian fit to the \feh\ distribution produces a
mean $\langle\feh\rangle=+0.03$~dex and dispersion 
$\sigma_{\rm [Fe/H]}=0.17$~dex. (Actually, the \feh\ 
distribution presents an asymmetric tail at
lower metallicities, which causes the straight mean of 
\feh\ to be $-0.04$~dex, i.e.\ slightly lower than the center 
of the Gaussian.)
This distribution is almost identical to the observed one 
(Fig.~\ref{fig_feh_hipp}), except for an offset of 
$+0.15$~dex.

The small \feh\ dispersion displayed in the lower panel
of Fig.~\ref{fig_helio_fehdist} is easily understood when we look
at the clump age distribution in the upper panel. 
Not surprinsingly (cf.\ the discussion in Sect.~\ref{sec_colour})
the mean age of nearby clump stars turns out to be 
$\langle t\rangle=2.5$~Gyr, whereas the peak of the 
distribution is at just 1~Gyr. At these ages, the local disc 
metallicity had already grown to $\feh\sim0.0$~dex (cf.\ 
Fig.~\ref{fig_sfr_helio}), and the bulk of clump stars is 
expected to have similar metallicities. Therefore, this 
main aspect of the metallicity distribution -- the 
small \feh\ dispersion -- is very well accounted for by the
models.

The reasons for the $+0.15$~dex offset 
are not clear, and may lie in some offset between the
high resolution spectroscopic \feh\ scale
by Cayrel de Strobel et al.\ (1997) and the one used by
Rocha-Pinto et al.\ (2000a), or in some inconsistency 
between the age scales used by Rocha-Pinto et al.\ (2000b)
and by our stellar models. 
As for the metallicity scale,
Rocha-Pinto et al.\ (2000a) used the \feh\ scale by Schuster \& Nissen
(1989) based on Stroemgren photometry. As discussed by 
Schuster \& Nissen (1989) this metallicity scale is on average $\simeq 0.06$
dex lower than direct high resolutions spectroscopic 
\feh\ determinations (see also Alonso et al.~1996).

Anyway, $+0.15$~dex may be a difference small enough 
to be allowed in the present work. To check this point,
we make a modified {\em Hipparcos} model by adopting
metallicities $0.15$~dex lower than those given by
Rocha-Pinto et al.\ (2000b) AMR. The clump
simulated in this way becomes bluer by about 0.1~mag, if compared
to the one shown in Fig.~\ref{fig_cmd_helio}, and then 
would better agree with the range of \vi\ colours
of the observed {\em Hipparcos} clump (see Fig.~\ref{fig_feh_hipp}). 
For this model, we obtain $M_I^{\rm RC}=-0.209$ with 
$\sigma_{M_I}=0.075$. Thus, a $-0.15$~dex change in \feh\ scale 
would cause just a $-0.04$~mag change in the reference value of 
$M_I^{\rm RC}$.

\begin{figure*}
	\begin{minipage}{8.3cm}
\psfig{file=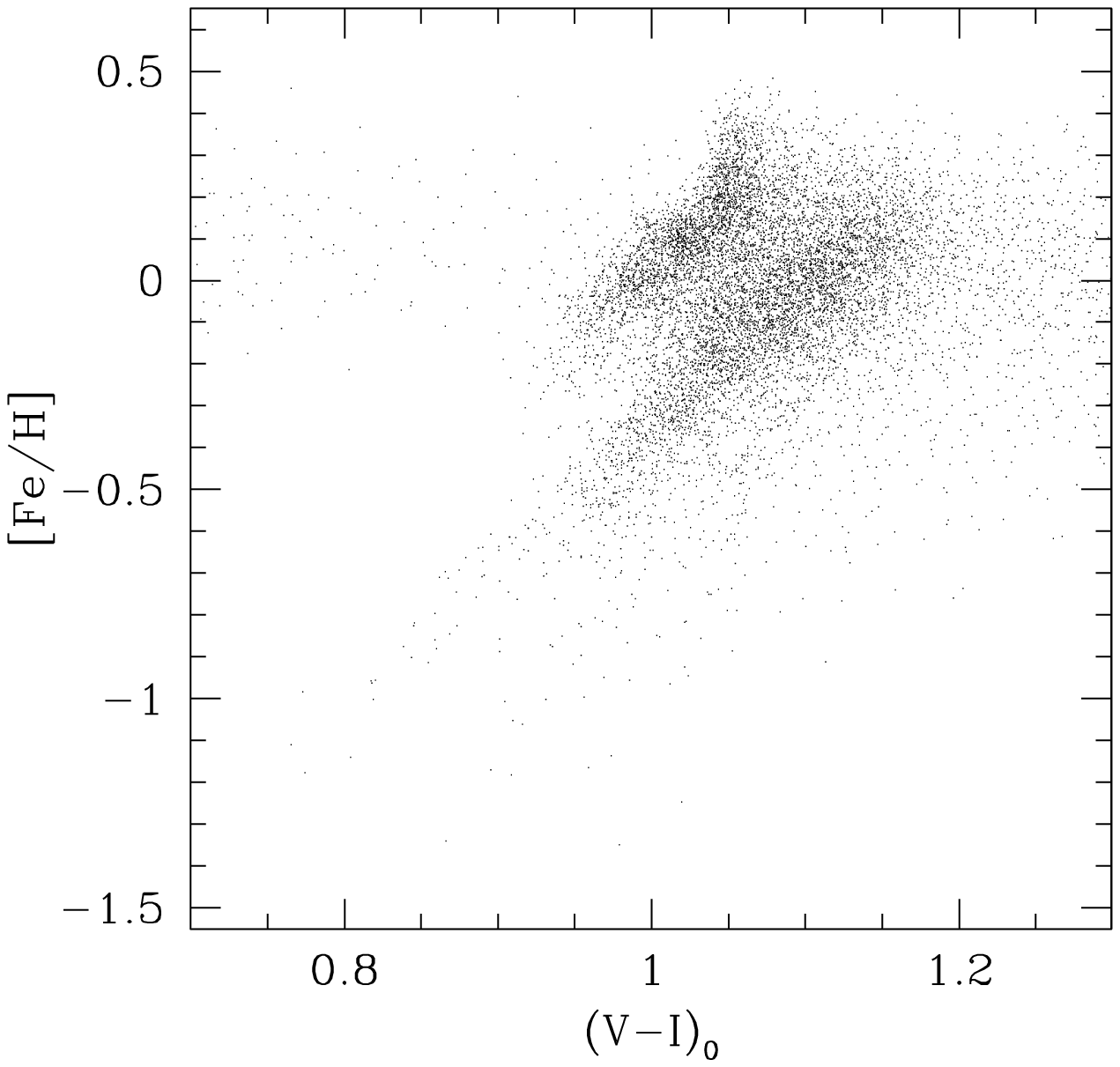,width=8.3cm}
	\end{minipage}
	\hfill
	\begin{minipage}{8.3cm}
\psfig{file=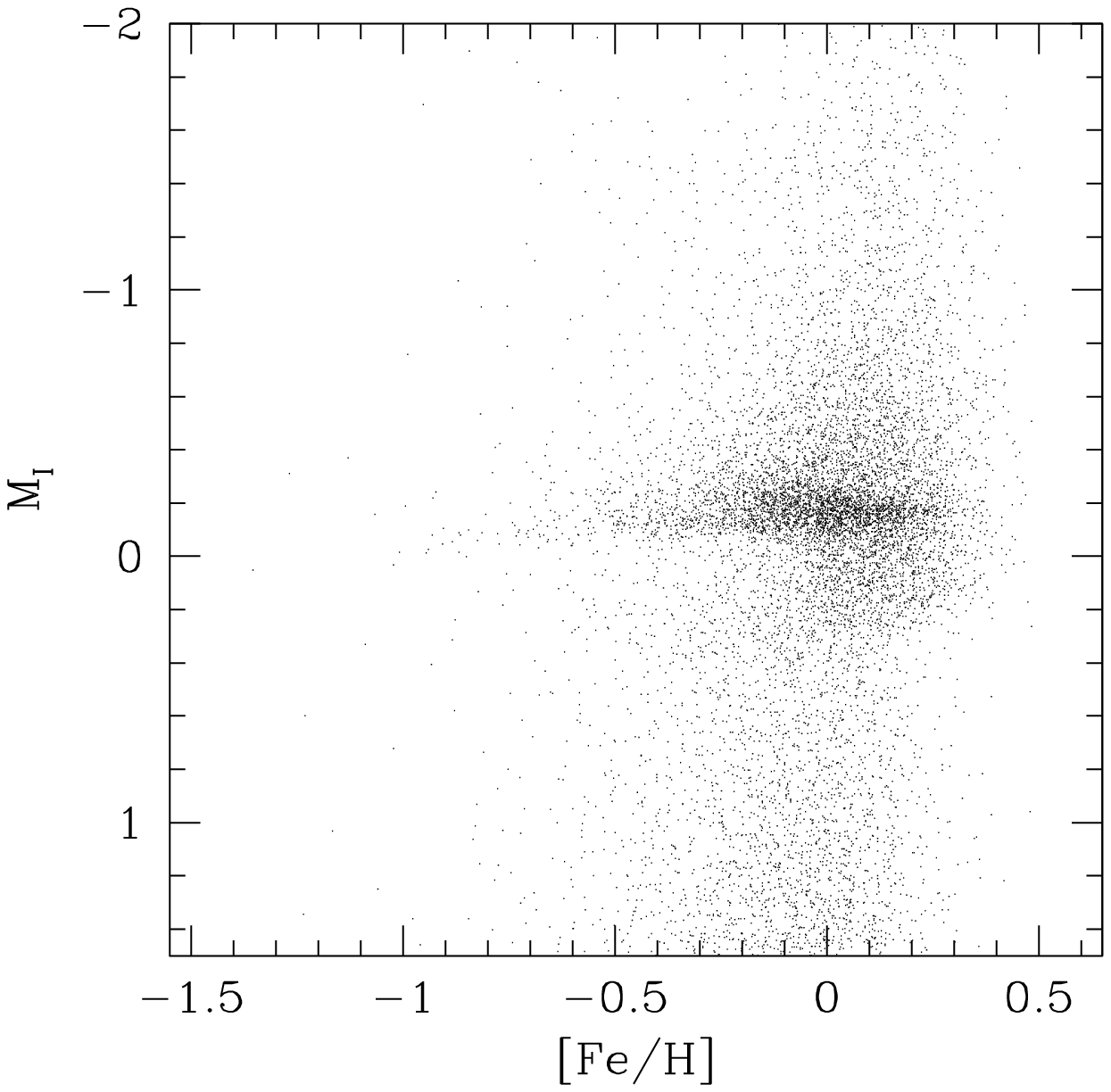,width=8.3cm}
	\end{minipage}
      	\caption{
The distribution of stars from our {\em Hipparcos} model
(see Fig.~\protect\ref{fig_cmd_helio}) 
in the \feh\ versus \vi\ (left panel)
and \mi\ versus \feh\ (right panel) planes. 
Compare with the observational distribution of 
Fig.~\protect\ref{fig_feh_hipp}. 
       }	 
	\label{fig_hipp_helio_feh} 
	\end{figure*}

Fig.~\ref{fig_hipp_helio_feh} shows the distribution 
of stars from our {\em Hipparcos} model in the \feh\ 
versus \vi\ (left panel) and \mi\ versus \feh\ (right panel) 
planes. These occupy the same regions of these planes 
as the data shown in Fig.~\protect\ref{fig_feh_hipp}, 
apart from the offsets of $+0.15$~dex in \feh, 
and $+0.1$~mag in \vi. Also 
the models do not show any significant correlation of
\feh\ with either \vi\ or \mi. The only particularity 
is the presence of two main sequences of clump stars in
the \feh\ versus \vi\ plane, which are due, essentially,
to two main groups of clump stars: the `old' ones which
follow the age-metallicity relation and span the \vi\ 
interval from 1.0 to 1.2, and \feh\ from 0.0 to $-0.6$;
and the youngest ones which have $\feh\ga0.0$ and
concentrate at $\vi\sim0.9-1.2$. With a reasonable 
distribution of observational errors (i.e., typical
errors on the \feh\ values are of about 0.15 dex), 
these two sequences
can give origin to a distribution in which no general 
\feh\ versus \vi\ relation is apparent.
%

Udalski (2000) used the data for local clump stars to
fit \mimax\ to data in different \feh\ bins. The results 
were then used to derive the `metallicity dependence' 
of \mimax. We can now check whether our models produce
a similar relation from synthetic data. There is however,
an important difference in our interpretation of this
relation: in the models, it
does not assume the character of a general relation,
as was intended to be measured in Udalski (2000). Instead, 
{\em the \mimax\ versus \feh\ relation derived in this
way cannot be considered universal, because it represents 
the result of a very particular distribution of clump 
ages and metallicities} (displayed in 
Fig.~\ref{fig_helio_fehdist}).   

\begin{table*}
\caption{$M_I^{\rm RC}$ for {\em Hipparcos} model, separated in 
\feh\ bins, and the derived slope of \mimax\ versus \feh.}
\label{tab_feh}
\begin{tabular}{lllllll}
\noalign{\smallskip}\hline\noalign{\smallskip}
Group     & \feh\ bin limits & \mimax & $\sigma$ & 
clump fraction & $\langle\feh\rangle$ & slope \\
\noalign{\smallskip}\hline\noalign{\smallskip}
1 & $-0.60, -0.25$ & $-0.135$ & 0.039 & 0.12 & $-0.38$ & $-0.07\pm0.06$ (a) \\
  & $-0.25, -0.05$ & $-0.176$ & 0.082 & 0.26 & $-0.14$ & $0.00$ (b) \\
  & $-0.05, +0.20$ & $-0.167$ & 0.049 & 0.58 & 0.07 & \\
\noalign{\smallskip}\hline\noalign{\smallskip}
2 & $-0.45, -0.10$ & $-0.221$ & 0.081 & 0.14 & $-0.22$ & $0.14\pm0.06$ (a) \\
  & $-0.10, +0.10$ & $-0.170$ & 0.059 & 0.46 & 0.01 & $0.24$ (b)\\
  & $+0.10, +0.35$ & $-0.164$ & 0.039 & 0.30 & 0.18 & \\
\noalign{\smallskip}\hline\noalign{\smallskip}
3 & $-0.55, -0.05$ & $-0.218$ & 0.072 & 0.18 & $-0.21$ & $0.13\pm0.05$ (a) \\
  & $-0.05, +0.15$ & $-0.167$ & 0.052 & 0.48 & 0.05 & $0.32$ (b) \\
  & $+0.15, +0.45$ & $-0.164$ & 0.038 & 0.21 & 0.23 & \\
\noalign{\smallskip}\hline\noalign{\smallskip}
\end{tabular} \\
Note: (a) Fit using all 3 bins; and (b) mean slope 
using only the metal-poor and intermediate bin. 
\end{table*}
%
Table~\ref{tab_feh} presents the \mimax\ values we derive 
for different \feh\ bins in the models. They are divided in
groups composed by `metal-poor', `intermediate', and 
`metal-rich' bins. For each group, the final slope of the
\mimax\ versus \feh\ relation is also presented, both for 
the cases in which all bins have been included, and
ignoring the metal-rich bin as in Udalski (2000). The
first group represents the same bin limits as in Udalski
(2000), and results in a very flat \mimax\ versus \feh\ 
relation. The second group has bins shifted by
$+0.15$~dex, in order to account for the offset 
in our models' metallicities; in this case, a marginal 
slope of $0.14\pm0.06$~mag/dex is detected, which 
increases to 0.24~mag/dex if we consider only the 
metal-poor and intermediate bins.

These two former groups present too few stars in the 
metal-poor bin.
We tried to improve upon this point, selecting bin limits 
such as to separate the peak and wings of the metallicity 
distribution shown in Fig.~\ref{fig_helio_fehdist}.
The result is a mean slope of $0.13\pm0.05$~mag/dex.
	\begin{figure}
	\psfig{file=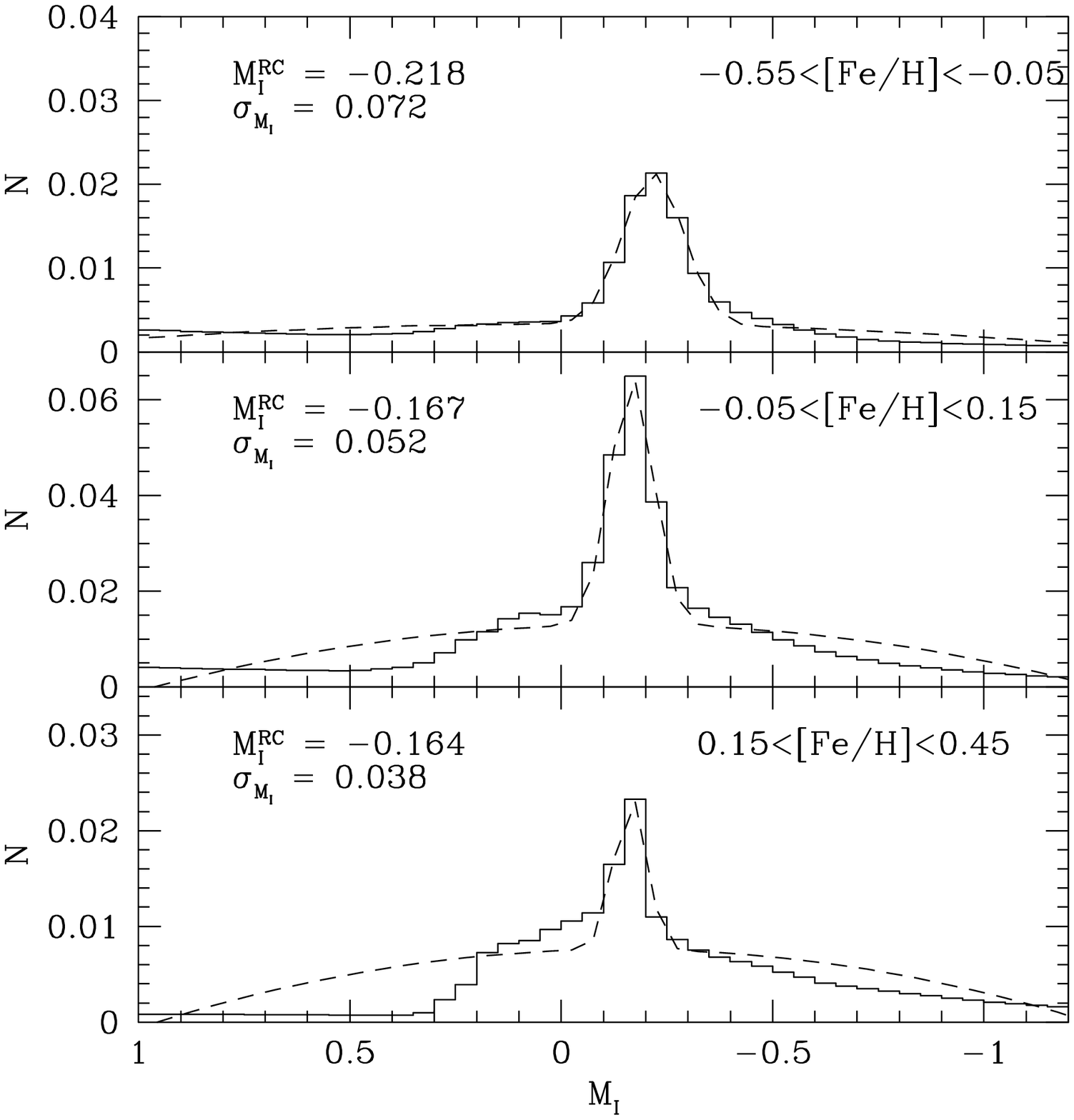,width=8.3cm}
      	\caption{
The LFs for clump stars in our {\em Hipparcos} model,
separated into 3 \feh\ bins (see text).
       }	 
	\label{fig_lf_hipp_helio_feh} 
	\end{figure} 

Udalski (2000) gets a slope 
for the \mimax\ versus \feh\ relation of about 0.2~mag/dex 
using all three bins in their fit, and of $0.13\pm0.07$~mag/dex 
considering only the metal-poor and intermediate bins.
Similar results (Table~\ref{tab_feh}) are obtained with our 
models which include an offset in the metallicity scale. 
However, we do not give any strong 
weight to the final slope resulting from the models. In fact, 
the present exercises convinced us that the obtained 
slope may depend somewhat on the way the bins are defined.
Moreover, the simulation of observational errors in the 
models, could probably lead to slightly different results.
What we have done in this section should be considered, 
rather than a model calibration, just a check of 
whether Udalski's results can be understood with present 
theoretical models of clump stars.

\section{The clump in star clusters}
\label{sec_clusters}

We have seen in the previous section how theoretical models
are able to reproduce the main observational features of the 
{\em Hipparcos} red clump. Another crucial test for the reliability
of theoretical models involves the use of star clusters.
Since star clusters are made of stars all with the same age and 
initial chemical composition, they constitute template 
single-burst stellar populations with which it is possible 
to compare the $\Delta M_I^{\rm RC}$ values derived from theory.
With a large sample of clusters of different ages and metallicities,
we can directly test the predicted
metallicity and age dependence of this key quantity.
Galactic open clusters are particularly useful in this respect, since
they are well studied objects with a large age range, and reasonably
accurate distance and age estimates.

To this aim, we have adopted the data (ages, $M_I^{\rm RC}$ and \feh) 
by Sarajedini (1999) for a sample
of 8 galactic open clusters, plus the data from Twarog et al.\
(1999) for NGC\,2506. The clusters span the age range
between 1.9 and 9.5~Gyr, while the metallicities range
between $\feh=-0.39$ and $\feh=0.15$.
Since we want to test the theoretical $\Delta M_I^{\rm RC}$ values,
for each cluster we have computed the observational 
$\Delta M_I^{\rm RC}$ value using 
$M_I^{\rm RC}({\rm Hipp.})=-0.23\pm0.03$ (Stanek \& Garnavich 1998)
for the local red clump.
The same quantity has been derived from the theoretical models, 
using for the local clump the $M_I^{\rm RC}$ value given in 
Table~\ref{tab_galaxies}. There is a further detail we considered;
since Sarajedini (1999) measured the luminosity of the 
peak of the red clump stellar distribution, and not its mean
value, we have derived from our simulations  
(see discussion in Sect.~\ref{sec_howto}) 
the $M_I$ values corresponding to the
peak of the red clump LF, for the relevant age and
metallicity range, rather than the mean values defined in
Sect.~\ref{sec_agemet}. We remark that the peak values
are systematically lower than mean values by about 0.06~mag.

	\begin{figure}
	\psfig{file=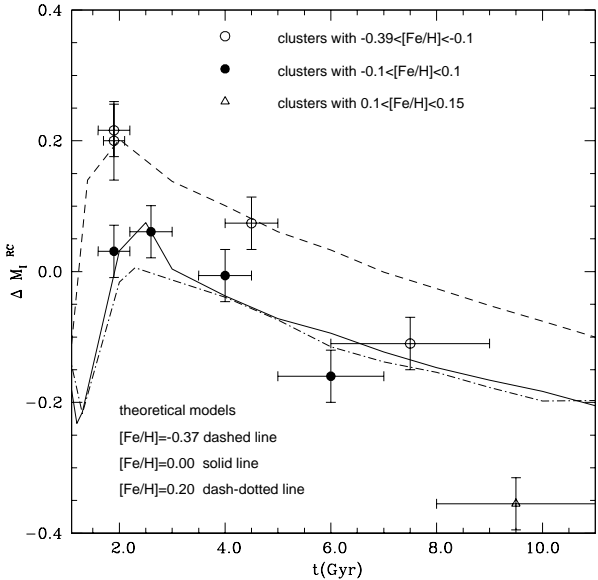,width=8.3cm}
      	\caption{
Comparison of theoretical $\Delta M_I^{\rm RC}$ values with
the observational counterpart in a sample of open clusters of various
metallicities and ages. 
       }	 
	\label{fig_clusters} 
	\end{figure} 
%
In Fig.~\ref{fig_clusters} we show the comparison between
observed and theoretical $\Delta M_I^{\rm RC}$ values.
It is immediately evident how the observations show a marked
dependence of $\Delta M_I^{\rm RC}$ on the clusters' age and 
\feh\ values. Moreover, it is clear that observations are well 
reproduced by theoretical models (with the sole exception of 
NGC\,6791, the oldest cluster in the plot). 
This lends further support to the use of 
theoretical models for computing $\Delta M_I^{\rm RC}$.

Before concluding this section we just mention that
the data by Sarajedini (1999) and Twarog et al.\ (1999) 
are inconsistent with the ones by Udalski (1998b),
who has shown how the $M_I^{\rm RC}$ 
value for a sample of 15 star clusters in the LMC and SMC 
is largely independent of age and metallicity, in the
age interval from 2 to 10 Gyr.
However, it is clear that the determination of metallicity and age
for LMC and SMC clusters is subject to larger uncertainties than
in the case of their Galactic counterparts. Moreover, additional
uncertainties due to depth effects and geometric corrections 
(values as high as $\pm0.2$~mag being applied for some SMC 
clusters) can also spuriously modify the observed relationship 
between red clump brightness, metallicity and age.

Udalski (1998b) has also noticed that in the
oldest clusters of his sample the clump is $0.3-0.4$~mag 
fainter than the mean in the $2-10$~Gyr interval. 
This observation could be simply reflecting the gradual 
fading of the clump, that occurs in the models for 
ages larger than 3~Gyr (see Fig.~\ref{fig_clusters}). 

We conclude that the empirical evidence for a negligible 
dependence of the clump magnitude on age (Udalski 1998b), 
is weak compared to the evidence that it depends on age
(Fig.~\ref{fig_clusters}). A larger sample of 
Magellanic Clouds cluster data may certainly improve 
upon the present-day results.

We also remark that in galaxies with recent star formation, 
a large fraction of the clump stars should have ages in 
the $1-2$~Gyr interval (as discussed in Sect.~\ref{sec_mass}). 
Clusters younger than 2~Gyr, however, are not present in
Udalski's (1998b) and Sarajedini's (1999) samples. It would be 
extremely interesting, in future empirical works, to test the 
age dependence of $M_I^{\rm RC}$ in samples containing also 
younger clusters.

\section{The clump in other galaxy systems}
\label{sec_galaxies}

In Sects.~\ref{sec_agemet} and \ref{sec_hipparcos}, 
the main factors determining the mean clump magnitude 
in a galaxy model have been extensively 
reviewed. In the present section we 
proceed computing $M_I^{\rm RC}$ for a series of models
representing nearby galaxy systems, whose mean clump magnitudes
have been used
in the past for distance determinations (by galaxy systems we mean 
composite stellar populations, whose stars cover  
relatively large intervals of age and metallicity).
All results from this section are summarized in 
Table~\ref{tab_galaxies}.
\begin{table*}
\caption{$M_I^{\rm RC}$ and mean $\feh$ values for the clump in
nearby galaxy systems.}
\label{tab_galaxies}
\begin{tabular}{lllllll}
\noalign{\smallskip}\hline\noalign{\smallskip}
System      & SFR$(t)$   & AMR &  $M_I^{\rm RC}$ & 
$\Delta M_I^{\rm RC}$ & ${\rm [Fe/H]}^{\rm RC}$ & comm. \\
\noalign{\smallskip}\hline\noalign{\smallskip}
Solar Neighbourhood  & Rocha-Pinto et al.\ (2000b) & 
	Rocha-Pinto et al.\ (2000a) & $-0.171$ & 0.000 & $-0.04$ & ($\star$) \\
({\em Hipparcos}) &  &  &  &  & \\
\noalign{\smallskip}\hline\noalign{\smallskip}
Baade's Window & $8-12$~Gyr old & McWilliam \& Rich (1994) & $-0.087$ & 
		$-0.084$ & $-0.22$ & ($\star$) \\
(solar-scaled) & Moll\'a et al.\ (2000) & Moll\'a et al.\ (2000)
		& $-0.063$ & $-0.108$ & $-0.36$ & (1) \\
\noalign{\smallskip}\hline\noalign{\smallskip}
Baade's Window & $8-12$~Gyr old & McWilliam \& Rich (1994) & $-0.161$ & 
		$-0.010$ & $-0.22$ & ($\star$) \\
($\alpha$-enhanced) & Moll\'a et al.\ (2000) & Moll\'a et al.\ (2000) 
		& $-0.148$ & $-0.023$ & $-0.36$ & (1) \\
\noalign{\smallskip}\hline\noalign{\smallskip}
Carina dSph & Hernandez et al.\ (2000a) & $\feh=-1.7$ & $-0.458$ 
	& $+0.287$ & $-1.7$ & ($\star$) \\
            & Hurley-Keller et al.\ (1998) best model &  
	$\feh=-1.7$ & $-0.271$ & $+0.100$ & $-1.7$ & (2) \\
\noalign{\smallskip}\hline\noalign{\smallskip}
SMC & Pagel \& Tautvaisiene (1998) & Pagel \& Tautvaisiene (1998) & 
	$-0.457$ & $+0.286$ & $-0.77$ & (3, $\star$)\\	
\noalign{\smallskip}\hline\noalign{\smallskip}
LMC bar & Holtzman et al.\ (1999; their fig.\ 2) & Pagel \& 
	Tautvaisiene (1998) & $-0.371$ & $+0.200$ & $-0.39$ & ($\star$) \\
        & Holtzman et al.\ (1999; their fig.\ 4) & Pagel \& 
	Tautvaisiene (1998) & $-0.373$ & $+0.202$ & $-0.39$ & \\
        & Holtzman et al.\ (1999; their fig.\ 11) & Pagel \& 
	Tautvaisiene (1998) & $-0.386$ & $+0.215$ & $-0.40$ & \\
LMC outer fields & Holtzman et al.\ (1999; their fig.\ 3) & Pagel \& 
	Tautvaisiene (1998) & $-0.360$ & $+0.189$ & $-0.37$ & \\
        & Holtzman et al.\ (1999; their fig.\ 5) & Pagel \& 
	Tautvaisiene (1998) & $-0.360$ & $+0.189$ & $-0.37$ & \\
        & Holtzman et al.\ (1999; their fig.\ 12) & Pagel \& 
	Tautvaisiene (1998) & $-0.396$ & $+0.225$ & $-0.38$ & \\
LMC northern fields & Dolphin (2000) & Dolphin (2000) & $-0.281$ & 
	$+0.110$ & $-0.88$ & (4) \\
\noalign{\smallskip}\hline\noalign{\smallskip}
\multicolumn{6}{l}{($\star$) The most representative or `preferred' 
value of $\Delta M_I^{\rm RC}$ for distance determinations.} \\
\multicolumn{6}{l}{(1) Theoretical model for the bulge.} \\
\multicolumn{6}{l}{(2) The clump LF is double, resulting in a bad fit of 
eq.~\protect\ref{eq_mu}. Their second and third best models 
give identical results.} \\
\multicolumn{6}{l}{(3) SFR comes from a theoretical model.} \\
\multicolumn{6}{l}{(4) This model produces a CMD with far too many 
old metal-poor stars.} 
\end{tabular}
\end{table*}

There are just two further technical details we need to 
mention before proceeding: (i) In all models, we assume the
relation $\feh=\log(Z/0.019)$, 
which is appropriate for populations with scaled-solar
distributions of metals. The case of stars with enhancement of 
$\alpha$-elements will be discussed in Sect.~\ref{sec_alpha}. 
(ii) For the few cases in which
tracks with metallicities lower than $Z=0.0004$, or higher than
$Z=0.03$ are required, we use these two
limiting values. This is done to avoid risky extrapolations of 
the model behaviours.
Anyway, apart from the case of the Carina dSph galaxy,
stars with such extreme metallicities represent 
just a tiny fraction of the clump stars in our models. 

\subsection{The Baade's Window clump} 

Determining the ages of Bulge stars is difficult, and so its
age distribution has been somewhat uncertain.
Whereas there are some indications
for the presence of intermediate-mass 
stars (see Rich 1999 for a review), studies of Baade's Window
population (e.g.\ Frogel 1988; Holtzman et al.\ 1993; 
Ortolani et al.\ 1995; Ng et al.\ 1996) generally indicate 
that the bulk of star formation occurred at old ages.
Only the very central regions of the Bulge show unequivocal
evidences of recent star formation 
(e.g.\ Frogel, Tiede \& Kuchinski 1999; Figer et al.\ 1999).

Let us then assume that the bulk of stellar populations
in the Bulge are older than 5~Gyr. After this age,
the clump magnitude fades slowly (only $\sim0.025$~mag/Gyr), and
$N_{\rm cl}(t)$ becomes almost flat (i.e.\ the age distribution
of clump giants become simply proportional to the SFR, as seen
in Fig.~\ref{fig_mcore}).
Therefore, the assumptions about the actual SFR history affect 
much less $M_I^{\rm RC}$ for such a population, than in the
case of galaxies with ongoing star formation.

\begin{figure}
	\psfig{file=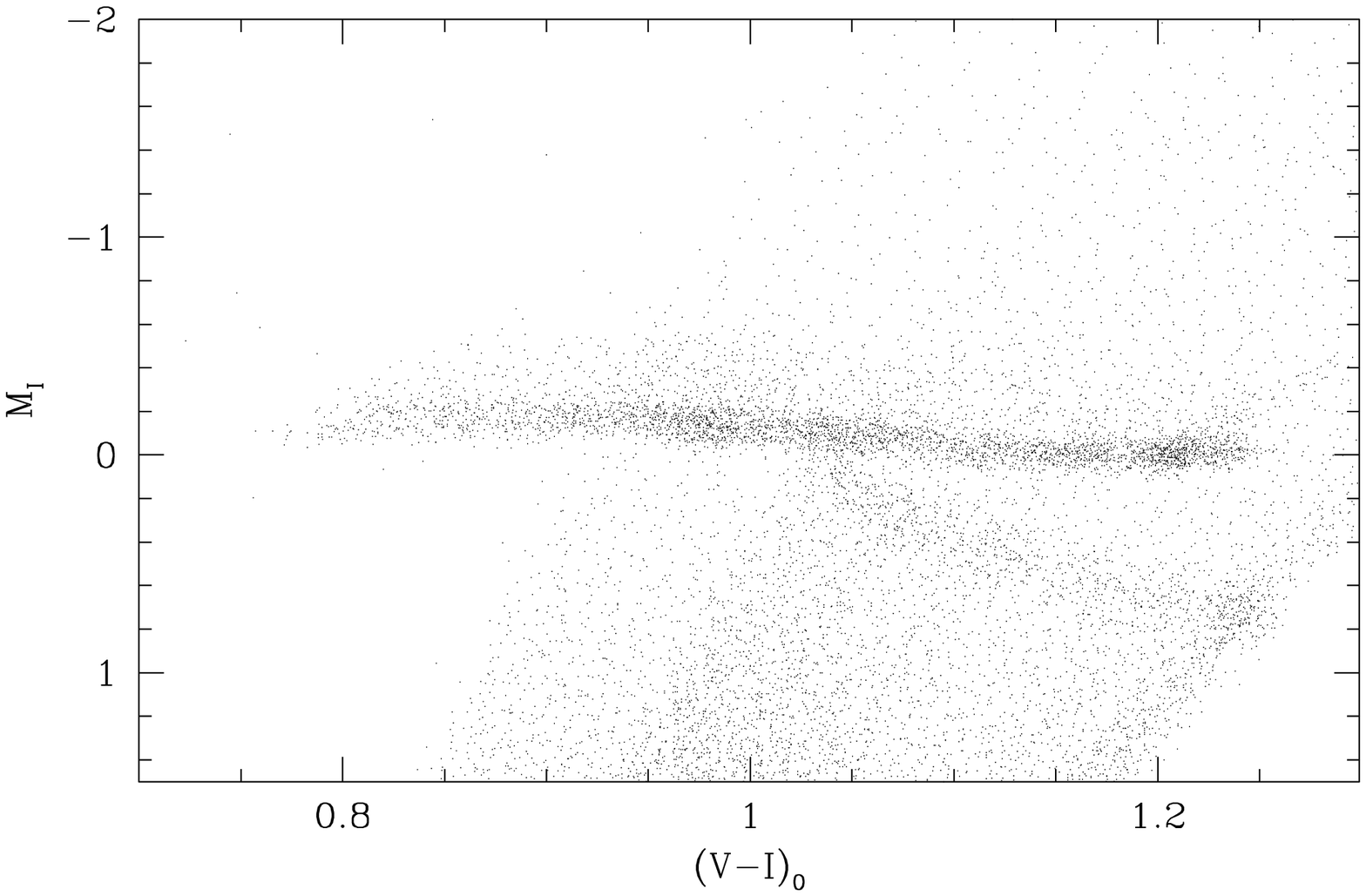,width=8.3cm}
	\psfig{file=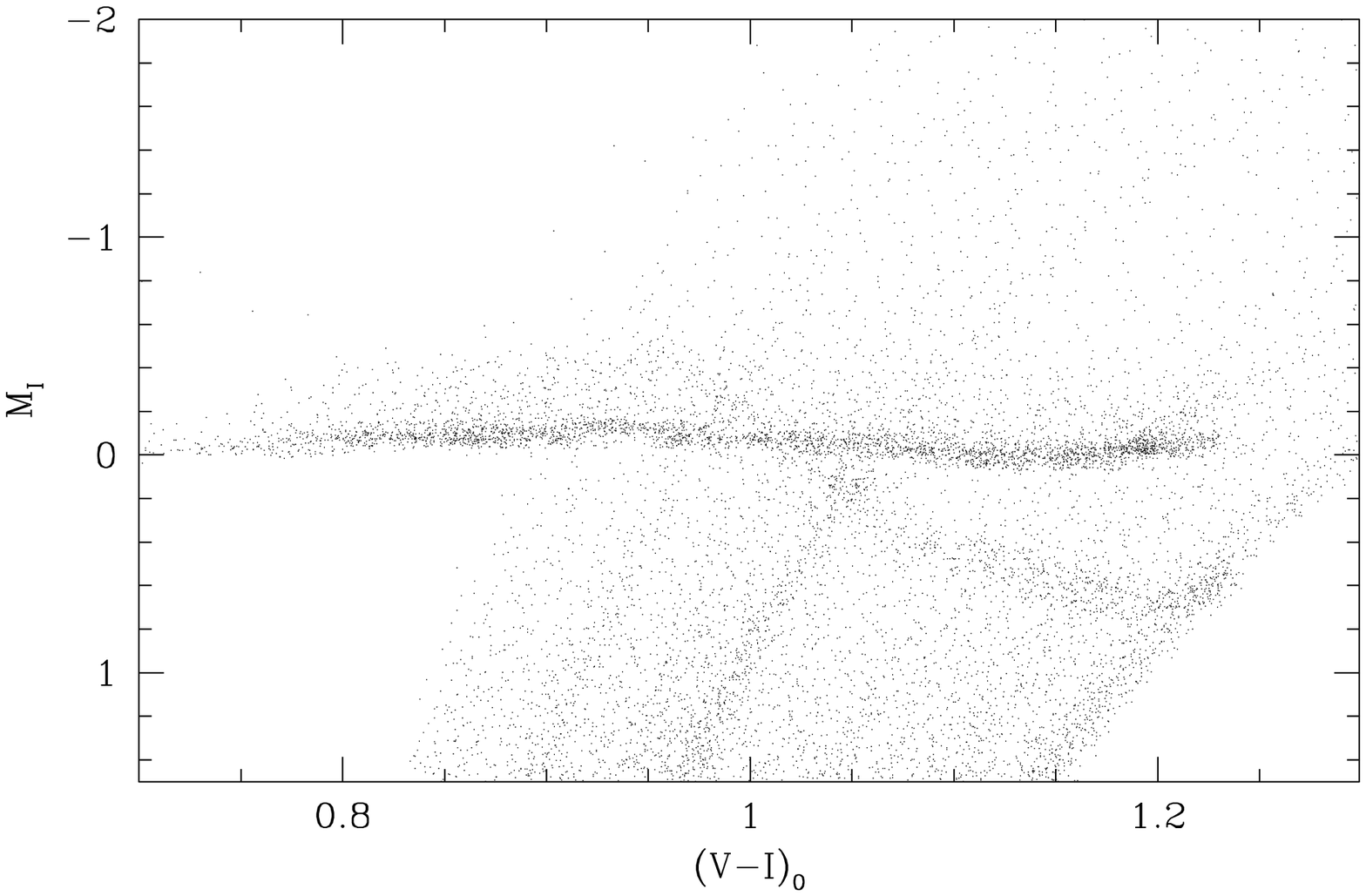,width=8.3cm}
      	\caption{
The same as Fig.~\protect\ref{fig_cmd_helio}, but for the
Bulge, as derived from two different models:
(upper panel) an `old' model with constant SFR between 8 and 12~Gyr 
and the McWilliam \& Rich (1994) \feh distribution at any age, 
and (lower panel) a model following the SFR and AMR from the
bulge theoretical model from Moll\'a et al. (2000).}	 
	\label{fig_cmd_bulge} 
	\end{figure}

We compute two models for the Bulge (or Baade's Window),
illustrated in Fig.~\ref{fig_cmd_bulge}:
The first assumes an `old' Bulge, with constant SFR between
8 and 12~Gyr ago, and the McWilliam \& Rich (1994) distribution
of \feh\ values at any age. The latter has been obtained from 
spectroscopic analysis of Bulge K giants. 
Our second model comes from the
recent `bulge model' by Moll\'a, Ferrini \& Gozzi (2000): 
it represents still a predominantly old Bulge, but with an 
ever-decreasing SFR going up to the present days. 
The $\feh(t)$ relation is derived from their chemical 
evolution models, and is shown (Moll\'a et al.\ 2000) to 
produce a distribution of \feh\ values very similar to the
observational one by McWilliam \& Rich (1994).
%
	\begin{figure}
	\psfig{file=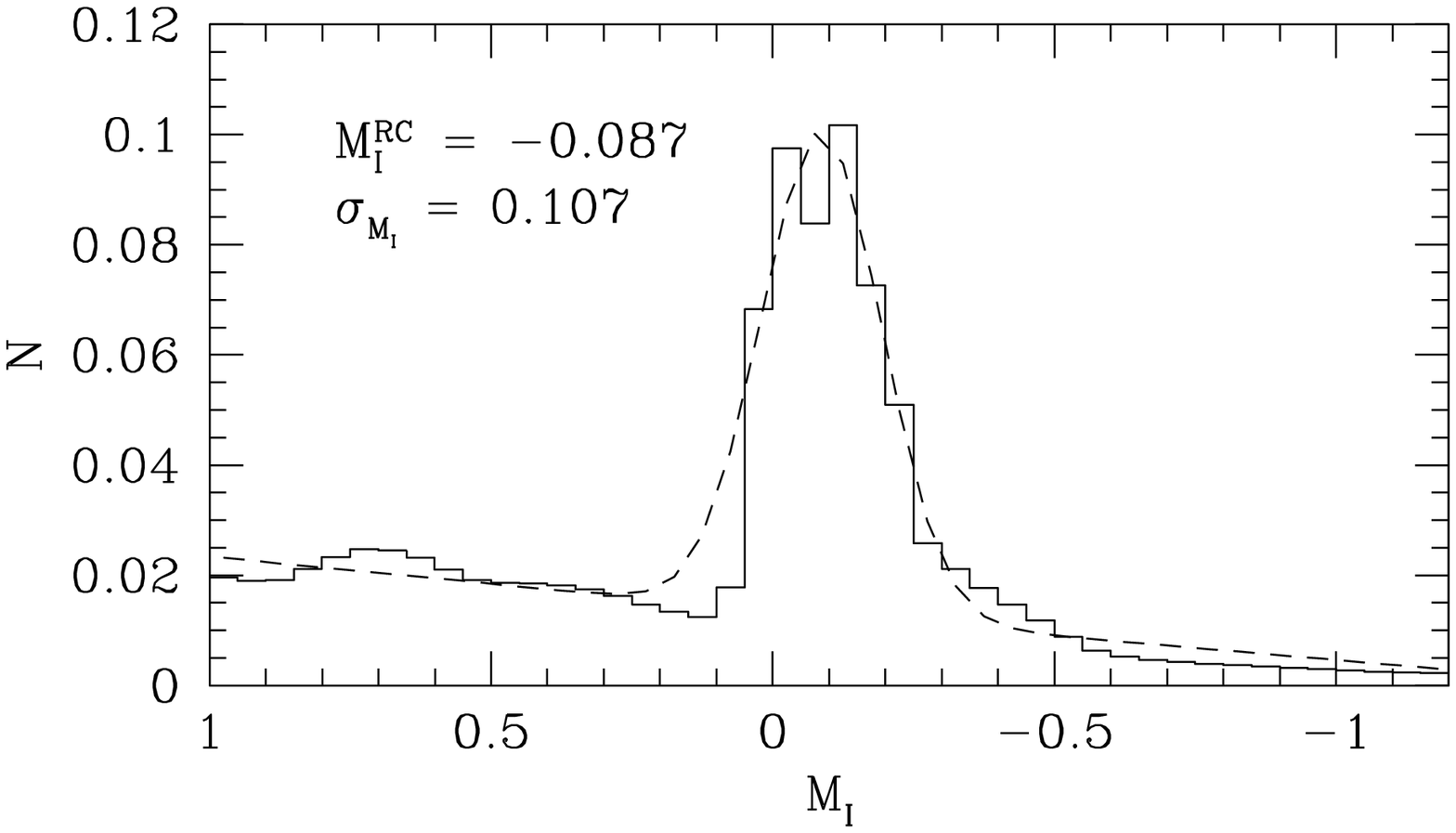,width=8.3cm}
	\psfig{file=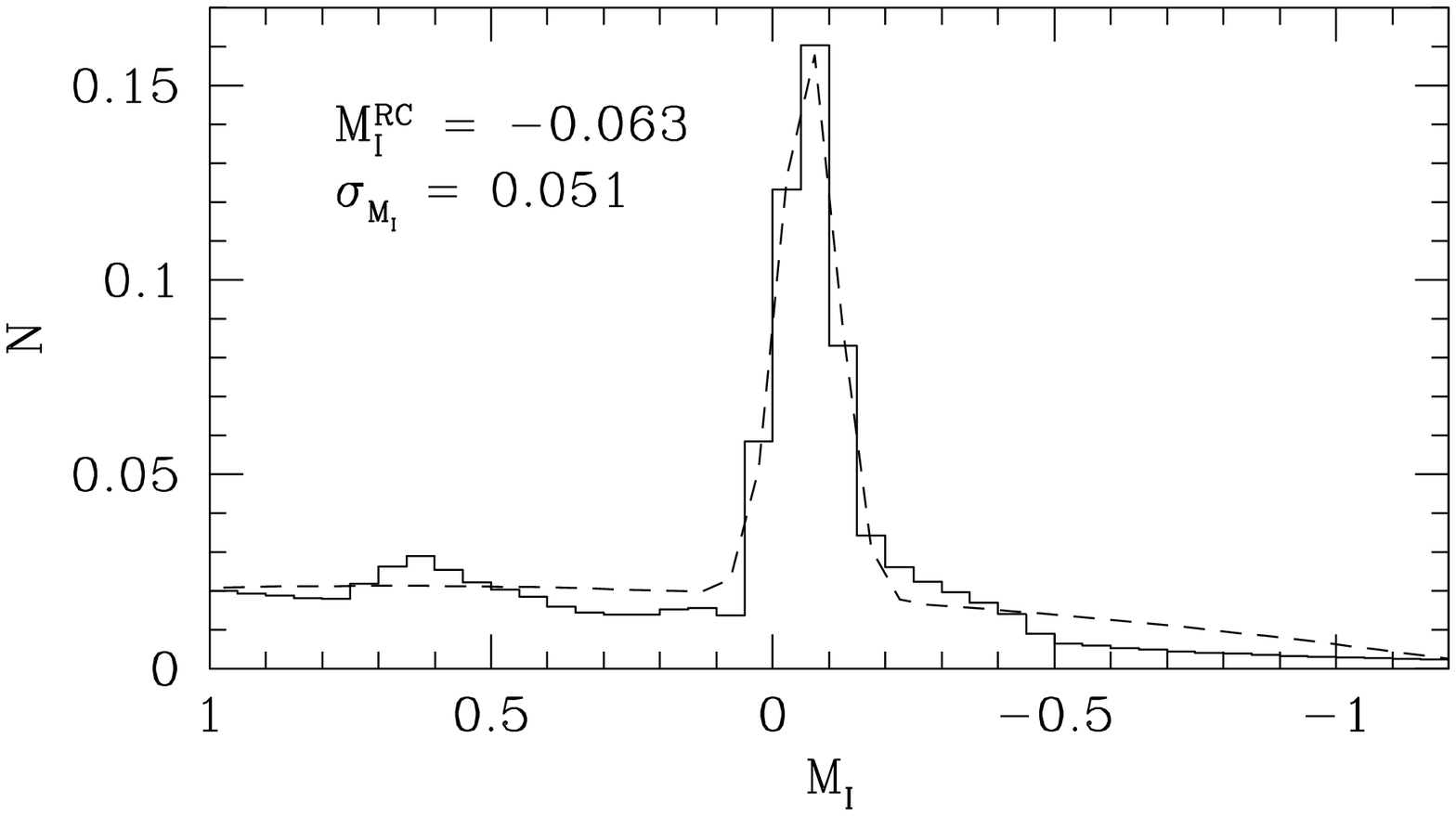,width=8.3cm}
      	\caption{
The same as Fig.~\protect\ref{fig_lf_helio}, but for the
Bulge models presented in 
Fig.~\protect\ref{fig_cmd_bulge}.}	 
	\label{fig_lf_bulge} 
	\end{figure} 

In both cases, Bulge models show an almost-horizontal clump --
in very good agreement with the observational CMD presented by
Paczy\'nski \& Stanek (1998) --
very extended in colour and almost evenly populated along the
colour sequence (Fig.~\ref{fig_cmd_bulge}). 
Also in both cases, Bulge $M_I^{\rm RC}$ values turn out to be 
very similar, and about 0.1~mag higher (i.e.\ fainter) than the 
local {\em Hipparcos} ones (Table~\ref{tab_galaxies}).  

Notice that the mean metallicity of clump stars 
is rather similar in the local and Bulge samples 
(Table~\ref{tab_galaxies}). 
Therefore, the fainter clump we find in Bulge models is, 
essentially, the result of their higher mean ages, compared to
local clump stars. 

\subsection{The clump in the Carina dSph}

The Carina dSph galaxy represents an interesting case, because 
of its very low metallicity, of mean
$\feh=-1.9$ and 1$\sigma$ dispersion of 0.2~dex
(Mighell 1997). Moreover, its CMD clearly 
indicates episodic star formation (Smecker-Hane et al.\, 1996),
with main bursts occurring $\sim15$, 7, and 3~Gyr ago
(Hurley-Keller, Mateo \& Nemec 1998).

Weighting these 3 different episodes of star formation in
different proportions, and with different durations, 
Hurley-Keller et al.\ (1998) find 
several solutions for the SFR history of Carina. 
We have tested their best model (first entry in their table~7),
assuming metallicity values equal to $Z=0.0004$ 
($\feh=-1.7$) at any age. 
Actually, we have tested Hurley-Keller et al.'s
three best models, obtaining always identical results for 
$\mimax$.

In addition, we have tested the SFR history derived by 
Hernandez, Gilmore and Valls-Gabaud (2000a). They apply an 
objective numerical algorithm to find the SFR which best fits the 
observed CMD, without imposing artificial or subjective
constraints on it. Their solution is characterized 
by periods of marked star formation separated by lower
(but not null) activity. Similar results have been obtained by
Mighell (1997), who also uses a non-parametric approach.

\begin{figure}
	\psfig{file=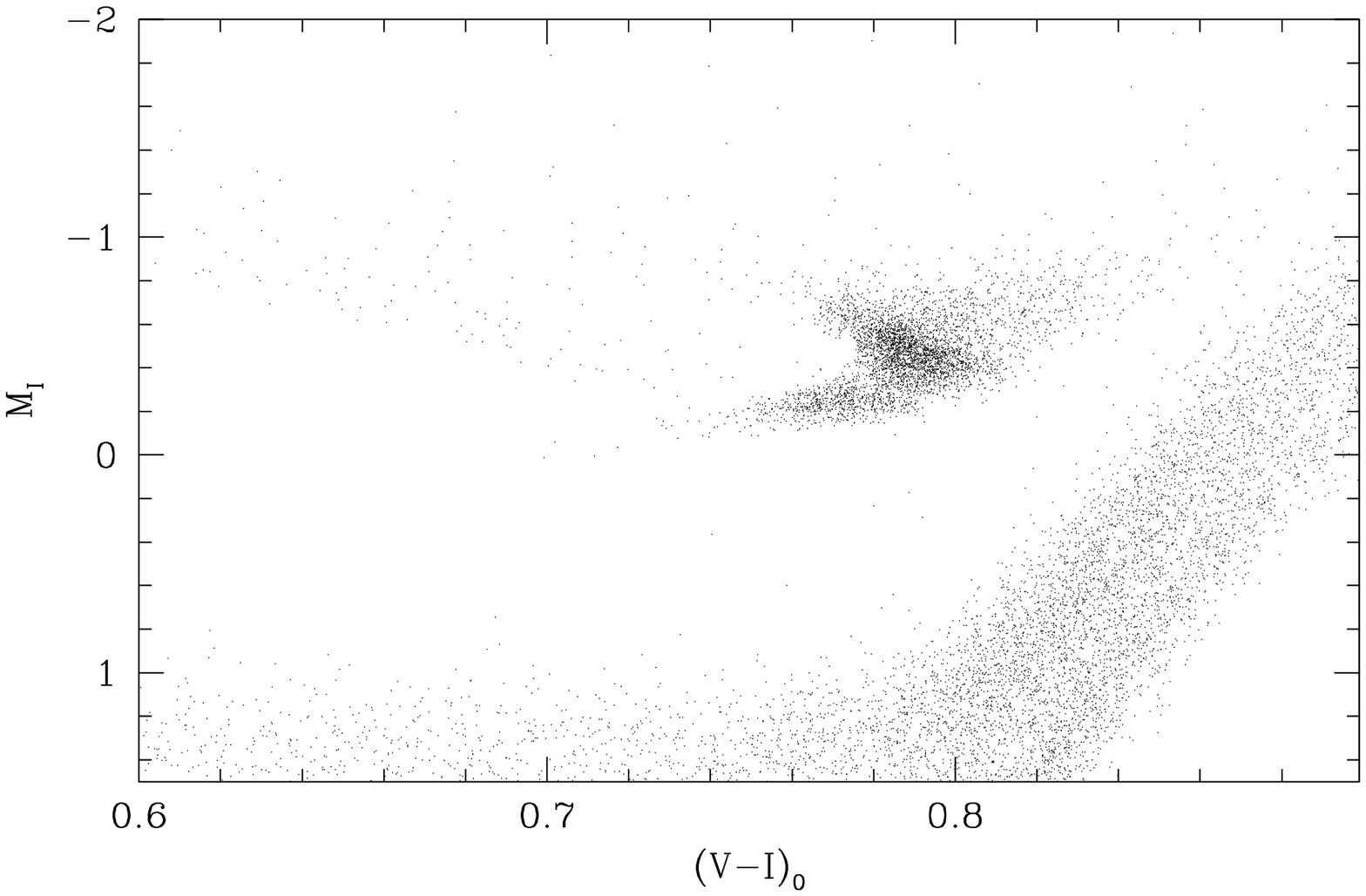,width=8.3cm}
	\psfig{file=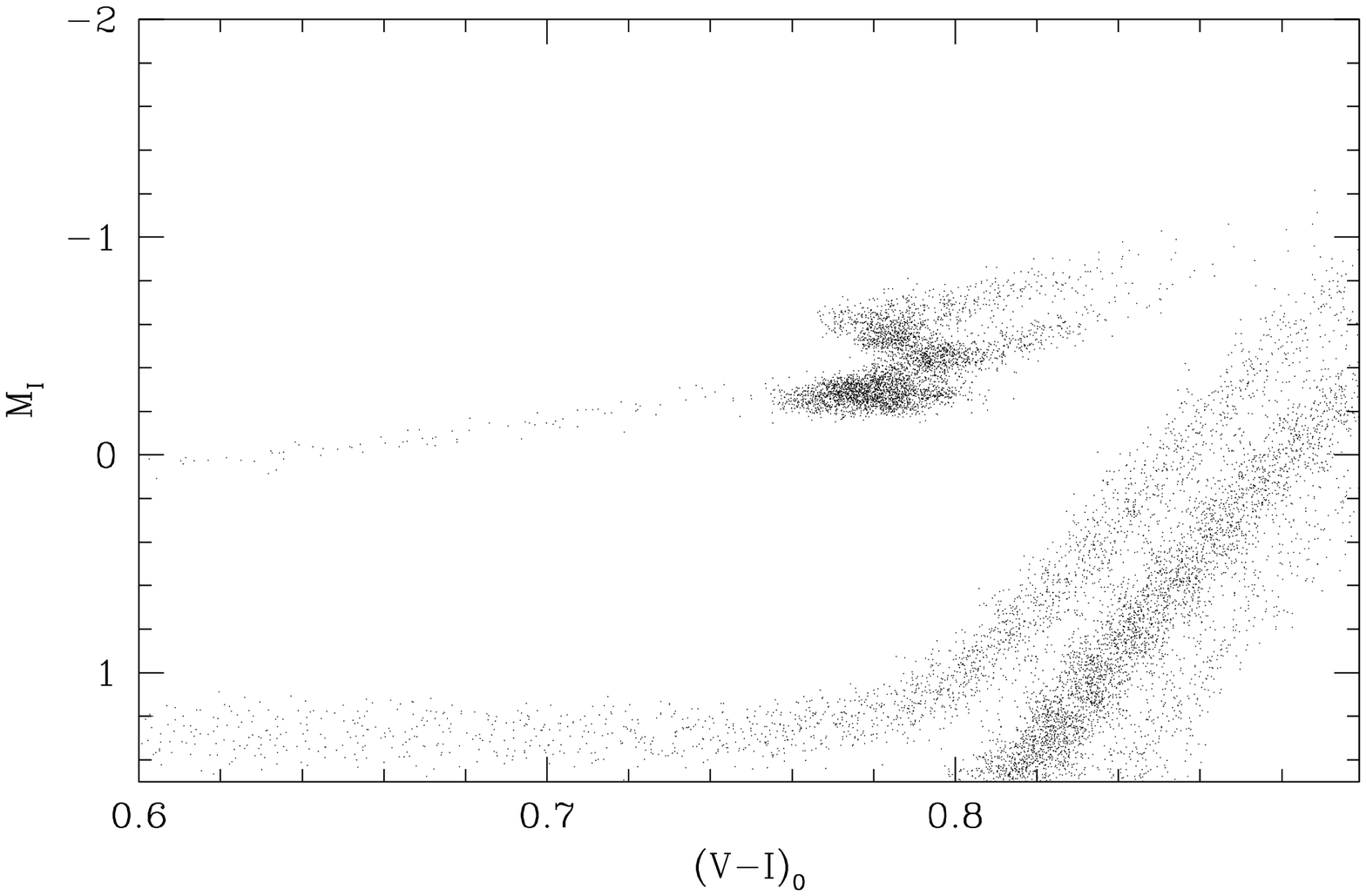,width=8.3cm}
      	\caption{
The same as Fig.~\protect\ref{fig_cmd_helio}, but for the
Carina dSph galaxy, as derived from the SFRs and AMRs
from (upper panel) Hernandez et al.\ (2000a), and
Hurley-Keller et al.\ (1998) best model (lower panel).
       }	 
	\label{fig_cmd_car} 
	\end{figure}

\begin{figure}
	\psfig{file=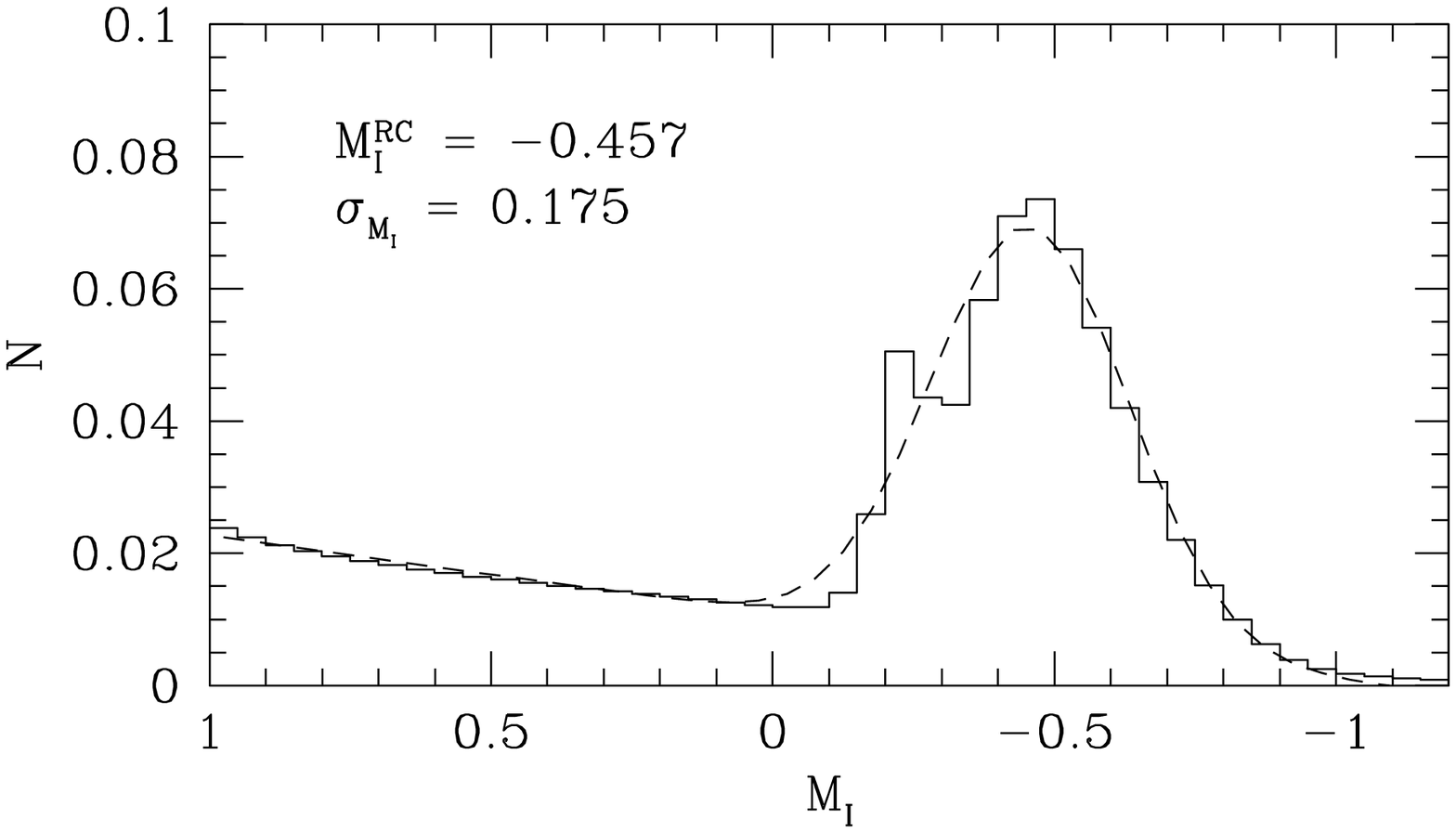,width=8.3cm}
	\psfig{file=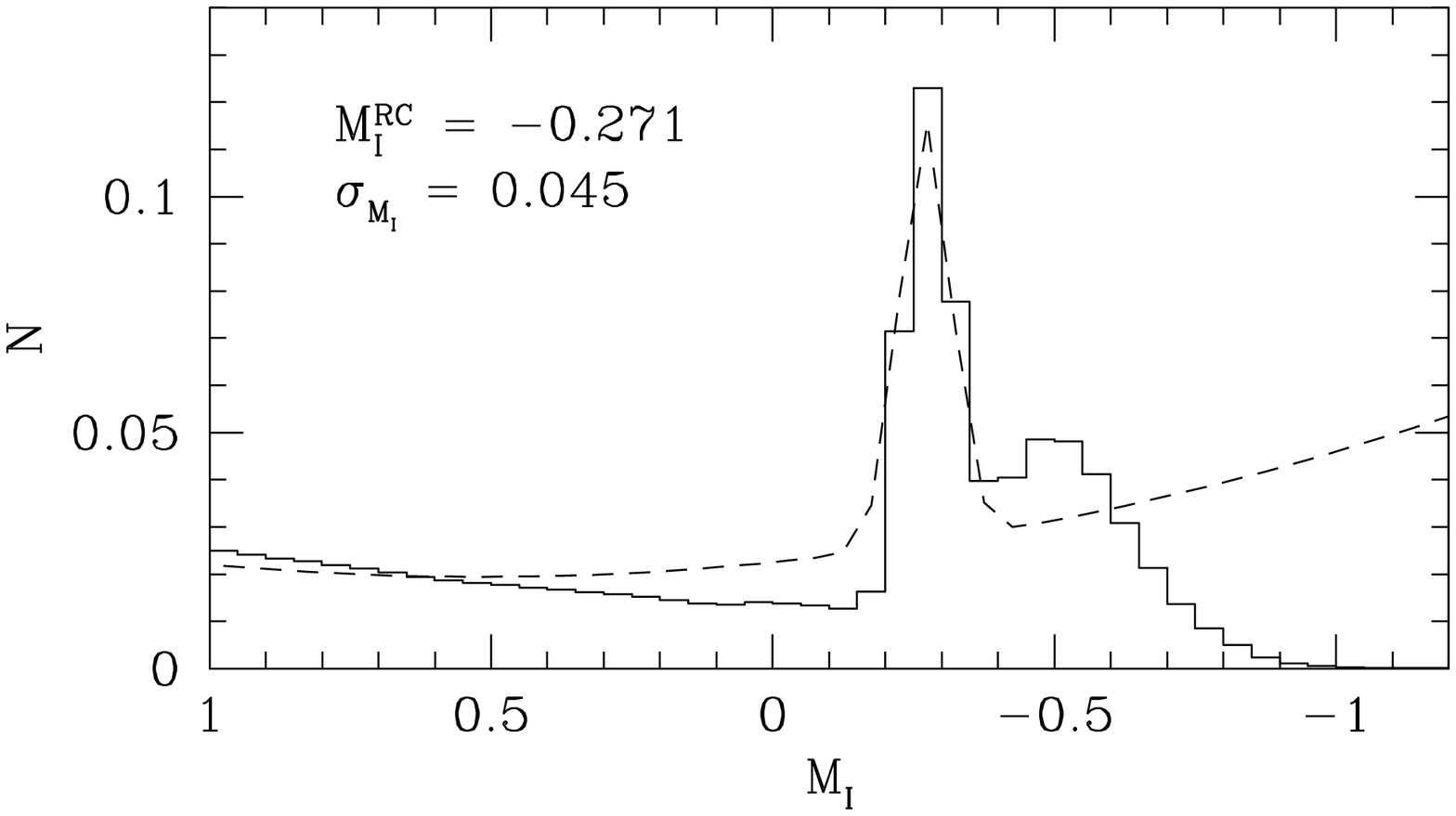,width=8.3cm}
      	\caption{
The same as Fig.~\protect\ref{fig_lf_helio}, but for the
Carina dSph galaxy models presented in 
Fig.~\protect\ref{fig_cmd_car}.
       }	 
	\label{fig_lf_car} 
	\end{figure}

We present, in Figs.~\ref{fig_cmd_car} and 
\ref{fig_lf_car} the synthetic CMDs and LFs, 
respectively, obtained from Hernandez et al.'s (2000a) solution,
and from Hurley-Keller et al.'s (1998) best model.
It can be noticed (Table~\ref{tab_galaxies}) that the two models provide
$M_I^{\rm RC}$ values differing by 0.18~mag. 
Actually, there is an obvious problem in the LF fit 
obtained from Hurley-Keller et al.'s model:
since it presents a kind of dual clump -- resulting from
their assumption of discrete bursts of
star formation separated by periods of quiescence -- the LF is
badly suited for a fit with a Gaussian function. Such a fit turns
out to favour the fainter clump, and clearly does not
represent in a satisfactory way the magnitude distribution
of clump stars. Moreover, there is no evidence of a dual clump
in the observational data (e.g.\ figure 6 in Udalski 1998a).
The clump obtained 
from Hernandez et al.'s (2000a) SFR, on the contrary, turns out to
be very compact in the CMD, in better agreement with the data.
For these reasons, this latter model is to be preferred in the present 
work.
%
%

\subsection{The clump in the SMC}

In the literature for the SMC, we did not find quantitative
assessments of the SFR, derived directly from SMC data. 
Qualitative descriptions can be found in e.g.\ 
Westerlund (1997) and Hatzidimitriou (1999).

Pagel \& Tautvaisiene (1998) describe both the SFR and 
AMR of the SMC population by means of a chemical evolution
model. Their AMR is shown to describe quite well
the data for SMC star clusters. Their SFR, however,
is not derived directly from stellar data
(as in the cases previously discussed), and hence should
be looked upon with some caution. It is characterised by strong
star formation in the last 4~Gyr of the SMC history, an
almost negligible SFR between 4 and 10 Gyr ago, and
more pronounced SFR at $10-12$~Gyr.

\begin{figure}
	\psfig{file=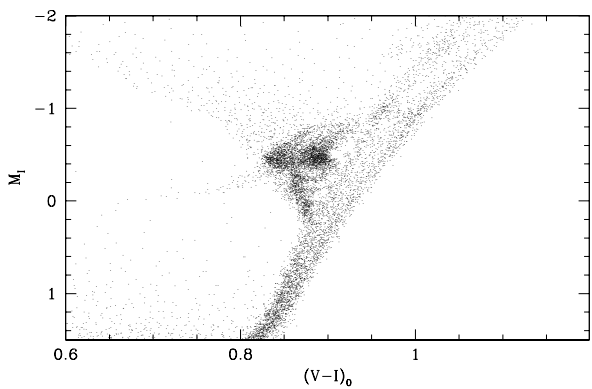,width=8.3cm}
      	\caption{
The same as Fig.~\protect\ref{fig_cmd_helio}, but for the
SMC galaxy, as derived from the SFRs and AMRs
from Pagel \& Tautvaisiene (1998).
       }	 
	\label{fig_cmd_smc} 
	\end{figure}

\begin{figure}
	\psfig{file=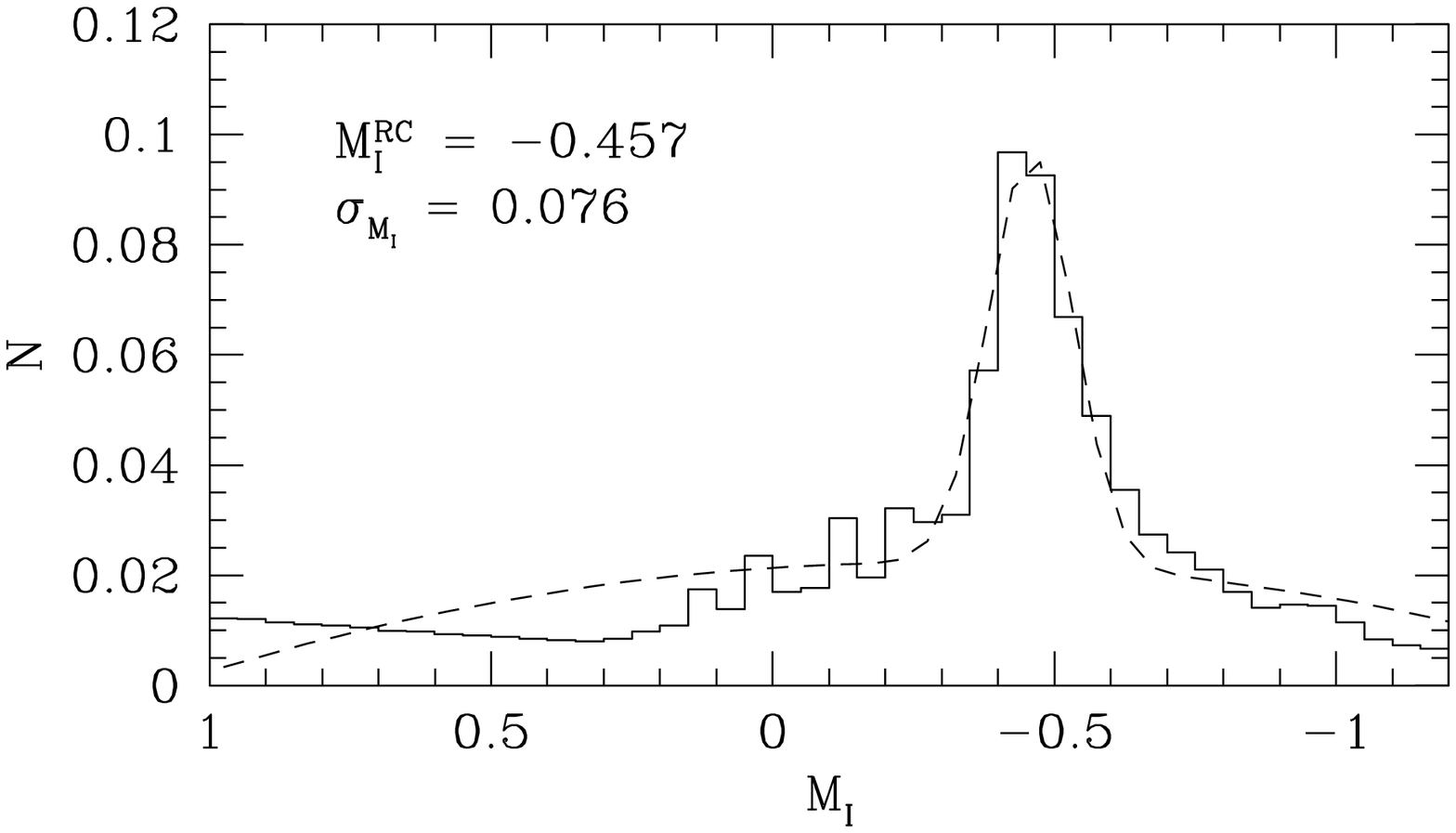,width=8.3cm}
      	\caption{
The same as Fig.~\protect\ref{fig_lf_helio}, but for
the SMC model presented in 
Fig.~\protect\ref{fig_cmd_smc}.
       }	 
	\label{fig_lf_smc} 
	\end{figure}

Simulations of the SMC clump, using Pagel \& Tautvaisiene (1998) 
results, are shown in Figs.~\ref{fig_cmd_smc} and 
\ref{fig_lf_smc}. This model produces a compact clump in the CMD,
but with some substructures which are due the discontinuous SFR
history. 
%
%

\subsection{The clump in the LMC}
\label{sec_lmc}

Quantitative determinations of the SFR in the LMC abound in the 
literature (Bertelli et al.\ 1992; Vallenari et al.\ 1996; 
Holtzman et al.\ 1997; Stappers et al.\ 1997; Elson, Gilmore
\& Santiago 1997; Geha et al.\ 1998).
The most recent determinations are generally
based on deep photometry of some few selected fields, 
and on fairly objective algorithms for reconstructing the 
SFR history (see e.g.\ Holtzman et al.\ 1999; Olsen 1999;
Dolphin 2000 and references therein).
A general result is that the SFR has increased in the last 
few Gyr (starting $2.5-4$~Gyr ago).
This increase roughly coincides with the start of a major
period of formation of star clusters, and with a major increase 
in the stellar mean metallicities (see Olszewski, Suntzeff \& Mateo 1996;
Dopita et al.\ 1997).

In the present work, we use the SFR results from Holtzman et al.\ 
(1999). Their results correspond to two LMC
regions (bar, and `outer') and three different assumptions in the
analysis. Namely, the following three cases
have been tested by Holtzman et al.:
(i) at any age, the metallicity follows a known AMR;
(ii) at any age, the metallicity is not constrained;
(iii) the metallicity follows a known AMR, and there has
been no star formation between 4 and 10~Gyr ago. 
For any of the 6 different SFRs from Holtzman et al.\ 
(1999), we have to assume some AMR; we take the
Pagel \& Tautvaisiene (1998) one, which is known to reproduce
reasonably well the AMR derived from LMC star clusters.
 
\begin{figure}
	\psfig{file=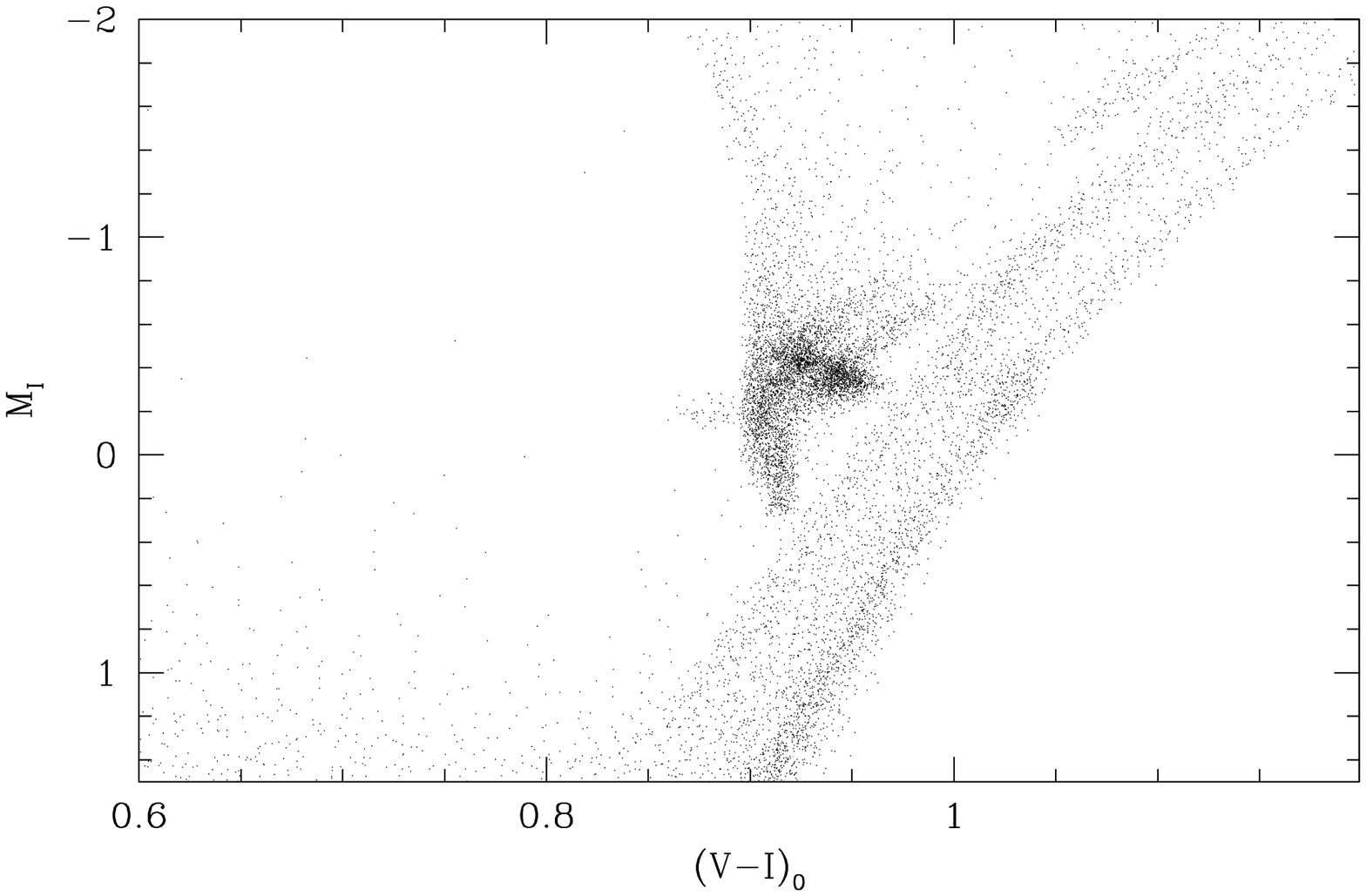,width=8.3cm}
	\psfig{file=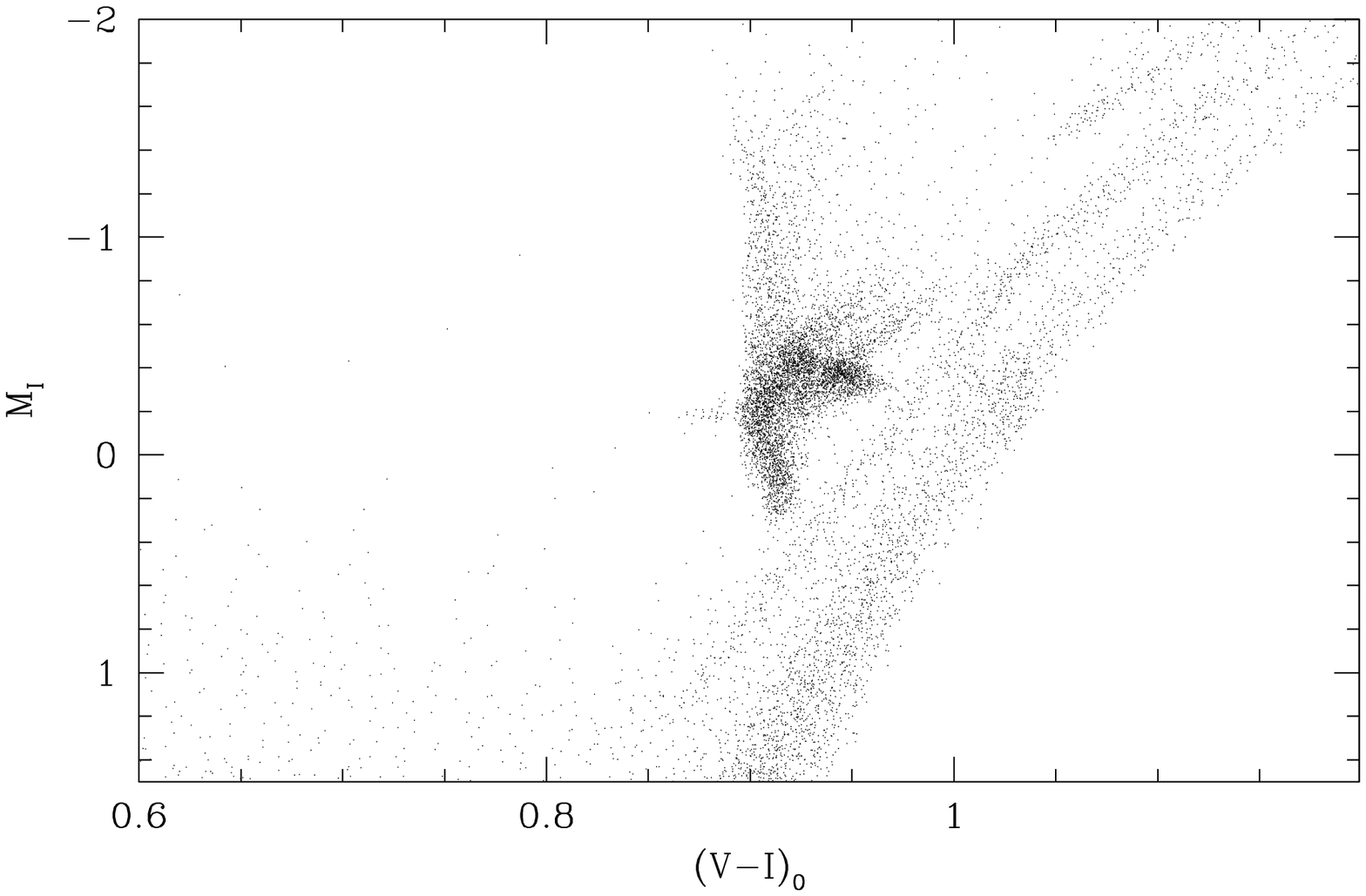,width=8.3cm}
	\psfig{file=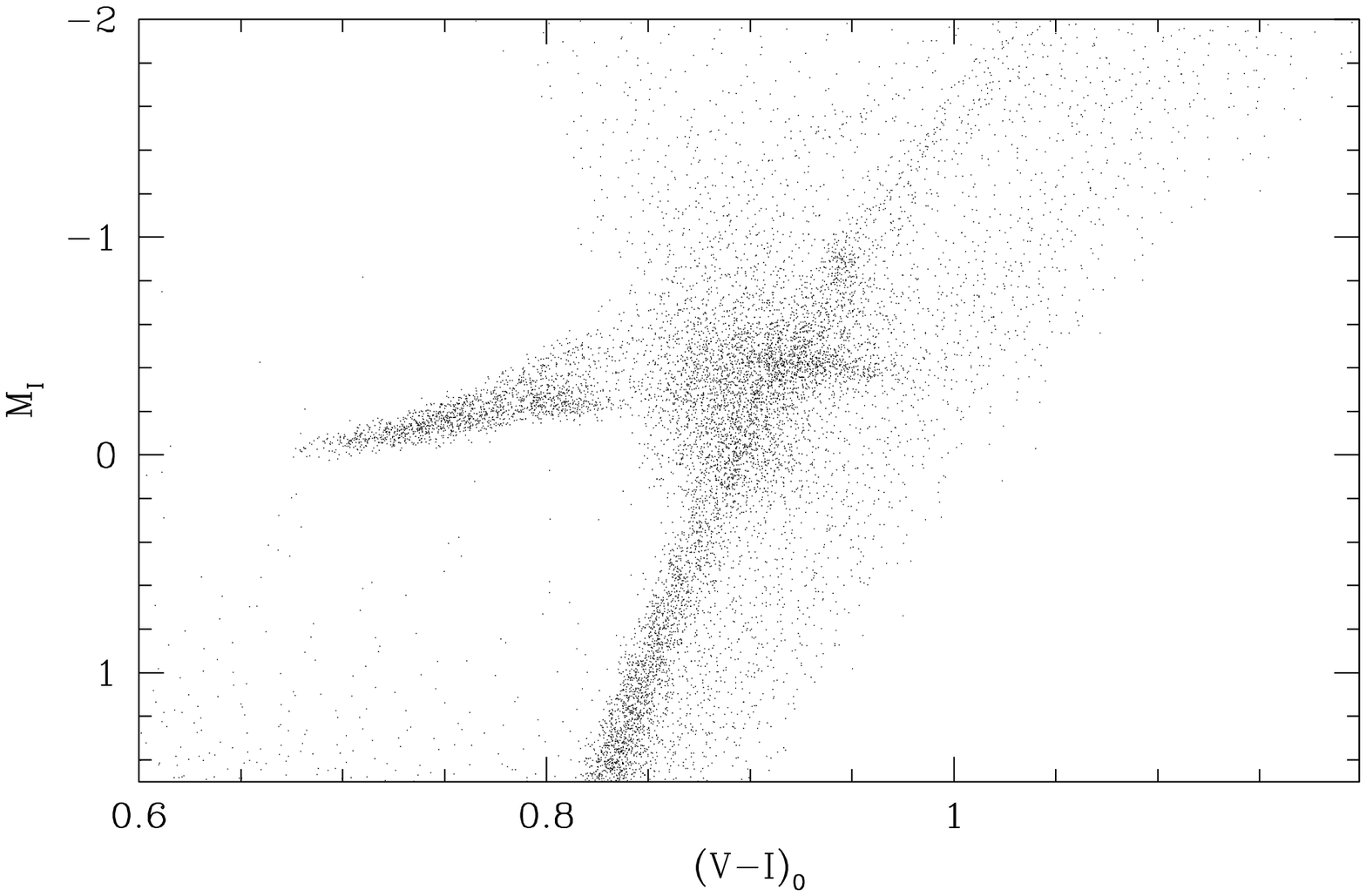,width=8.3cm}
      	\caption{
The same as Fig.~\protect\ref{fig_cmd_helio}, but for the
LMC, as derived from Holtzman et al.'s (1999; their figure 2) 
bar field (upper panel), 
from Holtzman et al.'s (1999; their figure 3) 
outer fields (middle panel), 
and from Dolphin (2000) outer fields 
(lower panel).
       }	 
	\label{fig_cmd_lmc} 
	\end{figure}

\begin{figure}
	\psfig{file=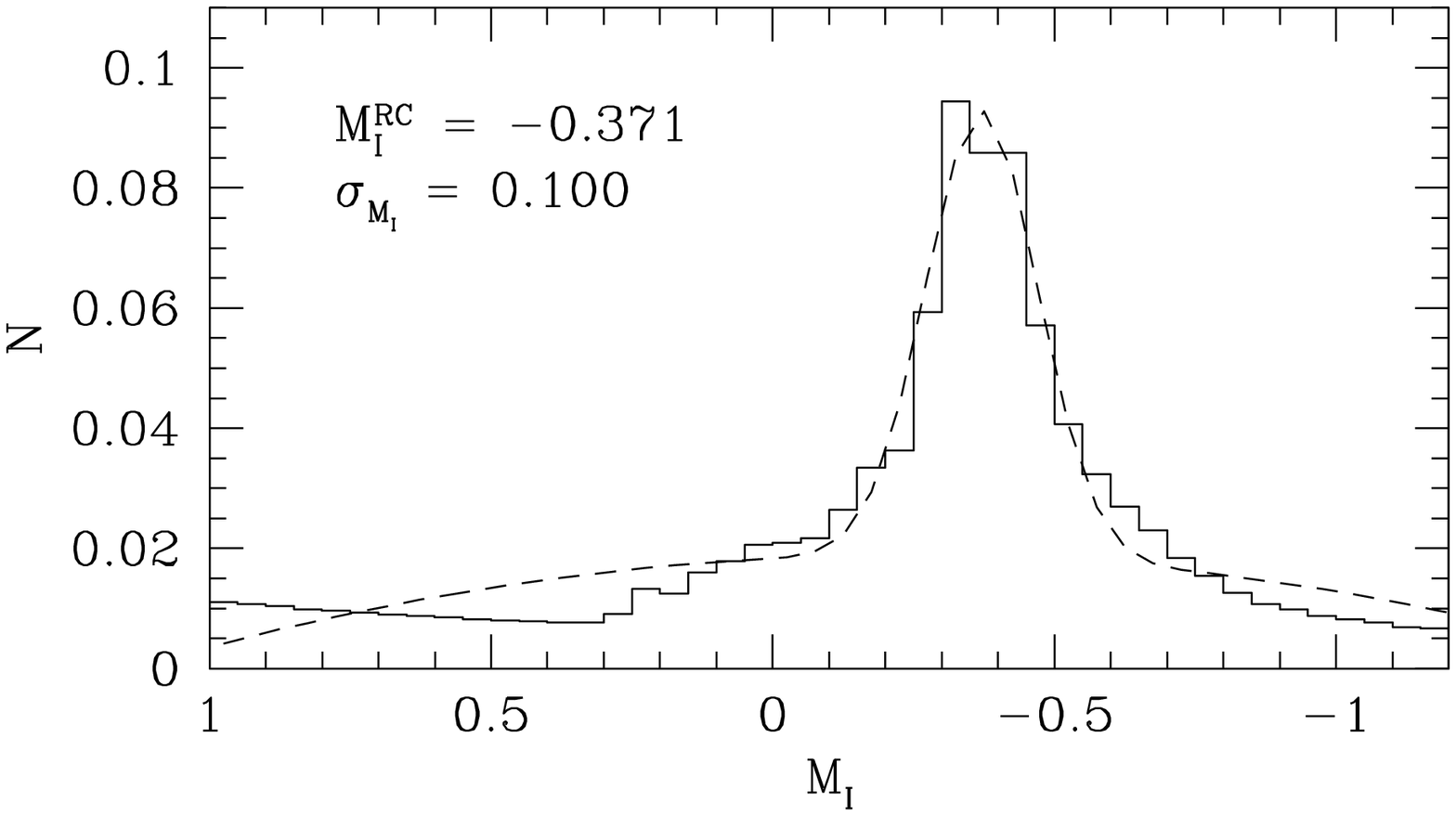,width=8.3cm}
	\psfig{file=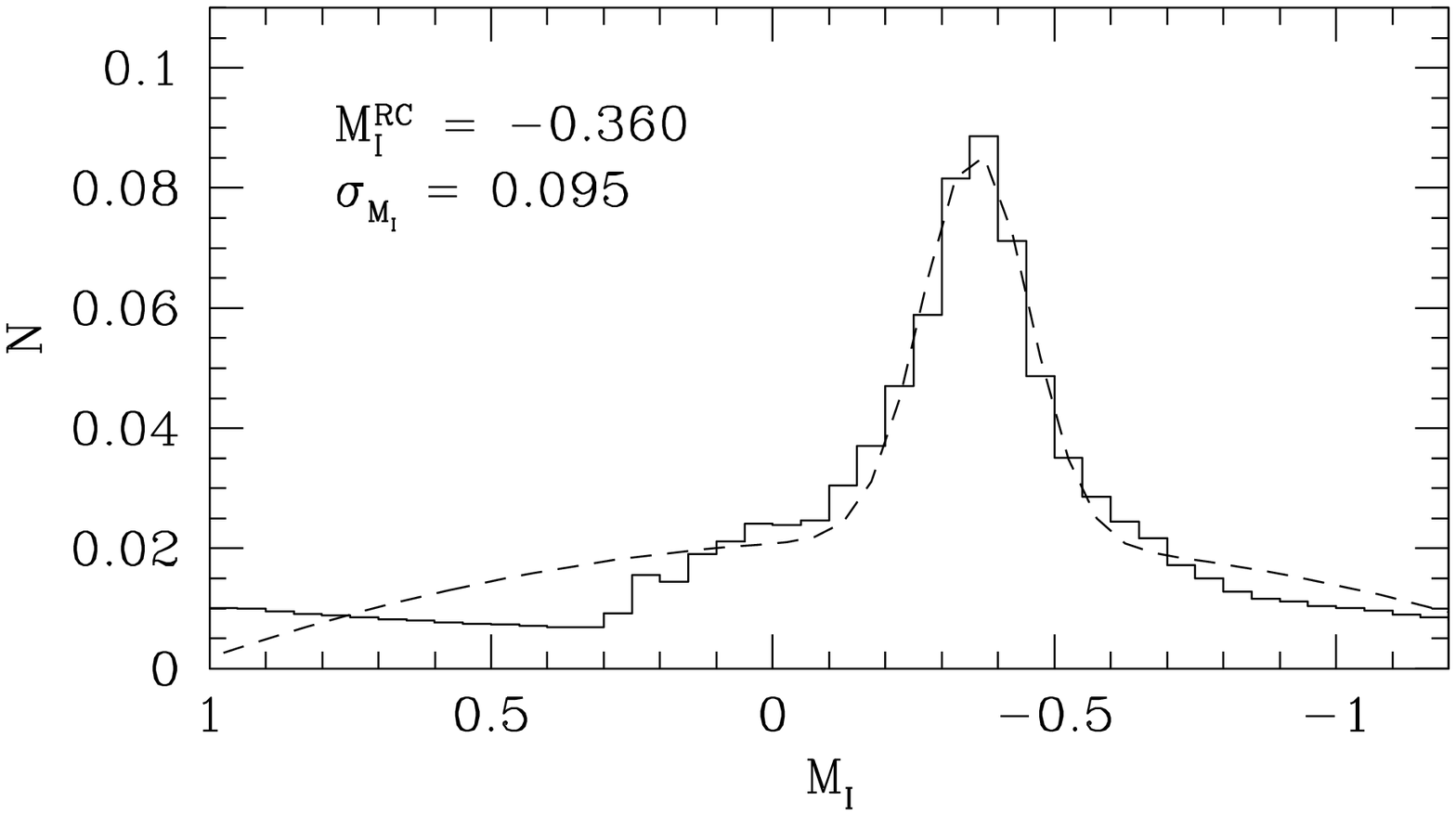,width=8.3cm}
	\psfig{file=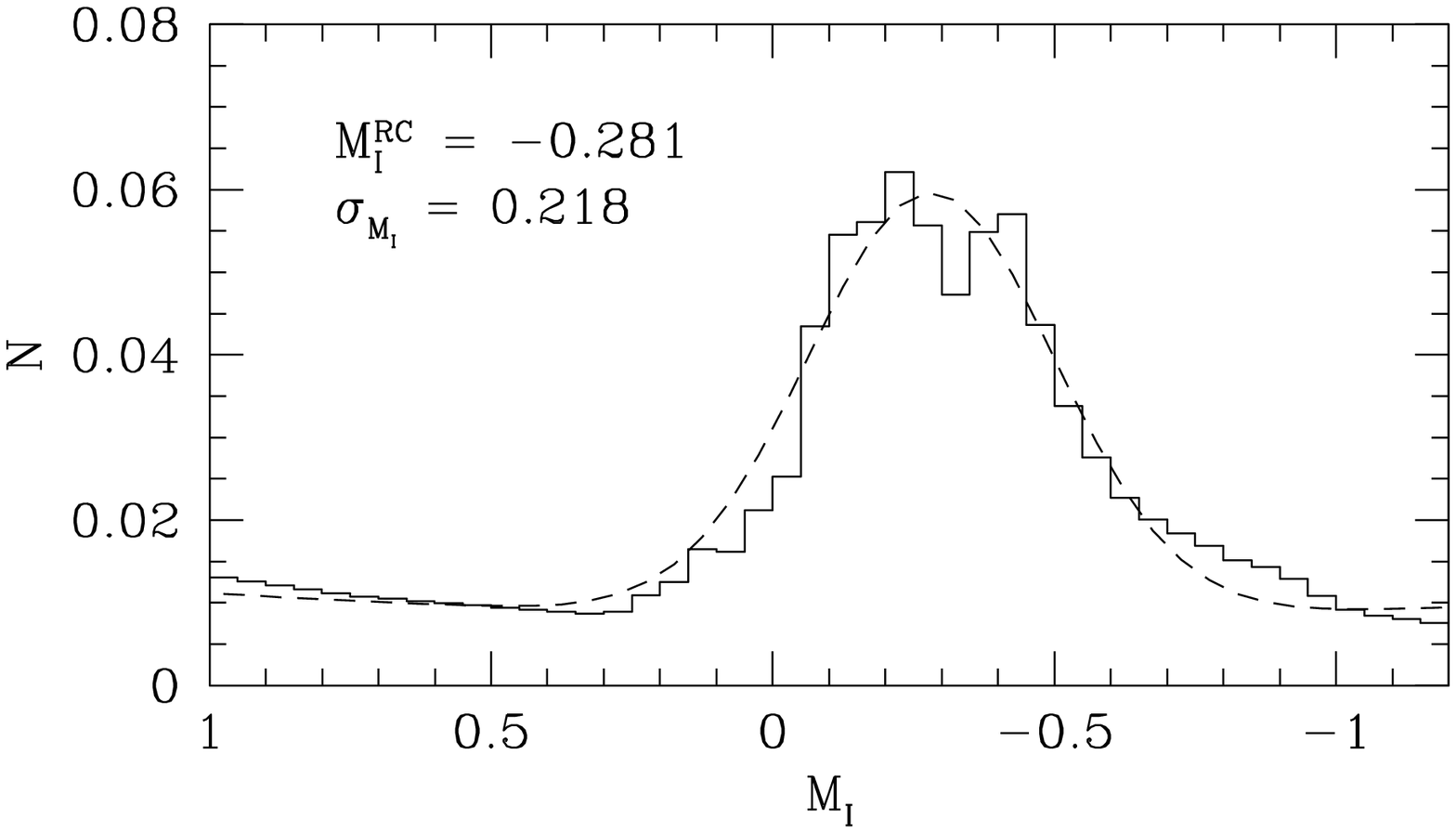,width=8.3cm}
      	\caption{
The same as Fig.~\protect\ref{fig_lf_helio}, but for the LMC
models presented in Fig.~\protect\ref{fig_cmd_lmc}.
       }	 
	\label{fig_lf_lmc} 
	\end{figure}

We show in Figs.~\ref{fig_cmd_lmc} and \ref{fig_lf_lmc} the
simulations of the bar and the outer field, with the
SFR derived according to item (i) above. For the same fields, 
(ii) and (iii) produce similar results (see 
Table~\ref{tab_galaxies}).
%
%

Remarkable in the CMD of Fig.~\ref{fig_cmd_lmc}, is
the complex structure of the predicted LMC clump. 
It presents a marked vertical structure on the blue side,
starting about $0.4-0.5$~mag below
the mean clump, and extending to higher luminosities. 
This feature of the models (the `secondary clump'; 
or `vertical structure')
has been extensively discussed by Girardi (1999), and 
has been clearly observed in some outer LMC fields by
Bica et al.\ (1998) and Piatti et al.\ (1999). 
Moreover, the simulated LMC clumps present a horizontal 
tail of stars departing to the blue, which is simply the 
beginning of the old and metal-poor horizontal branch.
It is interesting to notice the extreme similarity between
these simulated clumps (Fig.~\ref{fig_cmd_lmc}),
and the detailed CMD of a large area in the northern LMC
presented by Piatti et al.\ (1999, their figure 4). 

Finally, we have also tested the SFR and AMR derived by
Dolphin (2000) from a field in the northern LMC.
Surprisingly, this simulation produces a double clump 
(Fig.~\ref{fig_cmd_lmc}), which is evidently the result
of a large population of old metal-poor stars in Dolphin's 
(2000) solution. The synthetic CMDs turn out not to 
reproduce the characteristics of the LMC clump, as noticed 
by Dolphin himself. Moreover, the mean \feh\ of clump 
stars is $-0.88$~dex for this model, which is 
far too low if compared with typical values found for LMC
field giants. For these reasons, we prefer not
to use the results from this latter model in our analysis.

\subsection{Considering the enhancement of $\alpha$-elements}
\label{sec_alpha}

All the models discussed above assume a scaled-solar distribution
of metals. However, it is well established
that in some stellar populations (e.g.\ the Galactic Halo)
the group of $\alpha$-elements (mainly O, Ne, Mg, Si, Ca, Ti) 
is overall enhanced with respect to Fe in comparison
with solar ratios (i.e.\ $[\alpha/{\rm Fe}]>0$).
This is probably the case for Bulge giants,
where measurements of Mg and Ti abundances provide an
enhancement by about $+0.4$~dex 
(see McWilliam \& Rich 1994; Barbuy 1999). This is usually
considered to be evidence for the chemical enrichment in the 
Bulge having occurred in a relatively short time scale.

A ratio $[\alpha/{\rm Fe}]=0.4$  
means that, at a given \feh\ value, the metal content 
$Z$ is a factor of about 2.5 larger than given by the 
scaled-solar relation $\feh=\log(Z/0.019)$. Moreover, for nearly-solar
metallicities, $\alpha$-enhanced models cannot be reproduced
by scaled-solar ones by simply modifying the relationship between $Z$
and \feh\ (Salaris \& Weiss 1998; Salasnich et al.\ 2000). Therefore, 
it is worth exploring how the Bulge $M_I^{\rm RC}$ 
would change if appropriate $\alpha$-enhanced models were adopted.
To this aim, we 
repeated our Bulge simulations using the isochrones
from Salasnich et al.~(2000), for both scaled-solar and 
$\alpha$-enhanced ([$\alpha/{\rm Fe}]\simeq0.35$~dex) cases. 
The difference between both $M_I^{\rm RC}$ values was then 
added to the value obtained from Girardi et al.'s (2000) 
scaled-solar models (see Table~4).

It turns out that 
$\alpha$-enhanced models produce a Baade's Window clump 
about $0.08$~mag brighter than the scaled-solar ones. This occurs
because of two competing effects.
For the same \feh\ distributions centered at
almost-solar values ($\feh\sim-0.2$~dex; see Table 4), 
$\alpha$-enhanced models have a much higher $Z$, and higher $Z$
causes lower clump brightness at a given age. However, since
our models assume a constant helium-to-metals enrichment ratio of 2.25,
much higher values of the helium content $Y$ are reached by the
$\alpha$-enhanced models; due to the fact that an increase of $Y$ 
(at fixed age and $Z$) causes an increase of the clump brightness,
the net effect of using $\alpha$-enhanced models is an increase 
of the clump luminosities with respect to the scaled-solar case. 
The final result is that Bulge models computed 
considering the enhancement of $\alpha$ elements, may have 
$M_I^{\rm RC}$ values very similar to those of the
{\em Hipparcos} sample. 

In the other galaxies we are considering, $\alpha$-enhancement
should not be as important as in the Bulge. 
In the cases of the Solar
Neighbourhood and the Magellanic Clouds, the bulk of clump 
giants are relatively young ($\la3$~Gyr), and such stellar 
populations are expected to have a scaled-solar metal distribution. 
This is confirmed by spectroscopic observations of: (i)
thin disk stars, that indicate almost scaled-solar ratios
(${\rm [Mg/Fe]}$ between 0.0 and $0.1$~dex) for stars with
$\feh>-0.5$ (Fuhrmann 1998); and (ii)
giants in LMC clusters younger than 3~Gyr
and with $\feh\sim-0.5$, which have $\rm [O/Fe]\sim+0.1$~dex 
(Hill et al.\ 2000). These low levels of $\alpha$-enhancement
can, at least as a first approximation, be ignored. 

The situation for the Carina dSph is not clear. The
clump giants in this galaxy are neither too young, nor have 
been formed in a short time interval as the Bulge ones. Thus,
it is not clear whether some degree of $\alpha$-enhancement 
should be expected, and present observations do not give 
information on this. Thus, we prefer not to consider the 
possibility of $\alpha$-enhancement for this galaxy.

\subsection{The clump -- RR Lyrae difference}

Udalski (1998a) measured the mean clump apparent magnitude 
$I_0^{\rm RC}$ in Baade's Window, LMC, SMC, and Carina dSph 
galaxy. For the same fields, RR Lyrae data has provided a 
reference magnitude to compare the clump with. Assuming that
RR Lyrae stars follow a $M_V^{\rm RR}=(0.18\pm0.03)\feh+{\rm const}$
relation and adopting empirical mean values for the metallicity of RR 
Lyrae in the different environments, Udalski (1998a) constructed
the $I_0^{\rm RC}-V_0^{\rm RRatGB}$ parameter, where  
$V_0^{\rm RRatGB}$ means the magnitude that RR Lyrae in each galaxy 
would have if they had the same \feh\ as the Bulge ones.
Udalski's (1998a) data for Baade's Window has been later revised 
by Popowski (2000). Following the results by Paczy\'nski et al.~(1999) 
about a systematic difference between OGLE-I and OGLE-II photometry,
he considers a clump dimmer by 0.035 mag with respect to Udalski
(1998a) data and RR Lyrae stars brighter by 0.021 mag; this 
produces a change of $I_0^{\rm RC}-V_0^{\rm RRatGB}$ by $+0.06$ mag.  
Moreover, in order to solve the 
so-called `\vi\ colour problem' of Baade's Window clump giants 
(Paczy\'nski 1998), Popowski (2000) modifies
the original extinction values and reddening curves for Baade's Window
by an amount which implies a final global revision by $+0.17$~mag for the
$I_0^{\rm RC}-V_0^{\rm RRatGB}$ value.

By construction, $I_0^{\rm RC}-V_0^{\rm RRatGB}$
provides the differential behaviour of $M_I^{\rm RC}$ in these
four stellar systems, and it is therefore interesting to compare it with
the $\Delta M_I^{\rm RC}= M_I^{\rm RC}({\rm Hipp}) - 
M_I^{\rm RC}({\rm galaxy})$ values derived from our simulations.
Before proceeding, we just mention that our ZAHB models of low
metallicity at $\log\Teff=3.85$ 
(a typical average temperature for RR Lyrae stars) have the same
slope of $\Delta M_V^{\rm RR}/\Delta\feh=0.18$~mag/dex that has been used
by Udalski (1998a) to correct his RR Lyrae data. This slope is also in good 
agreement with the results from independent evolutionary models 
(see, e.g., Salaris \& Weiss 1998) and from the main sequence fitting 
technique (Gratton et al.\ 1997).

\begin{figure}
	\psfig{file=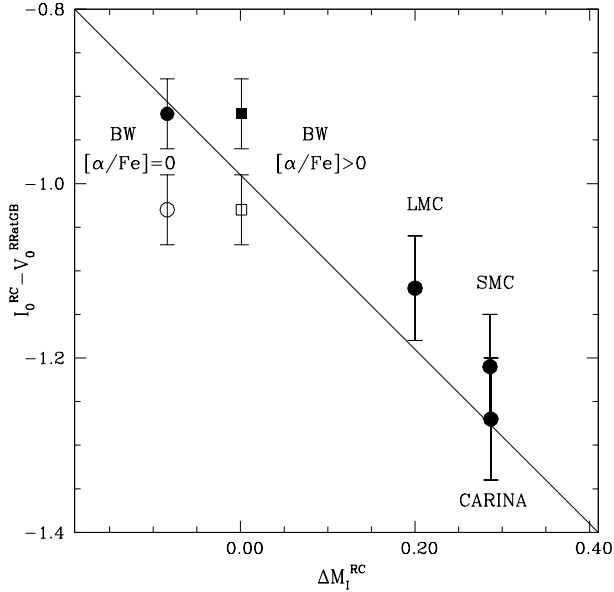,width=8.3cm}
      	\caption{
Observed $I_0^{\rm RC}-V_0^{\rm RRatGB}$ as a function of 
the modelled $\Delta M_I^{\rm RC}$ values for the stellar systems of 
Table~\protect\ref{tab_galaxies} (full circles). 
$I_0^{\rm RC}-V_0^{\rm RRatGB}$ 
comes from Udalski (1998a), but for the Baade's Window point that has 
been revised upward in magnitude by Popowski (2000). The full line
represents a straight line of slope $-1$.
The open symbols show the Baade's Window data corrected only by 
the zero point difference between OGLE-I and OGLE-II $V$- and
$I$-band magnitudes (see text for details). 
Squares represent models computed for the Baade's Window
adopting an $\alpha$-enhanced metal distribution. 
}
	\label{deltaMi__I@V} 
	\end{figure}

The result of the comparison is shown in Fig.~\ref{deltaMi__I@V}.
We have used the Udalski (1998a) data and error bars,
and the correction by Popowski (2000) for the Baade's Window 
population. The $\Delta M_I^{\rm RC}$ values come from 
Table~\ref{tab_galaxies}. The solid line indicates a 
straight line of slope $-1$ (increasing  $\Delta M_I^{\rm RC}$ 
means decreasing $M_I^{\rm RC}({\rm galaxy})$, which should correspond
to a decrease of $I_0^{\rm RC}-V_0^{\rm RRatGB}$).
It is evident that the differential behaviour of the 
theoretical $\Delta M_I^{\rm RC}$ values for these four systems
is in good agreement with the empirical behaviour of $M_I^{\rm RC}$ as 
provided by the $I_0^{\rm RC}-V_0^{\rm RRatGB}$ parameter.
This occurrence, together with the agreement between theoretical and
empirical $\Delta M_I^{\rm RC}$ absolute values for the open clusters
of Fig.~\ref{fig_clusters}, provides strong 
observational support to the 
the population corrections predicted by stellar models.

We have also tested if significant changes in this comparison are
introduced when using different measures of either $I_0^{\rm RC}$ or 
$V_0^{\rm RR}$. In particular, we have considered the determinations of 
the LMC $I_0^{\rm RC}$ by Romaniello et al.~(2000),
$V^{\rm RR}$ for Carina from Kuhn, Smith \& Hawley (1996) --  
dereddened using the same reddening as in 
Udalski (1998a) -- and the revised values of $V_0^{\rm RR}$
for both LMC and SMC from Udalski et al.~(1999).
The satisfactory agreement between the differential behaviour of the 
theoretical $\Delta M_I^{\rm RC}$ and $I_0^{\rm RC}-V_0^{\rm RRatGB}$
is not sensibly altered.

	\begin{figure}
	\psfig{file=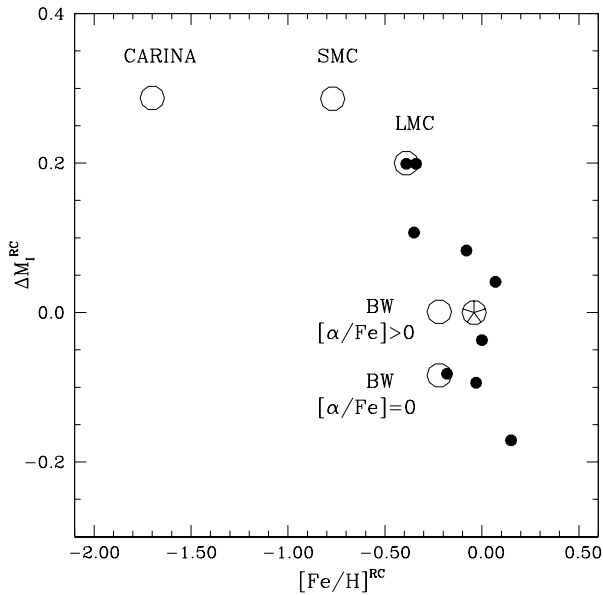,width=8.3cm}
      	\caption{
$\Delta M_I^{\rm RC}$ as a function of the red clump mean
[Fe/H] for the stellar systems in Tab.~\ref{tab_galaxies} 
(large open circles) and
the open clusters displayed in Fig.~\ref{fig_clusters} (full circles).
The point corresponding to the {\em Hipparcos} red clump (starred circle)
is also plotted. 
Two data points are shown for the Baade's Window clump, corresponding
to the results obtained with either scaled-solar or 
$\alpha$-enhanced models.
}
	\label{deltaMi__Fe} 
	\end{figure} 

In Fig.~\ref{deltaMi__Fe} we show the run of $\Delta M_I^{\rm RC}$
as a function of the mean red clump [Fe/H] for the stellar systems
in Tab.~\ref{tab_galaxies}, the local {\em Hipparcos} red clump 
and the open clusters of Fig.~\ref{fig_clusters}.
If we consider, for example, metallicities around 
$\feh=-0.1$ one can easily notice the large dispersion of the 
$\Delta M_I^{\rm RC}$ values. 
On the basis of all the evidences discussed 
in this paper it is easy to understand that this dispersion is due
to the different SFR and AMR of the stellar populations we are
considering.
This should also warn against the use of empirical linear
relationships for deriving $\Delta M_I^{\rm RC}$ as a function of 
[Fe/H]. $\Delta M_I^{\rm RC}$ depends in a complicated way on 
the properties of the stellar populations under scrutiny and any
empirical calibration of this quantity is not universal, but reflects
the particular SFR and AMR of the calibrating sample. Moreover, there
is no physical reason at all for linear relationships
to hold for $\Delta M_I^{\rm RC}$.

If we consider only the stellar systems of 
Tab.~\ref{tab_galaxies} and try to fit a linear relationship to
the points displayed in Fig.~\ref{deltaMi__Fe} (open circles)
we would derive a slope of about 0.18, in agreement
with the slope of the empirical corrections ($0.19\pm0.05$) derived by
Popowski (2000) using the $I_0^{\rm RC}-V_0^{\rm RRatGB}$
values as a function of [Fe/H] for the same sample of objects; 
but we want to stress the point 
that this slope has no meaning whatsoever. It is just
an accident that for this particular sample 
of galaxies there is a relationship close to
a linear one between the clump brightness and the metallicity.
In distance determinations,
the individual values of $\Delta M_I^{\rm RC}$
derived from population synthesis simulations
must be used, and not an average relation obtained from 
a linear fit to the real corrections.

\section{Conclusions about the clump distance scale}
\label{sec_conclu}

In the previous two sections we have shown how theoretical
models of stellar populations are able to reproduce most of the 
relevant observational features regarding red clump stars
in different environments.
In the following we will redetermine the red clump distances to the
galactic and extragalactic systems previously discussed
using $I_0^{\rm RC}$ values taken from the literaure.
It is not our intention to exhaustively discuss the 
uncertainties and conflicting results about the distances to these
stellar systems; we only want to show the changes of the distance
estimates using red clump stars when one is using the 
$\Delta M_I^{\rm RC}$ corrections predicted by stellar models.

\subsection{The Bulge -- Magellanic Clouds -- Carina dSph 
distance scale}

By applying the red clump method (see Eq.~\ref{eq_mu}) 
and the population corrections $\Delta M_I^{\rm RC}$
displayed in Table~\ref{tab_galaxies}, we derive here
the distances to the Galactic Bulge, LMC, SMC and Carina dSph.
The values for 
$I^{\rm RC}$ and $A_I$ come from the literature, and
$M_I^{\rm RC}=-0.23\pm0.03$ (Stanek \& Garnavich 1998) has been 
empirically obtained from local {\em Hipparcos} red clump stars.

By using a dereddened $I_0^{\rm RC}=14.32\pm0.04$ 
for the Galactic Bulge
(Udalski 1998a) and $\Delta M_I^{\rm RC}=-0.087$ 
(from solar-scaled models; Table~\ref{tab_galaxies}),
using the McWilliam \& Rich (1994) AMR,
we get $\mu_0^{\rm GB}=14.47\pm0.05$. Analogous value is obtained 
using the SFR and AMR by Moll\'a et al.\ (2000).
It corresponds to a linear distance of
$7.8\pm0.2$~Kpc, in good agreement with the generally accepted value
of the distance to the Galactic center $8.0\pm0.5$~Kpc (Reid 1993).
Our estimate, however, is slightly lower than the
values obtained by Paczi\'nsky \& Stanek (1998) and Stanek \& 
Garnavich (1998), simply because we include a population effect
which was then assumed to be negligible. 
If instead we correct $I^{\rm RC}$ by $+0.035$~mag
as suggested by Paczy\'nski et al.\ (1999)
and Popowski (2000), and modify $A_I$ by $-0.11$~mag
as suggested by Popowski (2000), we get
$\mu_0^{\rm GB}=14.62\pm0.05$ ($8.4\pm0.2$~Kpc).
These results slightly change if we consider the $\alpha$-enhanced
model for the Bulge. In this case the population effect is negligible 
and we obtain, respectively, 
$\mu_0^{\rm GB}=14.55\pm0.05$ ($8.1\pm0.2$~Kpc)
and $\mu_0^{\rm GB}=14.70\pm0.05$ ($8.7\pm0.2$~Kpc).

In the case of the LMC there is still much debate about the
observational value of the red clump $I_0^{\rm RC}$ (Zaritsky 1999; 
Romaniello et al.\ 2000; Udalski 2000), the main reason being the
extinction correction (see also Sec.~\ref{sec_intro}). 
Romaniello et al.\ (2000) obtained
$I_0^{\rm RC}=18.12\pm0.02$, a value in agreement also with the results
by Zaritsky (1999), while Udalski (2000) 
obtained $I_0^{\rm RC}=17.94\pm0.05$.
When adopting Romaniello et al.'s (2000) result, together with 
$\Delta M_I^{\rm RC}=0.200$ (considering the SFR from figure 2 of
Holtzman et al.\ 1999; see Table~\ref{tab_galaxies}), 
we get a LMC distance modulus 
$\mu_0=18.55\pm0.05$ ($51.3\pm1.1$~Kpc). The other 
SFR prescriptions by Holtzman et al.\ (1999)
displayed in the Table~\ref{tab_galaxies} do not significantly
modify $\Delta M_I^{\rm RC}$. Our $\mu_0^{\rm LMC}$ value is in good
agreement with the so-called `long' distance scale.
In case of assuming the Udalski (2000) dereddened red clump
brightness one obtains $\mu_0=18.37\pm0.07$ ($47.2\pm1.5$~Kpc).
Notice that this value is `longer' by 0.13~mag with respect to the
result obtained by Udalski (2000) using the same red clump 
brightness but his empirical correction for metallicity effects.

As for the SMC Udalski (1998a) gives 
$I_0^{\rm RC}=18.33\pm0.05$ which, together with 
$\Delta M_I^{\rm RC}=0.286$ obtained from Table~\ref{tab_galaxies}, 
provides $\mu_0^{\rm SMC}=18.85\pm0.06$
($58.9\pm1.6$ Kpc). 

The distance to Carina turns out to be $\mu_0^{\rm Car}=19.96\pm0.06$
($98.2\pm2.7$~Kpc)
when using $I_0^{\rm RC}=19.44\pm0.04$ from Udalski (1998a)
and $\Delta M_I^{\rm RC}=0.287$ from Table~\ref{tab_galaxies} 
(using the Hernandez et al.\ 2000a SFR).

\subsection{Final comments}

In this paper, we use an extended set of stellar models,
standard population synthesis algorithms, and
independent data about the distributions of stellar ages and 
metallicities, to derive the behaviour of the clump magnitude in 
different stellar systems. We also provide the basic equations and 
tables for a straightforward computation of the red clump mean
brightness for any stellar system.

We are able to reproduce quite well a number of 
observational features of the clump in nearby galaxy 
systems. The most striking are: 
\begin{enumerate} 
\item {\em For the {\em Hipparcos} clump}:
a) the distribution in the \mi\ versus \vi\ diagram (a colour 
shift of 0.1~mag being probably due to a mismatch between two 
different empirical metallicity/age scales);
b) the narrow and Gaussian-like \feh\ distribution; 
c) the absence of a correlation between \vi\ colour and \feh;
d) the approximate slope of the empirical $M_I^{\rm RC}$ versus 
[Fe/H] relation.
\item {\em For the Baade's Window clump}, the wide and nearly
horizontal clump in the \mi\ versus \vi\ diagram.
\item {\em For the LMC}, the striking vertical 
structure (fainter secondary clump plus bright tail) 
at the blue side of the clump, and a blue plume of 
horizontal branch stars.
\item {\em For the SMC and Carina dSph}, the compact clump structure.
\item {\em For galactic open clusters older than 2~Gyr}, the rate of
change of the clump brightness with both age and metallicity.
\item {\em For the Bulge, Magellanic Clouds, Carina dSph}, the
approximate slope of the empirical $M_I^{\rm RC}$ versus [Fe/H] relation.
\end{enumerate}

We have shown that the   
models predict a complex dependence of the red clump magnitude on 
age, metallicity, and star formation history, which cannot be 
expressed by relations such as:
(i) a linear $M_I^{\rm RC}$ versus \feh\ relation, or
(ii) a linear (or constant) $M_I^{\rm RC}$ versus age relation.
{\em Present empirical linear $M_I^{\rm RC}$ versus \feh\ relations, 
used to describe the dependence of the red clump $M_I^{\rm RC}$ 
on the metallicity, are misleading, since they are originated by the 
particular age and metallicity distributions of the objects 
included in the calibrating sample, and do not have a general 
validity.} Using such linear relations may produce spurious
results, even when statistically good fits to the calibrating data 
are obtained.

To summarize, there are four main features indicated by the models, 
that cannot be expressed by present empirical relations:
\begin{enumerate}
\item $M_I^{\rm RC}$ depends on both metallicity and age, and then
on the underlying age-metallicity relation;
\item for a given metallicity, the $M_I^{\rm RC}$ versus age relation 
is complex and not monotonic;
\item at a given age, $M_I^{\rm RC}$ generally increases with \feh,
but not necessarily in a linear way;
\item stars of different ages have very different weights in 
determining $M_I^{\rm RC}$ in a galaxy, younger stars 
(if present) being dominant.
\end{enumerate}
These features are, nowadays, better predicted by models, than 
expressed by empirical calibrations. 

The results summarized above have been obtained using, essentially, 
models with scaled-solar metal abundances.
However, the entire problem of describing the clump behaviour 
with respect to age and metallicity gets even more complicated 
if we take into account that some stellar
populations may be characterized by different 
initial metal distributions. We have demonstrated
the sensitivity of the clump brightness to the metals
relative abundances with our simulations of the Baade's 
Window clump using $\alpha$-enhanced evolutionary tracks. 
This adds a further variable -- degree of $\alpha$-enhancement 
-- to the problem. 
Moreover, if the metal content $Z$ is above solar -- a situation
that can be met even for $\feh=0$ if $\alpha$-enhancement is
present -- also the helium content $Y$ becomes important in 
determining the absolute clump magnitude (see GGWS98). 
Therefore, in these cases a fourth variable is to be
considered: the helium-to-metal enrichment ratio.

The correct approach to using the red clump as a distance indicator is
therefore to evaluate the population corrections $\Delta M_I^{\rm RC}$ 
-- using population synthesis models -- for each individual object,
provided that evaluations of the SFR and AMR do exist.
In addition, informations on [$\alpha$/Fe] and reasonable assumptions 
about the helium-to-metal enrichment ratio are needed.
These occurrence, however, raises a fundamental 
question about the accuracy of the red clump as distance indicator. 
Udalski (1998b) emphasizes the advantages
of the red clump with respect to other widely used standard candles 
such as
Cepheid and RR Lyrae stars, these being mainly: (1) the existence of 
larger samples of red clump stars
in galaxies with respect to Cepheids and RR Lyrae; (2) the very precise
calibration of the absolute red clump brightness for the Solar
Neighbourhood; (3) the existence of only a weak and empirically 
calibrated dependence of the red clump brightness on the metallicity.
However, now that one has demonstrated that these empirical
relations have no general 
validity, the red clump does not seem anymore to be
a very reliable standard candle. At least, 
any determination of red clump distances requires the 
critical evaluation of the population effects, then implying
that the red clump method cannot be meaningfully
applied to objects for which there are no determinations of the 
SFR and AMR, {\em unless errors up to $\approx$0.3 mag are to be accepted}.  

On the other hand, stellar evolution and population synthesis theory 
provide potentially important tools for the interpretation of clump 
data in nearby galaxies. We hope this
paper has provided convincing examples of this.

\section*{Acknowledgments}
Thanks are due to S.\ Cassisi, M.\ Groenewegen, P.\ Marigo, 
and A.\ Weiss, for their useful comments and suggestions.
We acknowledge the referee, B.\ Pagel, for the suggestions that
improved the final presentation of this paper.
X.\ Hernandez, M.\ Moll\'a, and H.\ Rocha-Pinto kindly provided us 
their results in computer-readable form. We gratefully acknowledge 
the hospitality of MPA during several visits.


\label{lastpage}

\end{document}